\documentclass[twocolumn,tighten]{aastex631}

\accepted{November 23, 2023}

\graphicspath{{./}{}}
\DeclareUnicodeCharacter{2212}{-}
\usepackage{multirow}
\usepackage{csvsimple,longtable,booktabs}
\usepackage{amsmath}
\begin{document}
% \turnoffedit
% \protect\renewcommand{\edit1[1]}{}
\renewcommand{\edit}[1]{}

\submitjournal{AJ}
\title{A Two-Component Probability Distribution Function Describes the mid-IR Emission from the Disks of Star-Forming Galaxies}

\shorttitle{Mid-IR PDFs of Galactic Disks}
\shortauthors{Pathak et al.}
\correspondingauthor{Debosmita Pathak}
\email{pathak.89@buckeyemail.osu.edu}

% Please add your name (and affiliation) here if you would like to join as a co-author! Thank you in advance :)

% -------------------

\author[0000-0003-2721-487X]{Debosmita Pathak}
\affiliation{Department of Astronomy, Ohio State University, 180 W. 18th Ave, Columbus, Ohio 43210}

\author[0000-0002-2545-1700]{Adam K. Leroy}
\affiliation{Department of Astronomy, Ohio State University, 180 W. 18th Ave, Columbus, Ohio 43210}
\affiliation{Center for Cosmology and Astroparticle Physics, 191 West Woodruff Avenue, Columbus, OH 43210, USA}

\author[0000-0003-2377-9574]{Todd A. Thompson}
\affiliation{Department of Astronomy, Ohio State University, 180 W. 18th Ave, Columbus, Ohio 43210}
\affiliation{Center for Cosmology and Astroparticle Physics, 191 West Woodruff Avenue, Columbus, OH 43210, USA}
\affiliation{Department of Physics, Ohio State University, 91 West Woodruff Ave
Columbus, Ohio 43210}

\author[0000-0002-1790-3148]{Laura A. Lopez}
\affiliation{Department of Astronomy, Ohio State University, 180 W. 18th Ave, Columbus, Ohio 43210}
\affiliation{Center for Cosmology and Astroparticle Physics, 191 West Woodruff Avenue, Columbus, OH 43210, USA}
\affiliation{Flatiron Institute, Center for Computational Astrophysics, NY 10010, USA}

% --- 1 ---

\author[0000-0002-2545-5752]{Francesco Belfiore}
\affiliation{INAF -- Osservatorio Astrofisico di Arcetri, Largo E. Fermi 5, I-50157, Firenze, Italy}

\author[0000-0003-0946-6176]{Médéric Boquien}
\affiliation{Instituto de Alta Investigación, Universidad de Tarapacá, Casilla 7D, Arica, Chile}

\author[0000-0002-5782-9093]{Daniel~A.~Dale}
\affiliation{Department of Physics and Astronomy, University of Wyoming, Laramie, WY 82071, USA}

\author{Simon C. O. Glover}
\affiliation{Universit\"{a}t Heidelberg, Zentrum f\"{u}r Astronomie, Institut f\"{u}r Theoretische Astrophysik, Albert-Ueberle-Stra{\ss}e 2,69120 Heidelberg, Germany}

\author[0000-0002-0560-3172]{Ralf S.\ Klessen}
\affiliation{Universit\"{a}t Heidelberg, Zentrum f\"{u}r Astronomie, Institut f\"{u}r Theoretische Astrophysik, Albert-Ueberle-Stra{\ss}e 2,69120 Heidelberg, Germany}
\affiliation{Universit\"{a}t Heidelberg, Interdisziplin\"{a}res Zentrum f\"{u}r Wissenschaftliches Rechnen, Im Neuenheimer Feld 205, 69120 Heidelberg, Germany}

\author[0000-0001-9605-780X]{Eric W. Koch}
\affiliation{Center for Astrophysics $\mid$ Harvard \& Smithsonian, 60 Garden St., 02138 Cambridge, MA, USA}

\author[0000-0002-5204-2259]{Erik~Rosolowsky}
\affiliation{Department of Physics, University of Alberta, Edmonton, AB T6G 2E1, Canada}

\author[0000-0002-4378-8534]{Karin M. Sandstrom}
\affiliation{Department of Astronomy \& Astrophysics, University of California, San Diego, 9500 Gilman Drive, La Jolla, CA 92093}

\author[0000-0002-3933-7677]{Eva Schinnerer}
\affiliation{Max Planck Institute for Astronomy, K\"{o}nigstuhl 17, D-69117, Germany}

% Rowan Smith (University of St Andrews)
\author[0000-0002-0820-1814]{Rowan Smith}
\affiliation{Jodrell Bank Centre for Astrophysics, Department of Physics and Astronomy, University of Manchester, Oxford Road, Manchester M13 9PL, UK}

\author[0000-0003-0378-4667]{Jiayi~Sun}
\affiliation{Department of Physics and Astronomy, McMaster University, 1280 Main Street West, Hamilton, ON L8S 4M1, Canada}
\affiliation{Canadian Institute for Theoretical Astrophysics (CITA), University of Toronto, 60 St George Street, Toronto, ON M5S 3H8, Canada}

\author[0000-0002-9183-8102]{Jessica Sutter}
\affiliation{Department of Astronomy \& Astrophysics, University of California, San Diego, 9500 Gilman Drive, La Jolla, CA 92093}

\author[0000-0002-0786-7307]{Thomas G. Williams}
\affiliation{Sub-department of Astrophysics, Department of Physics, University of Oxford, Keble Road, Oxford OX1 3RH, UK}

% --- 2 ---

\author[0000-0003-0166-9745]{Frank Bigiel}
\affiliation{Argelander-Institut f\"{u}r Astronomie, Universit\"{a}t Bonn, Auf dem H\"{u}gel 71, D-53121 Bonn, Germany}

\author[0000-0001-5301-1326]{Yixian Cao}
\affiliation{Max-Planck-Institut f\"{u}r Extraterrestrische Physik (MPE), Giessenbachstr. 1, D-85748 Garching, Germany}

\author[0000-0002-5235-5589]{J\'er\'emy Chastenet}
\affiliation{Sterrenkundig Observatorium, Ghent University, Krijgslaan 281-S9, 9000 Gent, Belgium}

\author[0000-0002-5635-5180]{M\'elanie Chevance}
\affiliation{Universit\"{a}t Heidelberg, Zentrum f\"{u}r Astronomie, Institut f\"{u}r Theoretische Astrophysik, Albert-Ueberle-Stra{\ss}e 2,69120 Heidelberg, Germany}
\affiliation{Cosmic Origins Of Life (COOL) Research DAO, coolresearch.io} 

\author[0000-0001-8241-7704]{Ryan Chown}
\affiliation{Department of Physics \& Astronomy, University of Western Ontario, London, ON N6A 3K7, Canada}

\author[0000-0002-6155-7166]{Eric Emsellem}
\affiliation{European Southern Observatory, Karl-Schwarzschild-Straße 2, 85748, Garching, Germany}
\affiliation{Univ Lyon, Univ Lyon1, ENS de Lyon, CNRS, Centre de Recherche Astrophysique de Lyon UMR5574, 69230, Saint-Genis-Laval, France}

\author[0000-0001-5310-467X]{Christopher M. Faesi}
\affiliation{Department of Physics, University of Connecticut, Storrs, CT, 06269, USA}

\author[0000-0003-3917-6460]{Kirsten L. Larson}
\affiliation{Space Telescope Science Institute, 3700 San Martin Drive, Baltimore, MD 21218, USA}

\author[0000-0003-0946-6176]{Janice C. Lee}
\affiliation{Space Telescope Science Institute, 3700 San Martin Drive, Baltimore, MD 21218, USA}

\author[0000-0002-6118-4048]{Sharon Meidt}
\affiliation{Sterrenkundig Observatorium, Universiteit Gent, Krijgslaan 281 S9, B-9000 Gent, Belgium}

\author[0000-0002-0509-9113]{Eve C. Ostriker}
\affiliation{Department of Astrophysical Sciences, Princeton University, Princeton, NJ 08544, USA}

\author[0000-0002-9190-9986]{Lise Ramambason}
\affiliation{Universit\"{a}t Heidelberg, Zentrum f\"{u}r Astronomie, Institut f\"{u}r Theoretische Astrophysik, Albert-Ueberle-Stra{\ss}e 2,69120 Heidelberg, Germany} 

\author[0000-0002-4781-7291]{Sumit K. Sarbadhicary}
\affiliation{Department of Astronomy, Ohio State University, 180 W. 18th Ave, Columbus, Ohio 43210}

\author[0000-0002-8528-7340]{David A. Thilker}
\affiliation{Department of Physics and Astronomy, The Johns Hopkins University, Baltimore, MD 21218, USA}

\suppressAffiliations

% \author[0000-0003-2721-487X]{Debosmita Pathak}
% \affiliation{OSU}

% \author{Adam K. Leroy}
% \affiliation{OSU}

% \author[0000-0003-2377-9574]{Todd A. Thompson}
% \affiliation{OSU}

% \author{Laura A.  Lopez}
% \affiliation{OSU}

% \author{+PHANGS}
% \affiliation{PHANGS}

\begin{abstract}
% s2ksumit: Check the word count. I think its generally 250 word max.

% Abstract here.
High-resolution JWST-MIRI images of nearby spiral galaxies reveal emission with complex substructures that trace dust heated both by massive young stars and the diffuse interstellar radiation field. We present high angular ($0\farcs85$) and physical resolution ($20{-}80$~pc) measurements of the probability distribution function (PDF) of mid-infrared (mid-IR) emission (7.7--2~$\mu$m) from $19$ nearby star-forming galaxies from the PHANGS-JWST Cycle-1 Treasury. The PDFs of mid-IR emission from the disks of all 19 galaxies consistently show two distinct components: an approximately log-normal distribution at lower intensities and a high-intensity power-law component. These two components only emerge once individual star-forming regions are resolved. Comparing with locations of \ion{H}{2} regions identified from VLT/MUSE H$\alpha$-mapping, we infer that the power-law component arises from star-forming regions and thus primarily traces dust heated by young stars. In the continuum-dominated 21~$\mu$m band, the power-law is more prominent and contains roughly half of the total flux. At 7.7--11.3~$\mu$m, the power-law is suppressed by the destruction of small grains (including PAHs) close to \ion{H}{2} regions while the log-normal component tracing the dust column in diffuse regions appears more prominent. The width and shape of the log-normal diffuse emission PDFs in galactic disks remain consistent across our sample, implying a log-normal gas column density $N$(H)$\approx10^{21}$cm$^{-2}$ shaped by supersonic turbulence with typical (isothermal) turbulent Mach numbers $\approx5-15$. Finally, we describe how the PDFs of galactic disks are assembled from dusty \ion{H}{2} regions and diffuse gas, and discuss how the measured PDF parameters correlate with global properties such as star-formation rate and gas surface density.

%Meanwhile the log normal component traces diffuse emission that fills the galaxies outside the star-forming regions, and we argue that it reflects the underlying distribution of dust column densities heated by a roughly constant interstellar radiation field. 
%We show that the mean of the log normal distribution varies systematically with radius and from region to region across our sample.
%for the underlying distribution of column density and turbulence in the ISM, and radiation pressure in \ion{H}{2} regions.

% - The PDFs of individual regions show complex shapes due to substructure and high resolution, but overall shape of PDFs show consistency.

% - The intensity of \ion{H}{2} regions remain roughly constant with radius, while the intensity of diffuse regions drives down the overall intensity in the outskirts. 

% - We also note differences in emission among the 4 bands

\end{abstract}

%--- Polycyclic aromatic hydrocarbons(1280) --- Dust continuum emission(412)

\section{Introduction} \label{sec:introduction}

The first high resolution mid-infrared (mid-IR) images of galaxies from JWST \citep[e.g.,][]{2022PONTOPPIDAN,2023LEE} show a stunningly detailed, and very complex, picture of the dusty interstellar medium (ISM).
The abundance of substructures in these images captures both the ``diffuse'' turbulent phase of the ISM as well as individual knots of young star-forming regions in galaxy centers and along the spiral arms. Characterizing the overall emission in the mid-IR from the dusty ISM at high physical resolution (20--80~pc), necessary to resolve individual star-forming regions from the surrounding ISM, is now possible for the first time outside the Local Group, and across the star-forming main sequence, using JWST. This paper presents the first high angular ($0\farcs269-0\farcs85$) and physical resolution ($20-80$~pc) measurements of the probability distribution function (PDF) of mid-IR emission (7.7--21~$\mu$m) from $19$ nearby star-forming galaxies, providing useful insights into the complex origins of the mid-IR emission in galaxies.

Most of the mid-IR emission from galaxies is produced from small dust grains, including polycyclic aromatic hydrocarbons (PAHs), with increasing contribution from larger dust grains at longer wavelengths \citep{2011DRAINE, 2018GALLIANO}. The physical origin of this emission is primarily from dust reprocessing of starlight from the UV and optical into the IR. At shorter wavelengths in the mid-IR, the stochastic heating of small grains leads to strong emission features from different vibrational modes of PAH molecules (especially in two of the JWST-MIRI filters we use: F770W and F1130W at 7.7 and 11.3~$\mu$m, respectively). %In very intense radiation fields, such as close to \ion{H}{2} regions, small grains are in thermodynamic equilibrium 
In addition, dust grains in thermodynamic equilibrium with very intense radiation fields, such as those close to \ion{H}{2} regions, can emit modified blackbody spectra. This contributes to continuum dust emission at longer wavelengths beyond 20$~\mu$m \citep[see for example][]{2007DRAINELI,2018GALLIANO}. In low-density regions far from strong heating sources, longer mid-IR wavelengths such as 21$~\mu$m also primarily trace stochastic heating of dust grains \citep{2001DRAINE&LI,2001LI&DRAINE}.

% - What information does the mid-IR emission from galaxies trace? -- starlight reprocessed into the IR by dust. mid-IR primarily traces (1) Stochastic heating of small dust grains at shorter wavelengths and (2) thermal emission from bigger grains at longer wavelengths. 
% - What is stochastic heating? What factors does it depend on? Why are PAHs important?
% - What is thermal emission? What factors does it depend on? Why are larger grains important?
% Ideally the above goes in one paragraph then you have a transition here: "

Since the processes above are sensitive to both the amount of dust as well as the strength of the radiation field incident on the dust, the distribution of mid-IR emission from galaxies should reflect both the underlying structure of the ISM and the location of strong heating sources \citep[see for example][]{2023SANDSTROM}. 
% - The mid-IR traces both the amount of dust as well as the strength of the ambient radiation field doing the heating (stars)
Because of its complex origin, the mid-IR has been used as both a star-formation rate tracer and a gas tracer \citep{2004PEETERS,2005CALZETTI,2007ARMUS,2021CHOWN}. Before JWST, distinguishing emission from regions dominated by recent star formation and the surrounding medium was challenging \citep{2015HUNT,2010LI} and was mainly possible in the Milky Way and the Local Group \edit1{(see for example, for galactic observations: \citealt{2009CHURCHWELL,2018BINDER}; Magellanic Clouds: \citealt{2009PARADIS,2010LAWTON}; M33: \citealt{2009RELANO,2018RELANO})}.

JWST at last offers the resolution to distinguish star-forming regions from the diffuse ISM in normal star forming galaxies beyond the Local Group ($d\sim10$ Mpc).
% - What previous observations tell us about dust in galaxies: (1) from Milky Way/Local Group observations, (2) from observations of nearby galaxies, (3) from observations of higher-z galaxies (much more difficult to resolve)
% - Why study nearby galaxies?
Studying nearby galaxies at such high resolution provides a global view of entire systems that is not possible in the Milky Way, while still being able to well resolve star-forming regions and the turbulent ISM. With the unprecedented resolving power that JWST provides in the mid-IR, we can study star-formation and baryonic feedback in the context of both its local and global environment. 
% - Our work extends this to nearby galaxies imaged at high resolution using dust emission in the mid-IR as a tracer of the gas in the ISM. We compare our new estimates of ISM properties with previous estimates of ISM properties from independent tracers of different ISM phases.

% some background on PDFs
% thank you Ralf!! :)
Measuring probability distribution functions (PDFs) of gas (column) density using multi-wavelength tracers is key to making a statistical characterization of different phases of the ISM.
Due to the availability of high-resolution data, most observational studies of the structure and distribution of different phases of the ISM using PDFs of the gas (column) density of various tracers have focused on the Milky Way and other Local Group galaxies (e.g. \citet{2008BERKHUIJSEN}, \citet{2015BERKHUIJSEN}, \citet{2016IMARA}, \citet{2018CORBELLI} and references therein for a galaxy-scale analysis, similarly \citet{2014KAINULAINEN} and \citet{2022SCHNEIDER} for nearby molecular clouds). This has been complemented by high-resolution numerical simulations which establish a clear physical relation between the properties of ISM turbulence and the width and shape of the density PDF \citep[see e.g.][]{2008FEDERRATH,2012MOLINA,2012HENNEBELLE,2014PADOAN,2016KLESSEN}. 

% - Talk about SF regions
To first order, the shape of the density PDF is set by gravity and turbulence \citep[see for example][]{2000aKLESSEN,2000bKLESSEN,2013FEDERRATH,2018BURKHART}. 
In addition, within young star-forming regions and other active parts of the galaxy, stellar feedback such as outflows can help shape the PDFs of emission \citep{2022APPEL}.
The PDFs of individual giant molecular clouds (GMCs) in extinction appear primarily log normal with a power-law, which mainly reflects the superposition of gas column distributions from both turbulent bulk gas and self-gravitating cores \citep[e.g. see][]{2018CHEN}. However, for PDFs of IR emission on large (galaxy) scales, a log normal and power-law shape is expected to be a convolution of the distribution of radiation intensity with the distribution of gas column \citep[see][for galactic IR SEDs and dust models]{2001DALE, 2007DRAINELI}. Both dust extinction and IR emission PDFs have been widely studied as tracers of gas (hydrogen) column density at very high resolution within a galactic context \citep[e.g ][and references therein]{2009GOODMAN,2017LADA,2022SCHNEIDER}. However, high resolution IR observations of external galaxies are necessary to extend this framework to galaxies with a diversity of global properties beyond the Local Group.
%Mid-IR emission PDFs on galaxy-scales are thus expected to reflect a convolution of turbulent gas column in the ISM and the luminosity and distribution of resolved structures in star-forming regions. 

%
% somewhere in here after you describe the advance you need a clear paragraph statement on what you do - 
In this paper, we measure the PDFs of mid-IR intensity in four JWST bands: 7.7, 10, 11.3, and 21~$\mu$m (F770W, F1000W, F1130W, and F2100W respectively). All four bands are expected to be dominated by dust emission, with PAH features contributing most of the emission at 7.7 and 11.3~$\mu$m, while the thermal dust continuum dominates at 10 and 21~$\mu$m \citep{2023WHITCOMB_RNAAS}. Our goal is to make a statistical characterization of the mid-IR emission from external galaxies when observed at high enough resolution to distinguish individual star-forming regions from the surrounding ISM. We do this by parameterizing the PDFs of mid-IR intensity and corresponding gas (column) density in the first 19 nearby galaxies from the PHANGS-JWST Cycle-1 Treasury survey that span a range of masses and star-formation rates across the star-forming main sequence of galaxies. 
% - What do we do with the PDFs
% Using our PDFs of intensity and derived quantities, we interpret the emission in the mid-IR in the context of global averaged gas column densities, sound speed in the turbulent ISM, radiation pressure and bolometric luminosity surface densities in young star-forming regions in nearby star-forming galaxies. Finally, we perform preliminary correlations with galaxy-averaged properties to set the framework for future work that will look at each of these correlations in a more local context within each galaxy. 
The future availability of 55 more galaxies from the PHANGS-JWST Cycle-2 GO Program (GO-3707) will aid more robust characterization of mid-IR PDFs and correlations with global properties of an unprecedented large sample of galaxies.

The paper proceeds as follows. In Section \ref{sec:data}, we describe the PHANGS-JWST Cycle-1 Treasury sample, mid-IR maps, as well as supporting VLT/MUSE data products (\ion{H}{2} region masks and H$\alpha$ coverage) that we use. In Section \ref{sec:PDFs}, we measure mid-IR intensity PDFs of each galaxy at each of the four filters. Section \ref{sec:makePDFs} describes how the overall PDFs are constructed, Section \ref{sec:centers} presents the PDFs of galactic centers, and Section \ref{sec:extended_disks} shows the PDFs of galactic disks after decoupling from the galactic centers. Section \ref{sec:HII_nonHII} decomposes the emission from the galactic disks into star-forming (\ion{H}{2}) regions and diffuse \footnote{Here, we use the label `diffuse' to refer to all gas outside of star-forming regions (\ion{H}{2} regions), which includes some contribution from molecular, cold neutral, and warm neutral gas in the ISM. Note that this is different from the typical definition of diffuse gas in a Milky Way context ($N(\rm{H})\lesssim 10^9 \rm{cm}^{-2}$). So while our label of `diffuse' mostly captures truly diffuse material, it likely also includes some bound/self-gravitating gas. For the purposes of the analysis included in this paper, distinguishing star-forming regions from the surrounding material is sufficient.} ISM and Section \ref{sec:characterizing_HII_diffuse} parameterizes each PDF component; Section \ref{sec:3ComponentContrast} compares and contrasts the mid-IR emission from the galactic centers, disk \ion{H}{2} regions, and diffuse ISM in the disk. Section \ref{sec:resolution} discusses the effect of changes in resolution on the PDFs; Section \ref{sec:globalprops} performs correlations between PDF parameters and galaxy-averaged properties of each target. In Section \ref{sec:HII} we focus on the mid-IR emission from the \ion{H}{2} region component, where Section \ref{sec:F2100W_HII_slope} discusses the power-law at 21$\mu$m, and Section \ref{sec:hii_luminosity_surface_dens} estimates the bolometric luminosity surface density and radiation pressure in star-forming regions. In Section \ref{sec:non-HII} we focus on the diffuse component, where Section \ref{sec:columndensity} measures high-resolution gas column density PDFs at native F770W resolution ($0\farcs269$), and Section \ref{sec:machnumbers} derives upper bounds for Mach numbers of isothermal turbulence from gas column PDFs in the diffuse ISM. Finally, Section \ref{sec:summary} summarizes our key results and conclusions.

% \newpage
% \clearpage
% \begin{afterpage}
% \startlongtable
\begin{deluxetable*}{lrrrrrcrccccc}[th!]
\tabletypesize{\small}
\tablecaption{Target specifications and global properties \label{tab:sample}}
\tablewidth{1\textwidth}

\tablehead{
\colhead{Galaxy} & \colhead{R.A.} &     \colhead{Dec.} &        \colhead{$i^a$} & \colhead{$D^b$} & \colhead{res.$^c$} &  \colhead{$A^c$}  & \colhead{$\log M_*^b$} & \colhead{$\log$ SFR$^b$} & \colhead{AGN$^d$} & \colhead{Type$^e$} &      \colhead{Bar$^e$}  & \colhead{Center$^f$}
}

\startdata
                  &  [deg]  &  [deg]   &  [deg]    &  [Mpc]    &  [pc]        & $10^4~$arcsec$^2$ &[$\log M_\odot$] & [$\log M_\odot$/yr] &  &  &  &  [$<c$ kpc] \\ \hline
NGC0628           &   24.17 &    15.78 &       8.9 &       9.8 &         40.5 &  1.6  &  10.3 &     0.2 &    - &   \edit1{SA(s)c                     }   &          - &       - \\
NGC1087           &   41.60 &    -0.50 &      42.9 &      15.9 &         65.3 &  2.7  &   9.9 &     0.1 &    - &   \edit1{SB(rs)c\_d\_pec            }   & \checkmark &       - \\
NGC1300           &   49.92 &   -19.41 &      31.8 &      19.0 &         78.3 &  1.7  &  10.6 &     0.1 &    - &   \edit1{(R')SB(s\_bl\_nrl)b        }   & \checkmark &     1.0 \\
NGC1365           &   53.40 &   -36.14 &      55.4 &      19.6 &         80.6 &  3.3  &  11.0 &     1.2 & S1.8 &   \edit1{(R')SB(r\_s\_nr)bc         }   & \checkmark &     3.0 \\
NGC1385           &   54.37 &   -24.50 &      44.0 &      17.2 &         71.0 &  3.2  &  10.0 &     0.3 &    - &   \edit1{SB(s)dm\_pec               }   & \checkmark &     3.0 \\
NGC1433           &   55.51 &   -47.22 &      28.6 &      18.6 &         76.8 &  1.1  &  10.9 &     0.1 &    - &   \edit1{(R\_1\_')SB(r\_p\_nrl\_nb)a}   & \checkmark &     1.0 \\
NGC1512           &   60.98 &   -43.35 &      42.5 &      18.8 &         77.6 &  3.3  &  10.7 &     0.1 &    - &   \edit1{(R\_L)SB(r\_bl\_nr)a       }   & \checkmark &     1.0 \\
NGC1566           &   65.00 &   -54.94 &      29.5 &      17.7 &         72.9 &  2.8  &  10.8 &     0.7 & S1.5 &   \edit1{(R\_1\_')SAB(rs\_r\_s)b    }   & \checkmark &     1.0 \\
NGC1672           &   71.43 &   -59.25 &      42.6 &      19.4 &         79.9 &  2.6  &  10.7 &     0.9 &    S &   \edit1{(R')SA\_B(rs\_nr)b         }   & \checkmark &     1.0 \\
NGC2835           &  139.47 &   -22.35 &      41.3 &      12.2 &         50.4 &  2.0  &  10.0 &     0.1 &    - &   \edit1{S/IB(s)m, SB(rs)c          }   & \checkmark &       - \\
NGC3351           &  160.99 &    11.70 &      45.1 &      10.0 &         41.0 &  1.9  &  10.4 &     0.1 &    - &   \edit1{(R')SB(r\_bl\_nr)a         }   & \checkmark &     1.0 \\
NGC3627           &  170.06 &    12.99 &      57.3 &      11.3 &         46.6 &  2.7  &  10.8 &     0.6 &   S3 &   \edit1{SB\_x\_(s)b\_pec           }   & \checkmark &     1.0 \\
NGC4254           &  184.71 &    14.42 &      34.4 &      13.1 &         54.0 &  2.3  &  10.4 &     0.5 &    - &   \edit1{SA(s)c\_pec                }   &          - &     3.0 \\
NGC4303           &  185.48 &     4.47 &      23.5 &      17.0 &         70.0 &  2.7  &  10.5 &     0.7 &   S2 &   \edit1{SAB(rs\_nl)b\_c            }   & \checkmark &     1.0 \\
NGC4321           &  185.73 &    15.82 &      38.5 &      15.2 &         62.7 &  2.7  &  10.8 &     0.6 &    - &   \edit1{SAB(rs\_nr\_nb)bc          }   & \checkmark &     1.0 \\
NGC4535           &  188.58 &     8.20 &      44.7 &      15.8 &         65.0 &  3.2  &  10.5 &     0.3 &    - &   \edit1{SAB(s)c                    }   & \checkmark &     1.0 \\
NGC5068$^\dagger$ &  199.73 &   -21.04 &      35.7 &       5.2 &         21.4 &  1.7  &   9.4 &    -0.6 &    - &   \edit1{SB(s)d                     }   & \checkmark &       - \\
NGC7496           &  347.45 &   -43.43 &      35.9 &      18.7 &         77.1 &  2.6  &  10.0 &     0.3 &   S2 &   \edit1{(R')SB\_x\_(rs)b           }   & \checkmark &     1.0 \\
IC5332$^\dagger$  &  353.61 &   -36.10 &      26.9 &       9.0 &         37.1 &  1.1  &   9.7 &    -0.4 &    - &   \edit1{S\_AB(s)cd                 }   & \checkmark &       - \\ \hline
\enddata
\tablecomments{$^a$ Galaxy inclinations adopted from \citet{2020LANG} and \citet{2021LEROY}.\\
$^b$ Galaxy distances adopted from \citet{2021ANAND}, and stellar masses and star formation rates adopted from \citet{2021LEROY} \edit1{and assume a \citet{2003CHABRIER} initial mass function (IMF).}\\
$^c$ The physical resolution for each galaxy corresponding to the presented common convolved angular resolution of 0.85", and the total area of the JWST-MIRI footprint with VLT-MUSE coverage over which analysis is performed for each target.\\
$^d$ AGN type, if present, for each galaxy from \citet{2010VERONcat}.\\
$^e$ \edit1{Comprehensive de Vaucouleurs revised Hubble-Sandage (CVRHS) types and flags noting the presence or absence of a morphologically and kinematically identified bar respectively in each galaxy \citep{1991DEVAUCOULEURS,2014MAKAROV,2015BUTA}.}\\
$^f$ Morphologically identified central regions used in this paper to measure PDFs for each galaxy, for the purposes of separating the central emission from the disk. This distinction is only when the center appears morphologically distinct from the disk in the MIR.\\
$^\dagger$ Dwarf galaxies in our sample.
}
\end{deluxetable*}
% \end{afterpage}

%   Sc
%   Sc
%  Sbc
%   Sb
%   Sc
%  SBa
%   Sa
% SABb
%   Sb
%   Sc
%   Sb
%   Sb
%   Sc
%  Sbc
% SABb
%   Sc
%   Sc
%   Sb
% SABc

% SA(s)c                                     
% SB(rs)c\_d\_pec                                     
% (R')SB(s\_bl\_nrl)b                                     
% (R')SB(r\_s\_nr)bc                                     
% SB(s)dm\_pec                                     
% (R\_1\_')SB(r\_p\_nrl\_nb)a                                     
% (R\_L)SB(r\_bl\_nr)a                                     
% (R\_1\_')SAB(rs\_r\_s)b                                     
% (R')SA\_B(rs\_nr)b                                     
% S/IB(s)m, SB(rs)c                                     
% (R')SB(r\_bl\_nr)a                                     
% SB\_x\_(s)b\_pec                                     
% SA(s)c\_pec                                     
% SAB(rs\_nl)b\_c                                     
% SAB(rs\_nr\_nb)bc                                     
% SAB(s)c                                     
% SB(s)d                                     
% (R')SB\_x\_(rs)b                                    
% S\_AB(s)cd                                      

\section{Data and Methods} \label{sec:data}

\subsection{PHANGS-JWST} \label{sec:data_PHANGSJWST}

We measure the distributions of mid-IR intensity for 19 nearby star-forming spiral galaxies observed by JWST as part of the PHANGS-JWST Cycle 1 Treasury program \citep[GO 2107, PI Lee;][]{2023LEE}. This program targets the same sample of galaxies observed by the PHANGS-MUSE program \citep{2022EMSELLEM}. Table \ref{tab:sample} summarizes the properties of our targets, which span a range of stellar masses (from $2.5 \times 10^9 M_\odot$ to $9.8 \times 10^{10} M_\odot$) and star-formation rates (from 0.3 to 16.9~$M_\odot~\mathrm{yr}^{-1}$) on the star-forming main sequence and capture a wide diversity in morphologies and distinct local physical environments. %This aligns with the original PHANGS-MUSE sample, from which the 19 PHANGS-JWST Cycle-1 galaxies were drawn \citep{2022EMSELLEM}.
The sample mostly consists of massive spiral galaxies, including 17 barred galaxies, 6 Seyfert galaxies, and two lower-mass dwarf spirals.  %This range of targets thus probe a wide range of galactic environments and morphology. 
% - Describe which wavelengths we look at
For each galaxy, we use JWST-MIRI data from four mid-IR filters F770W, F1000W, F1130W, and F2100W, centered respectively at 7.7, 10, 11.3, and~21 $\mu$m. %Together these capture continuum emission from small grains (F1000W and F2100W) and two of the brightest PAH bands (F770W and F1130W, F1000W to some extent). 
Strong PAH emission complexes peak at 7.7 and 11.3~$\mu$m, while the emission at 21~$\mu$m is expected to better trace the dust continuum emission \citep{2007SMITH, 2007DRAINELI}. The 10~$\mu$m band is known for silicate absorption \citep{2005HAO,2007SPOON} and shows some intermediate behavior between PAH-tracing bands and the continuum, as noted later in Section~\ref{sec:3ComponentContrast}. 

\citet{2023LEE} describe the observations, reduction, and processing of the JWST-MIRI data, with updates to the processing and data validation described in \edit1{T. Williams et al. (in preparation)}. Relevant to this paper, the MIRI data are run through the standard JWST pipeline calibration, which is expected to yield an intensity scale accurate to better than $5\%$ \citep{2023RIGBY}. The MIRI images are astrometrically aligned with corresponding NIRCam observations of the PHANGS-JWST targets using bright point sources, and the NIRCam observations are in turn registered against Gaia DR3 astrometry. As a result, we expect the overall astrometric accuracy of the observations to be significantly better than the resolution at F2100W, our lowest resolution band. The MIRI images are well-aligned with the VLT/MUSE comparison data, which have astrometry derived indirectly through broad-band imaging alignment \citep{2022EMSELLEM}. During data reduction, each MIRI frame is flux-corrected using an off-galaxy observation and pointings toward different fields are mosaicked together by the pipeline. Because of the lack of large signal-free regions near the galaxy, the overall background level for each image is set using a simultaneously recorded off-axis MIRI background frame and previous wide field mid-IR imaging, either from WISE or \textit{Spitzer}. After this procedure, the overall background level is uncertain to better than $\pm 0.05$~MJy~sr$^{-1}$ \citep[for more details see][and T. Williams et al. in preparation]{2023LEROY,2023LEE}.

\subsection{Convolution to Matched Resolution} \label{sec:data_Convolution}

In order to compare the emission across different wavelengths, we convolve all four filters to a Gaussian PSF following \citet{2011ANIANO} to share a common angular resolution of $0\farcs85$. We use this slightly coarser resolution than the native F2100W ($0\farcs674$), which would be the sharpest possible common resolution, in order to gain signal-to-noise.
Because F2100W is our lowest-resolution and noisiest band, and since the noise level drops rapidly with smoothing, this small convolution yields a significant improvement in the signal-to-noise in this band. To do this, we use PSFs appropriate for each filter generated following \citet{2014PERRIN} and convolve following the approach described in \citet{2011ANIANO}. 
%\footnote{Code available at \url{https://github.com/PhangsTeam/jwst_scripts}}. 
Targeting this fixed angular resolution results in slightly different physical resolutions depending on the distance of each target (20--80 pc), as included in Table \ref{tab:sample}. As we describe in Section \ref{sec:resolution}, the results at the native resolution of each band largely resemble those derived at our common resolution of $0\farcs85$.

Several of our targets include some ``empty'' galaxy-free regions in the image, which we use to estimate the statistical noise in the data at our common resolution. At our working resolution of $0\farcs85$, the maps show typical noise values of 0.03, 0.04, 0.05, 0.08~MJy~sr$^{-1}$ at F770W, F1000W, F1130W, and F2100W, respectively (for more details see T. Williams et al. in preparation). Because all galaxies have the same exposure time and observing procedure, we expect, and mostly observe, the statistical (random) noise to be relatively uniform (i.e., we find similar dispersion in ``empty'' galaxy-free regions) across different galaxies. In detail, the noise varies slightly across each image due to the varying detector response, the dither pattern, and overlap among tiles, but these values still provide a good point of reference. 

% previously some text on the centers - now moved to its own section downstairs

\subsection{Decomposing Galaxies into Centers, Disks, and Star Forming Regions} \label{sec:data_decomposing}

In the following sections, we construct and analyze the probability density functions of emission for each target. We systematically decompose each galaxy into three components. First, we identify galactic centers that show extreme mid-IR emission \edit1{characterized by coherent peaks in emission (e.g., from unresolved point sources, saturation artifacts, etc.) from the central 1-3 kpc of galaxies that contribute to higher order peaks at least 1 dex brighter the mean intensity of the overall PDF}. The radial extent of nuclear structures present in each target are included in Table \ref{tab:sample}, and \edit1{their selection is discussed in detail} in Section \ref{sec:centers}.

After removing the extreme central regions, \edit1{we separate the galactic disks into} star-forming and diffuse regions using masks that identify the locations and footprint of \ion{H}{2} regions based on the observed H$\alpha$ emission observed by VLT/MUSE. The MUSE data are described by \citet{2022EMSELLEM} and the \ion{H}{2} region catalog creation and mask constructions are described in \citet[][and see also \citealt{2022SANTORO}]{2023GROVES}. 
\edit1{In brief, the \ion{H}{2} region masks are derived from H$\alpha$ emission from the PHANGS–MUSE survey which includes corrections for internal extinction using the Balmer decrement method that contrasts H$\alpha$ and H$\beta$. A slightly altered version of HIIphot as first described in \citet{2019KRECKEL} is used to identify and characterize \ion{H}{2} regions with irregular morphology with a detection threshold in $\Sigma_{H\alpha}$ at least $3\sigma$ above $1 \times 10^7 \mathrm{erg} ~ \mathrm{s}^{-1} ~ \mathrm{cm}^{-2} ~ \mathrm{arcsec}^{-2}$ \citep{2023GROVES}.}
This work uses \ion{H}{2} region masks derived from ``convolved and optimized'' versions of these maps from the PHANGS–MUSE data release.
%internal data release version 2.2. %that have been described in \textcolor{red}{Belfiore et al. (2022a,b) and Pessa et al. (2021, 2022)}, reprojected onto the field of view for each galaxy mosaic.
In these masks, individual \ion{H}{2} regions are classified using emission line diagnostics, and each pixel is identified as either belonging to an \ion{H}{2} region or not \citep{2023GROVES}, which we use to distinguish between diffuse and star-forming \ion{H}{2} regions in our analysis. The angular resolution of the convolved and optimized maps span from $0\farcs56{-}1\farcs25$, median $0\farcs92$, almost the same as that of our convolved MIRI data. Therefore, we treat the masks as effectively matched in resolution to our observations and apply them as is with no further processing.

\section{PDFs of Mid-IR Emission in Galaxies} \label{sec:PDFs}

\subsection{Constructing PDFs of Emission} \label{sec:makePDFs}

\begin{figure*}[htbp]
    \centering
    \includegraphics[width=1\textwidth]{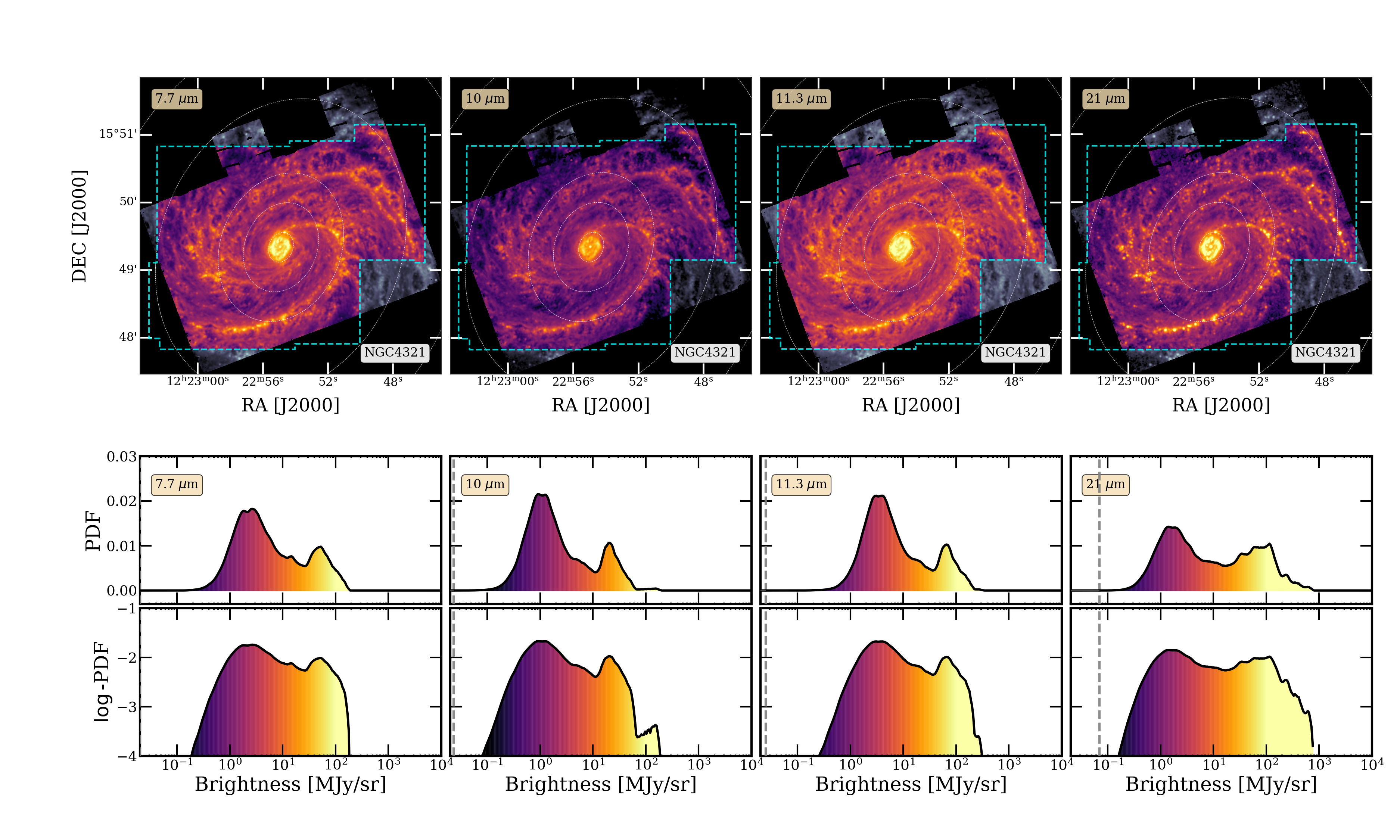}
    \caption{Images and PDFs of emission at 7.7~$\mu m$, 10~$\mu m$, 11.3~$\mu m$, and 21~$\mu m$ for NGC4321 at $0\farcs85$ (60 pc) resolution. \textbf{Top:} JWST images for NGC4321 in each MIRI filter, restricted to the footprint of JWST-MUSE coverage ($\sim 10$ kpc, dashed blue). \textbf{Middle:} PDFs of inclination-corrected intensity from each image. The colors in the PDF map to the colorbar in intensity used to display each image. Vertical dashed lines indicate the RMS noise level in each filter at $0\farcs85$. \textbf{Bottom:} The same PDFs as the middle row, now with the y-axis on a logarithmic stretch to highlight features at the high-intensity end.}
    \label{fig:overall_ngc4321}
\end{figure*}

% - How were the PDFs constructed?
We characterize the emission from each galaxy in each filter by constructing and parameterizing the probability distribution functions (PDFs) of inclination-corrected intensities at each pixel measured in units of MJy~sr$^{-1}$.
%These are analogous to PDFs of luminosity, modulo the solid angle and distance to each target. Figure \ref{fig:overall_ngc4321} shows the PDFs of intensity for an example galaxy from our sample, NGC4321. 
%Figure \ref{fig:overall_ngc4321} shows the PDFs of intensity for NGC4321 (as an example) at each filter. 
The PDFs at each wavelength $\lambda$ are constructed by first dividing the mid-IR maps into logarithmically spaced bins of intensity. We then sum the total intensity $I_{\lambda}$ (i.e., the map values) within each bin and normalize by the total intensity in the image. That is, for any intensity bin $k$, 
\begin{align} \label{eqn:xy} 
\displaystyle
   % x &= I_\lambda, \\
   \text{PDF} (I_{\lambda_k}) &= \frac{1}{\sum I_{\lambda}\cos{i}} \sum\limits_{I_{\lambda_k}-\Delta I_{\lambda}/2}^{I_{\lambda_k}+\Delta I_{\lambda}/2} I_\lambda \cos{i},
\end{align}
where the logarithmic bin step used to construct the histograms is typically chosen as $\Delta I_\lambda = 0.025$~dex, and $i$ is the inclination of the galaxy (see Table \ref{tab:sample}). The PDF is in units of the fraction of total intensity, so $\sum\limits_k I_{\lambda_k}\cos{i} = 1$. %The x-axis values are inclination-corrected intensities at each pixel measured in MJy~sr$^{-1}$. 
Figure \ref{fig:overall_ngc4321} shows the PDFs of intensity for an example galaxy from our sample, NGC4321. The PDFs show some variation in shape and intensity range across the four filters, which we discuss further in the following sections. The mid-IR images and overall PDFs of our full sample are included in Appendix \ref{sec:all_images}.

%Note that our PDFs express the distribution of luminosity as a function of intensity. 
Note that the intensity-weighted PDFs essentially multiply the un-weighted (area-weighted) PDF by the mean intensity captured in each bin.  
Hence, intensity-weighted PDFs better capture the contribution from regions that
%is essentially an intensity-weighed pixel-wise PDF, which allows us to trace high S/N regions giving 
give rise to most of the emission, compared to pixel-wise un-weighted PDFs. 
%in galaxies, and is 
% They are less sensitive than un-weighted PDFs to large areas of low-intensity emission. 
As a result they are also less affected by the presence of extended galaxy-free regions in the image, uncertainties in the %outskirts, 
background level, or contributions from low signal-to-noise (S/N) regions. We prefer this luminosity-weighted PDF because most of the emission in our maps comes from regions \edit1{brighter than 0.1 MJy sr$^{-1}$} that we detect at good S/N, as is evident from Figure \ref{fig:overall_ngc4321}. Transforming our results back into an area-weighted PDF is a simple algebraic operation (in the limit of very fine bins, simply divide our PDF by the x-axis value), and in that case, the low-intensity end will be uncertain. 

In the following sections, since our PDF intensities are detected at good S/N, we fit the binned PDF with different models.
Only for display, the PDFs shown in Figure \ref{fig:overall_ngc4321} and throughout this paper include a convolution with a smoothing kernel that averages over four adjacent intensity bins. However, all fit parameters reported throughout the paper (Table \ref{tab:decomposed}, etc.) use the original PDFs without the smoothing kernel.

% Suggesting to move the centers paragraph down into the centers subsection.

\subsection{Galactic Centers} \label{sec:centers}

\begin{figure*}[htbp]
    \centering
    \includegraphics[width=1\textwidth]{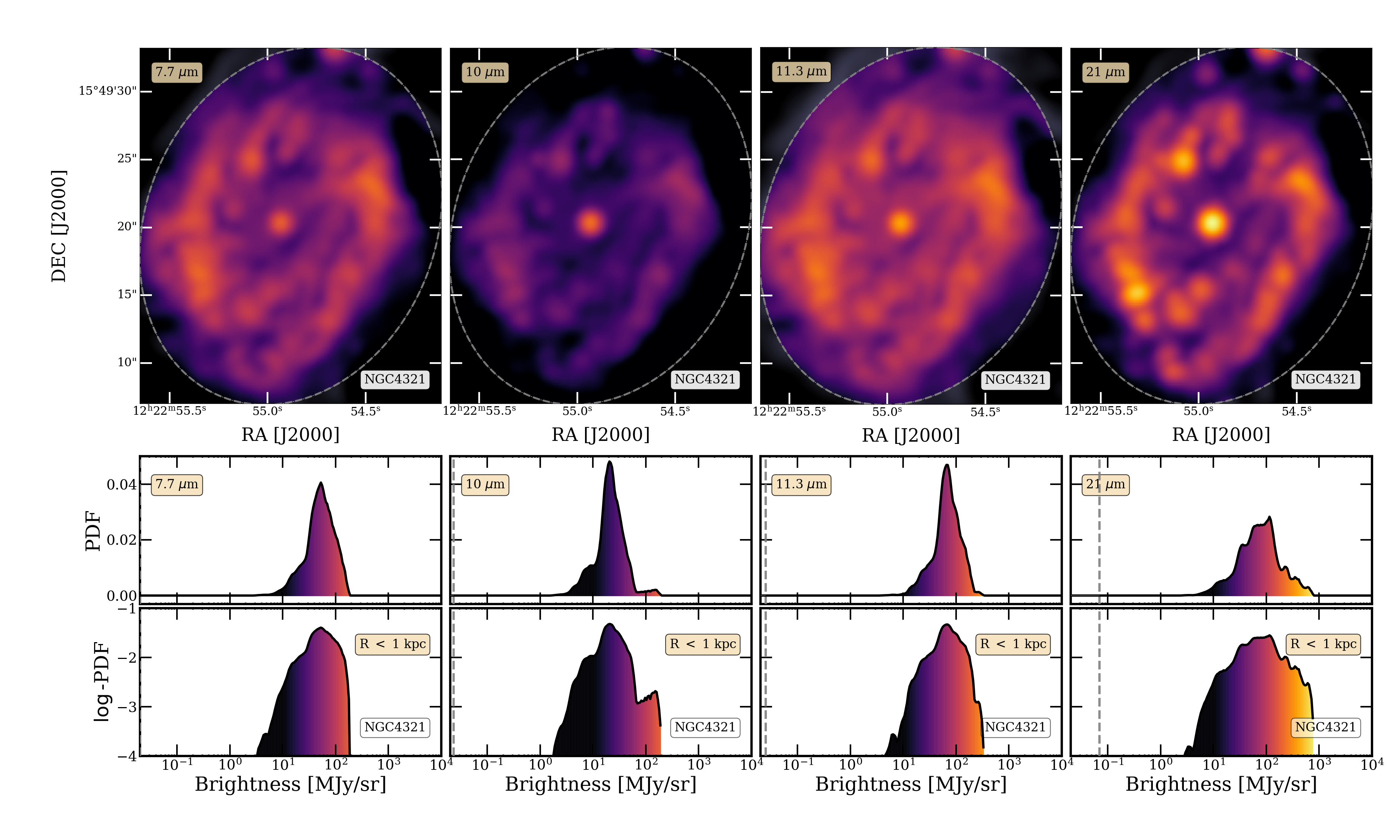}
    \caption{Images and PDFs of emission from the central 1 kpc at 7.7~$\mu m$, 10~$\mu m$, 11.3~$\mu m$, and 21~$\mu m$ for NGC4321. Similar to Figure \ref{fig:overall_ngc4321}, \textbf{Top:} Images of the central 1~kpc in NGC4321 in each MIRI filter. \textbf{Middle:} PDFs of inclination-corrected intensity from each image. The colors reflect the common colorbar in intensity used to display each image. \textbf{Bottom:} The same PDFs as the middle row, now with the y-axis on a logarithmic stretch to highlight features at the high-intensity end.}
    \label{fig:center_ngc4321}
\end{figure*}

% Several galaxies host central features that contribute intense mid-IR emission, including AGN, nuclear star clusters, and circum-nuclear rings of star formation. 
%We treat these regions separately from the extended disks of our targets. To do this, we identify central regions from the MIRI images using cuts in radial distance selected by hand for each target. 
% In addition to nuclear regions that are physically distinct from the rest of the galactic disk, some central regions also include saturated pixels and regions strongly affected by the wings of the emergent JWST PSF due to extremely MIR-bright central sources. Since these extreme centers host distinct physical environments compared to the rest of the galaxy and often occupy a distinct, high intensity regime compared to the disk, we decouple the extreme centers from the rest of the galactic disk and present them separately.

As illustrated in Figure \ref{fig:overall_ngc4321} and Appendix \ref{sec:all_images}, PDFs of overall intensity constructed using the whole JWST image tend to show both an approximately log normal component at lower intensities and distinct, complex features at high intensity. These complex features in the PDF at high intensities can be traced back to the resolved substructure, extended JWST-MIRI PSF, and occasionally from saturated pixels in the central regions. For example, in Figure \ref{fig:overall_ngc4321}, NGC4321 shows a first `smooth' and broad peak at low intensity due to the disk and a second, less smooth, high intensity peak due to emission from the center. The shapes of these high-intensity features vary dramatically across galaxies with different central sources.
%This suggests that different morphologies in the galactic centers give rise to different high-intensity features in the overall PDFs.

These distinct features reflect the stark difference in physical conditions between the centers and disks found in some galaxies. Six out of 19 galaxies in our sample host AGNs (see Table \ref{tab:sample} for morphological classification and AGN classes). Several galaxies host intense central star-forming environments such as nuclear star clusters and bar-fed circum-nuclear rings of star formation \citep[see for example][]{2021QUEREJETA,2023SCHINNERER,2023SORMANI,2023HOYER}.
%, as well as extended diffuse gas disks.
In order to isolate the emission from these extreme central environments,  
%in our galaxies
we measure the PDFs of emission from only the central 1$-$3~kpc of each target, as summarized in Table \ref{tab:sample}. 

The extent of the central region is selected based on the mid-IR in order to isolate extreme mid-IR emission and significant contamination from the JWST-MIRI PSF. 
\edit1{The extent of the central region is selected based on the distribution of mid-IR emission for each galaxy with the goal of both enclosing bright emission associated with the center and isolating any artifacts associated with central sources. A number of the centers contain bright sources such as an AGN or star-forming regions in a central starburst ring. These bright compact or point sources both appear physically distinct from the disk and sometimes cause strong PSF-related artifacts, especially at F2100W. These can significantly impact our analysis (see e.g. the center of NGC7496 at F2100W in Figure \ref{fig:ngc7496}). These centers sometimes also contain areas with imperfect saturation correction \citep[e.g. at the AGN and in the starburst ring of NGC1365][and S. Hannon, priv. comm.; see Figure \ref{fig:ngc1365} for MIRI image]{2023LIU}. To exclude these regions, we set a radial cut that both encompasses the area where PSF wings are visible at F2100W, our lowest resolution filter, and that includes all bright emission associated with the galaxy center at this resolution. This has the risk of including some disk emission in the central regions, but we prefer this conservative approach to provide the cleanest possible characterization of the diffuse and HII region PDFs in the galactic disks. Conversely, because we weigh the PDFs by intensity, the effect of a small amount of disk contamination in the central regions will be modest.}
Figure \ref{fig:center_ngc4321} includes zoomed-in images and the corresponding PDFs of the central 1~kpc of NGC4321 (see Appendix \ref{sec:all_images} for the full sample). While the PDFs of the central 1~kpc of NGC4321 show a single peak in brightness, different targets show different central PDF shapes. These extreme centers can often contribute a significant fraction of the total flux in galaxies, as discussed further in Section \ref{sec:3ComponentContrast}.

In addition to capturing physically distinct environments, these central regions contain the majority of known image imperfections in our targets. The high brightness can lead to complications such as saturated pixels and diffraction spikes from the PSF associated with very bright compact sources. These complicate the interpretation of the PDFs and tend to only be present when the centers show bright emission distinct from the rest of the galactic disk. Both because of these artifacts and because the centers appear to represent a physically distinct regime, for most of this paper, we separate the center from the disk and exclude these central regions from the main analysis.

In Section \ref{sec:3ComponentContrast}, we measure the flux fractions and percentiles of the intensity PDF associated with the centers, but otherwise largely exclude the central regions from more sophisticated analysis of the PDF. Note that we do not exclude centers when the galactic centers largely resemble the rest of the galactic disks in the mid-IR. We allow this when the PDFs of the centers do not contribute additional peaks centered at $\geq 1$ dex of the peak of the overall PDF of the galaxy. \edit1{This translates to a criteria for excluding the galactic centers when the centers have a median intensity more than 1 dex higher than the median intensity in the diffuse disk component (the 'peak' in the overall disk PDF). The median intensities are also reflected later in Figure \ref{fig:intensity_contrasts}.} This is the case in NGC0628, NGC1087, NGC2835, NGC4254, NGC5068, and IC5332. Future \edit1{works} will evaluate morphological differences between the centers and disks of these targets in the optical and mid-IR.

\subsection{Galactic Disks} \label{sec:extended_disks}

\begin{figure*}[htbp]
    \centering
    \includegraphics[width=1\textwidth]{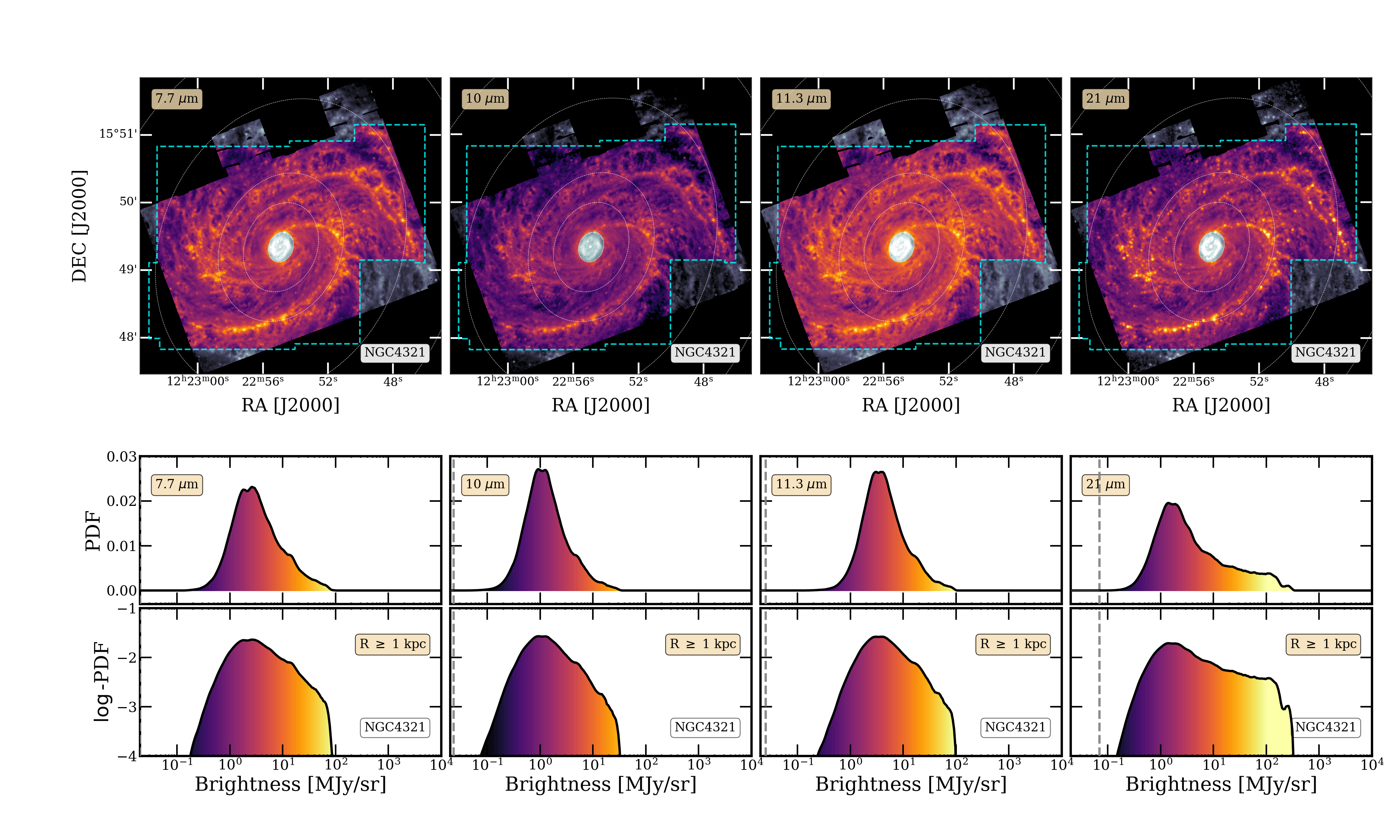}
    \caption{PDFs of emission of the galactic disk ($r\geq1$~kpc) after masking the center, at 7.7~$\mu m$, 10~$\mu m$, 11.3~$\mu m$, and 21~$\mu m$ for NGC4321. Similar to Figure \ref{fig:overall_ngc4321}, \textbf{Top:} Images of the galactic disk of NGC4321 in each MIRI filter, with the masked central region shaded. \textbf{Middle:} PDFs of inclination-corrected intensity from each image excluding the masked central region. The colors in the PDF reflect the colorbar in intensity used to display each image. Vertical dashed lines indicate rms noise levels in each filter. \textbf{Bottom:} The same disk-only PDFs shown in the middle row, now with the y-axis in log-stretch to highlight features at the high-intensity end. The PDFs of the disk show a consistent log normal component that peaks at lower intensities, as well as a high-intensity tail that can be roughly approximated to a power-law.}
    \label{fig:disk_ngc4321}
\end{figure*}

\begin{figure*}[htbp]
    \centering
    \includegraphics[width=0.95\textwidth]{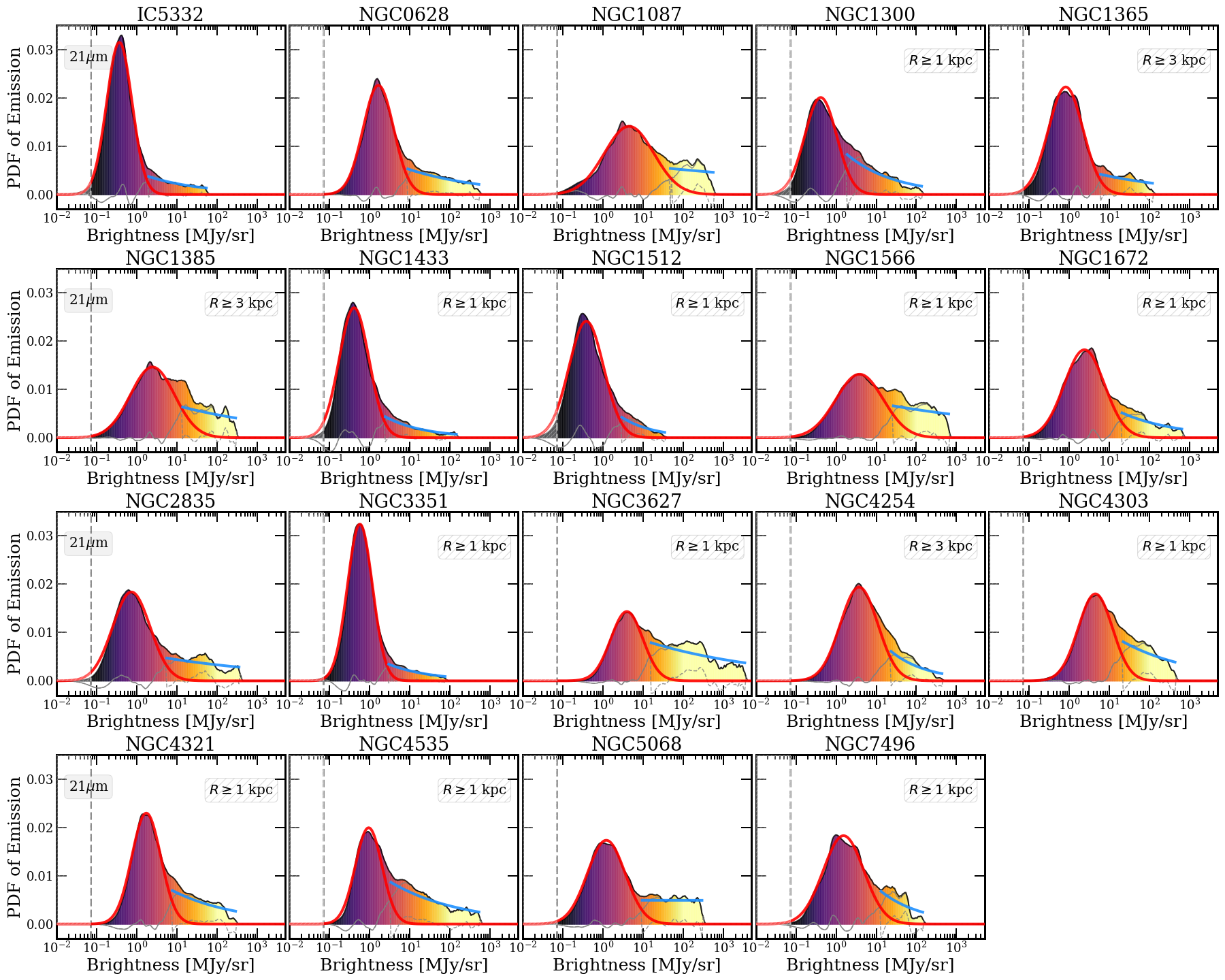}
    \caption{The PDFs of the galactic disk (in black) for our full sample at 21~$\mu m$. Vertical dashed lines show the rms noise levels for 21~$\mu$m at $0\farcs85$. The PDFs at each filter show an overall distribution well-fit by a log normal (red) at lower intensity (residuals in gray), with high intensity residuals (dashed gray) which are well-fit by a power-law (blue).}
    \label{fig:PDF}
\end{figure*}
% Saved figure comment:
% jessica.sutter93: Gray solid/dashed lines seem fairly difficult to see to me.  Could they be made thicker?

%Masking the centers decouples the emission in the disks from the extreme radiation fields and high column densities in the centers of galaxies hosting central bar-fueled starburst rings and AGNs. Figure \ref{fig:disk_ngc4321} shows the PDFs of the disk ($r \geq 1$ kpc) of NGC4321. 
%The PDFs across all four filters show a log normal peak at lower intensity, and a roughly power-law tail at high intensities. The power-law tail is more pronounced at 21 $\mu$m. 

Masking the central zone removes the high-intensity features, revealing the PDF of the disk. Figure \ref{fig:disk_ngc4321} shows the disk PDFs for NGC4321 (see Appendix \ref{sec:all_images} for disk PDFs of the full sample). In all four filters, these disk PDFs show a broad peak of emission at low intensity, which appears approximately log normal, with a continuous tail extending towards high intensity. %The complex high-intensity features from the center are now gone.
%consistent shapes in all four filters without any complex high-intensity features from the center. 
%Figure \ref{fig:PDF} shows these disk PDFs for our full sample.
%of galaxies are compiled in Figure \ref{fig:PDF}.
%, and overall brightness covered in our sample. 
The PDFs at F1000W and F2100W peak at lower intensities compared to the PAH-tracing F770W and F1130W filters. This is expected due to the presence of strong PAH-emission features at 7.7 and 11.3~$\mu$m. \edit1{The PDFs of NGC4321 at F770W, F1000W, and F1130W show a more prominent log-normal component, while the high-intensity tail appears more prominent and extended in the continuum-dominated F2100W filter}. Nested between two strong PAH emission features, F1000W could be a tracer of either PAH bands or continuum emission. \edit1{In the PDFs seen here, F1000W appears to resemble the PAH-tracing bands, featuring a more log normal shape with lower high-intensity emission \citep[see for example][for similar F1000W behavior]{2023BELFIORE, 2023LEROY, 2023SANDSTROM}. However, the F1000W PDFs differ from the PAH-tracing bands in some key ways - the low-intensity tail at F1000W is much more similar to the F21000W PDF, and the fall at high intensities happens at significantly lower values than PAH-tracing filters. The 9.7~$\mu$m silicate absorption feature may be a contributing factor to this intermediate behaviour exhibited by the F1000W filter, as discussed in Section \ref{sec:characterizing_HII_diffuse}. Silicate absorption is most likely to be particularly important in regions of very high column density such as the galactic centers and central molecular zones \citep[see e.g.,][for silicate features in the center of NGC7469]{2023LAI}.}
% Finally, the luminosity associated with the power-law component increases dramatically at F2100W, which shows shallower power-law slopes and higher range in intensities. 
The pronounced high intensity tail seen in the F2100W PDF is consistent with the increased thermal emissivity of the dust continuum in stronger radiation fields beyond 20~$\mu$m \citep[e.g,][]{2007DRAINELI, 2008BENDO, 2011DRAINE, 2023EGOROV}, as discussed further in Section \ref{sec:HII}. The depression of the power-law tail at shorter wavelengths is consistent with %lower dust emissivity combined with 
the destruction of small dust grains (PAHs) in intense radiation fields \citep[see e.g.,][]{2023CHASTENET,2023EGOROV}.

% Since the luminosity function of young star clusters is a power-law, and mid-IR emission is primarily reprocessed young star light, the power-law shape of the high-intensity tail of the PDF (where dust heated by the young stars dominate) is consistent. In addition, from turbulence theory, power-law tails are expected in regions where gravity takes over and dominates the dynamics and spatial distribution of gas (and hence dust) \textcolor{red}{(e.g., Klessen \& Glover, 2016; Girichidis et al. 2020)}. 

Expanding to the full sample, Figure \ref{fig:PDF} compiles all PDFs of galactic disks at 21~$\mu$m (for brevity, see Figures \ref{fig:PDF_HII} and \ref{fig:logPDF_HII} for PDFs using the full filter set which all show similar shapes). It is evident that the PDFs of the galactic disks for each galaxy consistently show the same two main features we saw in NGC 4321 (Fig.~\ref{fig:disk_ngc4321}), a log normal component that peaks at lower intensities and an approximately power-law tail extending to high intensities. % and across the full radial extent in our fields of view.
As a first check for this log normal plus power-law model, we fit a simple two-component model to the disk PDFs at 21~$\mu$m, the filter in which the power-law is most pronounced. Due to variability in the strength of the power-law, we attempt a two-part fit as
\edit1{
    \begin{align}
        \text{Log normal: } & {\rm PDF} (I_{\lambda_k}) = \dfrac{1}{\sqrt{2 \pi \sigma} I_{\lambda_k}} e^{- \dfrac{(\ln I_{\lambda_k} - \mu)^2}{2 \sigma^2}}, \\
        \text{Power-law: } & {\rm PDF} (I_{\lambda_k}) = C I_{\lambda_k}^m .
    \end{align}
    }
% We fit the PDFs using a two-component model. For the two-component model, 
To do this we fit the low-intensity side of the peak using a log normal, which is the dominant component. Then we reflect and subtract this best-fit log normal from the overall PDF. This reveals the residuals at the high-intensity end. This residual second component is well-modeled by a power-law distribution, suggesting a composite shape similar to the PDF shapes of $A_V$ in local Milky Way clouds \citep[e.g.,][]{2015LOMBARDI}. This suggests that more sophisticated models, like the Pareto log normal \citep[e.g.,][]{2018HOFFMANN}, could be an appropriate model to explore in future work.
% fit best by a power-law, compared to potential functional forms such as log normal, exponential, and Pareto log normal. 
The 21~$\mu$m fits and associated residuals are included in Figure \ref{fig:PDF}. 

The errors associated with the observations are small, as reported in Section \ref{sec:data}. The RMS noise estimate for each filter is indicated on each PDF as \edit1{vertical dashed lines} at the low-intensity end. Since these noise estimates are small, the uniformity of the log normal part of the distribution across bands and galaxies does not reflect noise or some other bias, but instead indicates that the mid-IR emission from diffuse regions shows a relatively universal shape across many galaxies and environments.

Following this proof-of-concept for describing the overall disk PDF, we devote the remainder of the paper to finding and using physically motivated independent multiwavelength ISM tracers to separate the power-law and log normal contributions to the PDFs, and performing a more rigorous parameterization of each component separately.

\startlongtable
\begin{deluxetable*}{ccrrrrrrrrrrr}

\tabletypesize{\small}
\tablecaption{Best-fit parameters for decomposed PDFs of galactic disks \label{tab:decomposed}}
\tablewidth{1\textwidth}

\tablehead{
\multicolumn{2}{c}{} & \multicolumn{4}{c}{\textsc{Hii Region MIR PDFs}$^a$} & \multicolumn{4}{c}{\textsc{Diffuse MIR PDFs}$^b$} & \multicolumn{3}{c}{\textsc{Diffuse N(H) PDFs}$^c$} \\
\colhead{  Galaxy  } & \colhead{$\lambda$} &     \colhead{$m$} &        \colhead{$c$} & \colhead{$\epsilon_m$} & \colhead{$\epsilon_c$} &   \colhead{$\mu$} & \colhead{$\sigma$} & \colhead{$\epsilon_\mu$} & \colhead{$\epsilon_\sigma$} & \colhead{$\mu_N$} & \colhead{$\sigma_N$}  &  \colhead{$\mathcal{M}_N$}
}
\startdata
                            &    $\mu$m &         & MJy/sr &          dex &          dex & MJy/sr &      dex &         dex &               dex &  cm$^{-2}$&          dex &         \\ \hline
\multirow[t]{4}{*}{IC5332}  &       7.7 &   -2.12 &   0.06 &         0.09 &         0.06 &   0.45 &     0.34 &          0.004 &             0.004 & 2.6$\times10^{20}$ &         0.36 &    8.12 \\
                            &        10 &   -1.58 &   0.04 &         0.08 &         0.04 &   0.19 &     0.34 &          0.002 &             0.002 &    &              &         \\
                            &      11.3 &   -1.93 &   0.05 &         0.06 &         0.05 &   0.73 &     0.30 &          0.004 &             0.004 &    &              &         \\
                            &        21 &   -0.51 &   0.01 &         0.01 &         0.01 &   0.34 &     0.27 &          0.002 &             0.002 &    &              &         \\ \hline
\multirow[t]{4}{*}{NGC0628} &       7.7 &   -1.29 &   0.08 &         0.06 &         0.08 &   2.10 &     0.31 &          0.002 &             0.002 & 1.3$\times10^{21}$ &         0.33 &    6.42 \\
                            &        10 &   -1.33 &   0.03 &         0.02 &         0.03 &   0.92 &     0.30 &          0.002 &             0.002 &    &              &         \\
                            &      11.3 &   -1.56 &   0.07 &         0.05 &         0.07 &   3.13 &     0.28 &          0.002 &             0.002 &    &              &         \\
                            &        21 &   -0.35 &   0.02 &         0.01 &         0.02 &   1.32 &     0.32 &          0.002 &             0.002 &    &              &         \\ \hline 
\multirow[t]{4}{*}{NGC1087} &       7.7 &   -0.75 &   0.15 &         0.08 &         0.15 &   3.14 &     0.41 &          0.006 &             0.006 & 2.8$\times10^{21}$ &         0.42 &   12.77 \\
                            &        10 &   -0.77 &   0.08 &         0.06 &         0.08 &   1.22 &     0.40 &          0.006 &             0.006 &    &              &         \\
                            &      11.3 &   -0.79 &   0.14 &         0.08 &         0.14 &   4.49 &     0.40 &          0.006 &             0.006 &    &              &         \\
                            &        21 &   -0.22 &   0.05 &         0.02 &         0.05 &   2.04 &     0.42 &          0.006 &             0.006 &    &              &         \\ \hline
\multirow[t]{4}{*}{NGC1300} &       7.7 &   -1.92 &   0.23 &         0.20 &         0.23 &   0.61 &     0.44 &          0.003 &             0.003 & 4.5$\times10^{20}$ &         0.49 &   21.31 \\
                            &        10 &   -1.52 &   0.15 &         0.14 &         0.15 &   0.30 &     0.41 &          0.002 &             0.002 &    &              &         \\
                            &      11.3 &   -2.38 &   0.25 &         0.19 &         0.25 &   0.97 &     0.42 &          0.001 &             0.001 &    &              &         \\
                            &        21 &   -0.70 &   0.06 &         0.04 &         0.06 &   0.42 &     0.43 &          0.005 &             0.005 &    &              &         \\ \hline
\multirow[t]{4}{*}{NGC1365} &       7.7 &   -1.81 &   0.09 &         0.07 &         0.09 &   1.00 &     0.38 &          0.002 &             0.002 & 1.1$\times10^{21}$ &         0.41 &   11.86 \\
                            &        10 &   -1.53 &   0.05 &         0.07 &         0.05 &   0.48 &     0.35 &          0.004 &             0.004 &    &              &         \\
                            &      11.3 &   -1.92 &   0.15 &         0.12 &         0.15 &   1.61 &     0.35 &          0.003 &             0.003 &    &              &         \\
                            &        21 &   -0.59 &   0.11 &         0.06 &         0.11 &   0.72 &     0.39 &          0.003 &             0.003 &    &              &         \\ \hline
\multirow[t]{4}{*}{NGC1385} &       7.7 &   -0.81 &   0.19 &         0.11 &         0.19 &   2.21 &     0.38 &          0.004 &             0.004 &   2$\times10^{21}$ &         0.41 &   11.86 \\
                            &        10 &   -0.99 &   0.14 &         0.12 &         0.14 &   0.85 &     0.37 &          0.004 &             0.004 &    &              &         \\
                            &      11.3 &   -0.91 &   0.22 &         0.12 &         0.22 &   3.13 &     0.36 &          0.003 &             0.003 &    &              &         \\
                            &        21 &   -0.34 &   0.06 &         0.03 &         0.06 &   1.46 &     0.41 &          0.004 &             0.004 &    &              &         \\ \hline
\multirow[t]{4}{*}{NGC1433} &       7.7 &   -1.65 &   0.07 &         0.07 &         0.07 &   0.54 &     0.34 &          0.002 &             0.002 & 3.2$\times10^{20}$ &         0.38 &    9.47 \\
                            &        10 &   -1.63 &   0.06 &         0.09 &         0.06 &   0.29 &     0.35 &          0.001 &             0.001 &    &              &         \\
                            &      11.3 &   -1.79 &   0.11 &         0.09 &         0.11 &   0.90 &     0.35 &          0.002 &             0.002 &    &              &         \\
                            &        21 &   -0.53 &   0.05 &         0.04 &         0.05 &   0.37 &     0.33 &          0.002 &             0.002 &    &              &         \\ \hline
\multirow[t]{4}{*}{NGC1512} &       7.7 &   -1.66 &   0.11 &         0.11 &         0.11 &   0.47 &     0.36 &          0.002 &             0.002 & 3.1$\times10^{20}$ &         0.41 &   11.86 \\
                            &        10 &   -1.21 &   0.06 &         0.08 &         0.06 &   0.25 &     0.36 &          0.001 &             0.001 &    &              &         \\
                            &      11.3 &   -1.73 &   0.12 &         0.13 &         0.12 &   0.77 &     0.36 &          0.002 &             0.002 &    &              &         \\
                            &        21 &   -0.63 &   0.04 &         0.04 &         0.04 &   0.33 &     0.37 &          0.003 &             0.003 &    &              &         \\ \hline
\multirow[t]{4}{*}{NGC1566} &       7.7 &   -1.75 &   0.22 &         0.12 &         0.22 &   3.11 &     0.43 &          0.004 &             0.004 & 2.2$\times10^{21}$ &         0.46 &   17.12 \\
                            &        10 &   -1.23 &   0.09 &         0.06 &         0.09 &   1.43 &     0.39 &          0.005 &             0.005 &    &              &         \\
                            &      11.3 &   -1.61 &   0.19 &         0.10 &         0.19 &   4.64 &     0.41 &          0.004 &             0.004 &    &              &         \\
                            &        21 &   -0.24 &   0.06 &         0.03 &         0.06 &   2.13 &     0.42 &          0.003 &             0.003 &    &              &         \\ \hline
\multirow[t]{4}{*}{NGC1672} &       7.7 &   -2.51 &   0.19 &         0.12 &         0.19 &   2.45 &     0.40 &          0.004 &             0.004 &   2$\times10^{21}$ &         0.43 &   13.75 \\
                            &        10 &   -1.41 &   0.09 &         0.06 &         0.09 &   1.05 &     0.37 &          0.004 &             0.004 &    &              &         \\
                            &      11.3 &   -1.96 &   0.14 &         0.08 &         0.14 &   3.63 &     0.38 &          0.004 &             0.004 &    &              &         \\
                            &        21 &   -0.47 &   0.04 &         0.02 &         0.04 &   1.70 &     0.41 &          0.004 &             0.004 &    &              &         \\ \hline
\multirow[t]{4}{*}{NGC2835} &       7.7 &   -1.52 &   0.05 &         0.04 &         0.05 &   0.85 &     0.35 &          0.002 &             0.002 & 7.2$\times10^{20}$ &         0.38 &    9.47 \\
                            &        10 &   -0.94 &   0.04 &         0.04 &         0.04 &   0.37 &     0.34 &          0.002 &             0.002 &    &              &         \\
                            &      11.3 &   -1.34 &   0.06 &         0.04 &         0.06 &   1.32 &     0.33 &          0.003 &             0.003 &    &              &         \\
                            &        21 &   -0.29 &   0.04 &         0.02 &         0.04 &   0.53 &     0.35 &          0.001 &             0.001 &    &              &         \\ \hline
\multirow[t]{4}{*}{NGC3351} &       7.7 &   -1.47 &   0.03 &         0.03 &         0.03 &   0.64 &     0.27 &          0.001 &             0.001 & 4.8$\times10^{20}$ &          0.3 &    5.02 \\
                            &        10 &   -1.44 &   0.02 &         0.04 &         0.02 &   0.34 &     0.26 &          0.002 &             0.002 &    &              &         \\
                            &      11.3 &   -1.66 &   0.02 &         0.02 &         0.02 &   0.99 &     0.28 &          0.001 &             0.001 &    &              &         \\
                            &        21 &   -0.52 &   0.01 &         0.01 &         0.01 &   0.52 &     0.28 &          0.001 &             0.001 &    &              &         \\ \hline
\multirow[t]{4}{*}{NGC3627} &       7.7 &   -0.98 &   0.05 &         0.03 &         0.05 &   4.49 &     0.33 &          0.002 &             0.002 & 5.3$\times10^{21}$ &         0.36 &    8.12 \\
                            &        10 &   -1.08 &   0.04 &         0.03 &         0.04 &   1.93 &     0.30 &          0.002 &             0.002 &    &              &         \\
                            &      11.3 &   -1.33 &   0.10 &         0.04 &         0.10 &   6.67 &     0.32 &          0.002 &             0.002 &    &              &         \\
                            &        21 &   -0.27 &   0.04 &         0.02 &         0.04 &   2.97 &     0.34 &          0.001 &             0.001 &    &              &         \\ \hline
\multirow[t]{4}{*}{NGC4254} &       7.7 &   -1.74 &   0.07 &         0.04 &         0.07 &   3.62 &     0.35 &          0.003 &             0.003 &   3$\times10^{21}$ &         0.38 &    9.47 \\
                            &        10 &   -1.88 &   0.08 &         0.06 &         0.08 &   1.47 &     0.33 &          0.003 &             0.003 &    &              &         \\
                            &      11.3 &   -2.02 &   0.08 &         0.05 &         0.08 &   5.10 &     0.33 &          0.003 &             0.003 &    &              &         \\
                            &        21 &   -0.89 &   0.08 &         0.04 &         0.08 &   2.31 &     0.36 &          0.003 &             0.003 &    &              &         \\ \hline
\multirow[t]{4}{*}{NGC4303} &       7.7 &   -2.17 &   0.23 &         0.12 &         0.23 &   4.65 &     0.34 &          0.002 &             0.002 & 3.4$\times10^{21}$ &         0.37 &    8.77 \\
                            &        10 &   -2.10 &   0.21 &         0.15 &         0.21 &   2.01 &     0.30 &          0.003 &             0.003 &    &              &         \\
                            &      11.3 &   -2.43 &   0.32 &         0.17 &         0.32 &   6.94 &     0.32 &          0.002 &             0.002 &    &              &         \\
                            &        21 &   -0.51 &   0.06 &         0.03 &         0.06 &   3.10 &     0.33 &          0.002 &             0.002 &    &              &         \\ \hline
\multirow[t]{4}{*}{NGC4321} &       7.7 &   -1.44 &   0.18 &         0.12 &         0.18 &   2.11 &     0.33 &          0.001 &             0.001 & 1.6$\times10^{21}$ &         0.36 &    8.12 \\
                            &        10 &   -1.35 &   0.05 &         0.04 &         0.05 &   0.96 &     0.29 &          0.001 &             0.001 &    &              &         \\
                            &      11.3 &   -1.29 &   0.06 &         0.04 &         0.06 &   3.17 &     0.29 &          0.001 &             0.001 &    &              &         \\
                            &        21 &   -0.55 &   0.09 &         0.05 &         0.09 &   1.52 &     0.32 &          0.001 &             0.001 &    &              &         \\ \hline
\multirow[t]{4}{*}{NGC4535} &       7.7 &   -1.30 &   0.08 &         0.05 &         0.08 &   1.20 &     0.36 &          0.004 &             0.004 & 9.2$\times10^{20}$ &          0.4 &   11.01 \\
                            &        10 &   -1.16 &   0.05 &         0.04 &         0.05 &   0.53 &     0.34 &          0.002 &             0.002 &    &              &         \\
                            &      11.3 &   -1.32 &   0.12 &         0.07 &         0.12 &   1.74 &     0.34 &          0.002 &             0.002 &    &              &         \\
                            &        21 &   -0.35 &   0.03 &         0.02 &         0.03 &   0.91 &     0.33 &          0.002 &             0.002 &    &              &         \\ \hline
\multirow[t]{4}{*}{NGC5068} &       7.7 &   -1.60 &   0.08 &         0.06 &         0.08 &   1.22 &     0.36 &          0.002 &             0.002 & 7.9$\times10^{20}$ &         0.39 &   10.21 \\
                            &        10 &   -0.91 &   0.03 &         0.03 &         0.03 &   0.52 &     0.34 &          0.002 &             0.002 &    &              &         \\
                            &      11.3 &   -1.45 &   0.07 &         0.05 &         0.07 &   1.92 &     0.34 &          0.002 &             0.002 &    &              &         \\
                            &        21 &   -0.08 &   0.02 &         0.01 &         0.02 &   0.73 &     0.36 &          0.001 &             0.001 &    &              &         \\ \hline
\multirow[t]{4}{*}{NGC7496} &       7.7 &   -1.38 &   0.22 &         0.17 &         0.22 &   1.46 &     0.36 &          0.002 &             0.002 & 8.6$\times10^{20}$ &         0.39 &   10.21 \\
                            &        10 &   -1.75 &   0.13 &         0.14 &         0.13 &   0.68 &     0.34 &          0.003 &             0.003 &                        &              &     \\
                            &      11.3 &   -1.36 &   0.23 &         0.16 &         0.23 &   2.29 &     0.33 &          0.003 &             0.003 &                        &              &     \\
                            &        21 &   -1.47 &   0.31 &         0.17 &         0.31 &   1.14 &     0.43 &          0.003 &             0.003 &                        &              &     \\
\hline\hline\enddata
\tablecomments{$^a$ Non-linear regression fit parameters for the \ion{H}{2} region power-law component at each filter for each galaxy: the power-law index $m$, logarithmic intercept $c$, and associated errors $\epsilon_m$, $\epsilon_c$, respectively, as discussed in Section \ref{sec:HII_nonHII} and \ref{sec:HII}.\\
$^b$ Fit parameters for the diffuse region log-normal component at each filter for each galaxy: the log-normal mean $\mu$, dispersion $\sigma$, and associated errors $\epsilon_\mu$, $\epsilon_\sigma$, respectively, as discussed in Section \ref{sec:HII_nonHII} and \ref{sec:non-HII}.\\
$^c$ Fit parameters for the derived N(H) gas column density PDFs in diffuse regions, as discussed in Section \ref{sec:non-HII}: the best fit log-normal mean $\mu$ and dispersion $\sigma$, as well as derived average upper limits of Mach numbers $\mathcal{M}$ of isothermal turbulence.\\
}
\end{deluxetable*}

% \begin{longtable*}[htbp]{lllllllllll}
% \caption[Decomposed parameters]{Fit parameters of decomposed PDFs of intensities in the extended disks of each of our targets.}
% \label{tab:decomposed}\\
% \toprule
%  Galaxy & $\lambda$ &     $m$ &        $c$ & $\epsilon_m$ & $\epsilon_c$ &   $\mu$ & $\sigma$ & $\epsilon_\mu$ & $\epsilon_\sigma$ &      $\mathcal{M}$ \\
% \midrule
% \endfirsthead
% \caption[]{Fit parameters of decomposed PDFs of intensities} \\
% \toprule
%  Galaxy & $\lambda$ &     $m$ &        $c$ & $\epsilon_m$ & $\epsilon_c$ &   $\mu$ & $\sigma$ & $\epsilon_\mu$ & $\epsilon_\sigma$ &      $\mathcal{M}$ \\
% \midrule
% \endhead
% \midrule
% \multicolumn{11}{r}{{Continued on next page}} \\
% \midrule
% \endfoot

% \bottomrule
% \endlastfoot
%         &    $\mu$m & dex/dex &     MJy/sr &          dex &          dex &  MJy/sr &      dex &            dex &               dex &          \\
% data
% \end{longtable*}

\subsection{\textsc{Hii} Regions and the Diffuse ISM in Galactic Disks} \label{sec:HII_nonHII}

\begin{figure*}[htbp]
    \centering
    \includegraphics[width=1\textwidth]{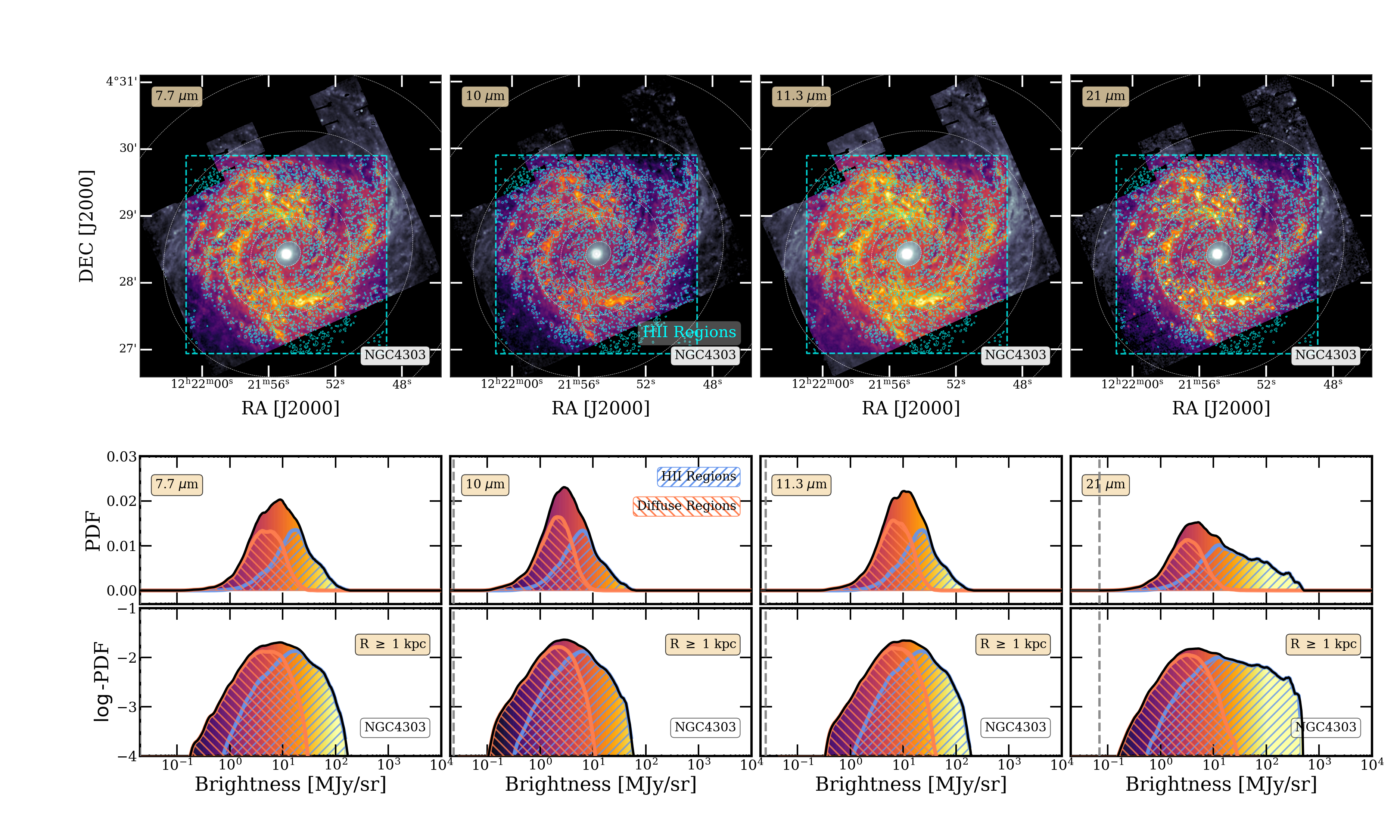}
    \caption{
    PDFs of emission of the galactic disk ($r\geq1$~kpc) after masking the center, at 7.7~$\mu m$, 10~$\mu m$, 11.3~$\mu m$, and 21~$\mu m$ for NGC4303, tagged by emission from within and outside of \ion{H}{2} regions. Similar to Figure \ref{fig:disk_ngc4321}, \textbf{Top:} Images of the galactic disk of NGC4303 in each MIRI filter, with the masked central region shaded. H$\alpha$-identified \ion{H}{2} regions are overlaid in blue contours. \textbf{Middle:} PDFs of inclination-corrected intensity from the disk (excluding the masked central region), colored by the intensity color bar used to display each image. The PDFs of emission from within \ion{H}{2} regions and outside of \ion{H}{2} region masks are included in hatched blue and orange, respectively. \textbf{Bottom:} The same disk-only PDFs shown in the middle row, now with the y-axis in log-stretch to highlight the power-law tail. It is evident that the emission from \ion{H}{2} regions builds up the power-law component, and the diffuse ISM outside of \ion{H}{2} regions shows a log normal PDF.}
    \label{fig:HIIdisk_ngc4303}
\end{figure*}
% saved figure comment
% rchown3: overlapping colours are a bit hard on the eyes. Since they're just hatched lines, you could make them black, with the HII PDF dashed and diffuse PDF dotted perhaps? could make the linewidths smaller so you still see the hatches

In Figure \ref{fig:PDF} all 19 galaxies show distinct log normal and high intensity power-law components in their disk PDFs. Therefore, we hypothesize that this two-component form may be a general feature of galactic disks in the mid-IR.

\edit1{These two components of the PDFs of galactic disks roughly map to two spatially distinct regions in the mid-IR images of galactic disks.} The high-intensity tail arises from bright spots of mid-IR emission that are often visible as compact regions near the spiral arms and bars of each galaxy. In the first four PHANGS-JWST targets, these bright point sources at 21$~\mu$m showed an excellent correspondence with the sites of recent star formation identified by H$\alpha$ emission \citep[e.g.,][]{2023HASSANI,2023EGOROV,2023LEROY}. This suggests that the high-intensity, power-law component can be naturally explained as reflecting mostly bright, compact star-forming regions.

To test this idea,
%that the power-law tail reflects mainly star-forming regions
we use the \ion{H}{2} region masks constructed in the optical from PHANGS-MUSE (Section \ref{sec:data}). %In these masks, each pixel is labeled as either belonging to an H$\alpha$-identified \ion{H}{2} region, and thus part of an optically visible star-forming complex, or not. 
As an illustration, Figure \ref{fig:HIIdisk_ngc4303} shows the disk of NGC4303 ($R \geq 1$~kpc) with \ion{H}{2} regions overlaid. The emission from the disk is then divided into emission from within (in blue) and outside of (orange) \ion{H}{2} regions.
%, and each component PDF is included
It is clear that the \ion{H}{2} regions identified in the optical successfully isolate the bright, approximately power-law component of the PDF, especially at the dust continuum-tracing 21$~\mu$m. In this case, the overall high-intensity component in the PDF reflects the sum of many individual star-forming regions, which together build up the observed approximately power-law tail. 
The mid-IR images and overlaid \ion{H}{2} region masks for all 19 galaxies are available in Appendix \ref{sec:all_images}. 

\begin{figure*}[htbp]
    \centering
    \includegraphics[width=0.9\textwidth]{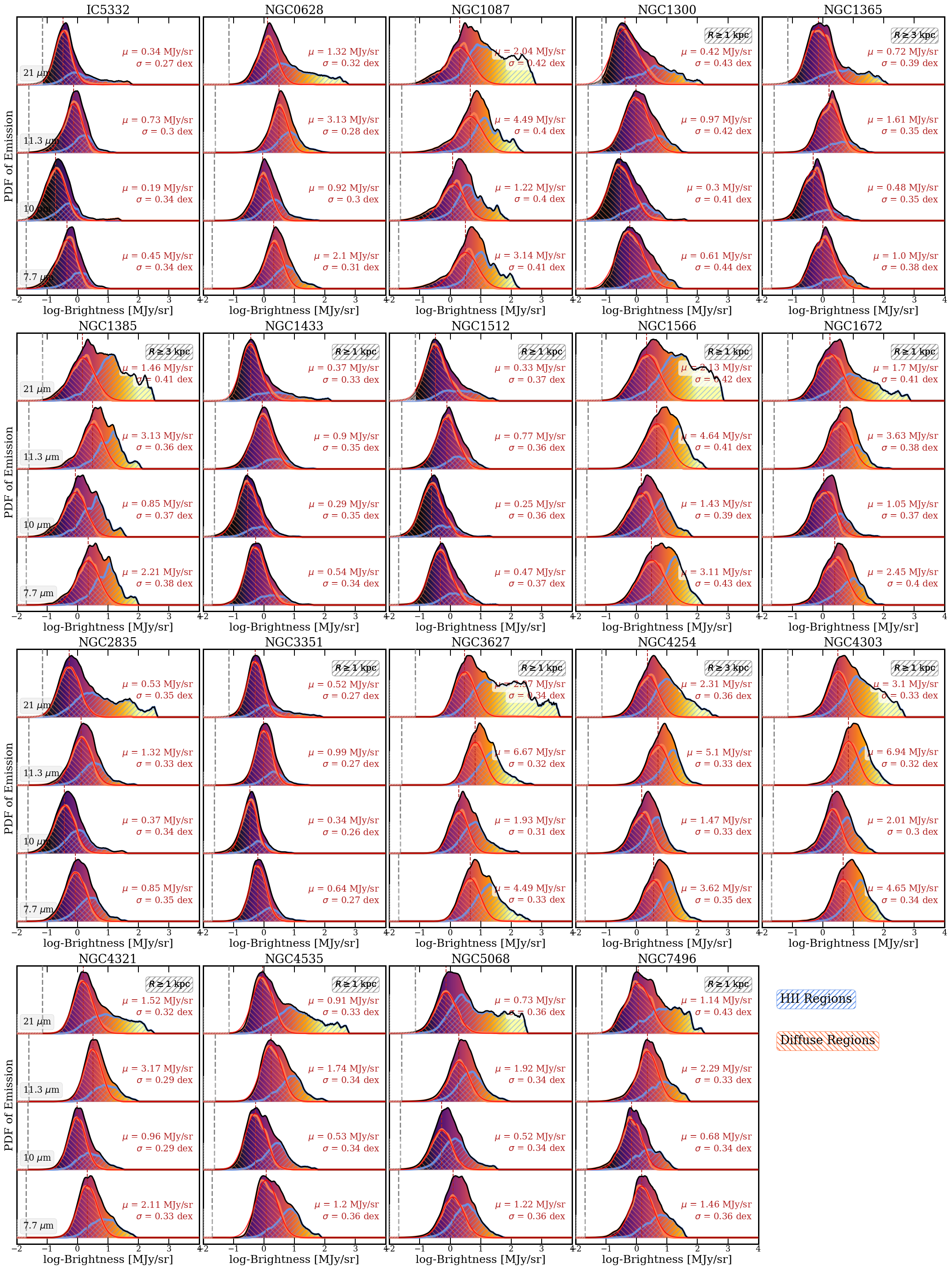}
    \caption{The galactic disk PDFs (in black) of our full sample of 19 galaxies at 7.7, 10, 11.3, and 21~$\mu$m show an overall log normal distribution with a power-law tail at higher intensities. The power-law component can be primarily attributed to emission from individual \ion{H}{2} regions (blue) and the log normal component comes from emission from outside \ion{H}{2} regions (red). Vertical dashed lines indicate the rms noise level in each filter for reference. The best-fit log normal mean $\mu$ and dispersion $\sigma$ are included for each PDF.}
    \label{fig:PDF_HII}
\end{figure*}

\begin{figure*}[htbp]
    \centering
    \includegraphics[width=0.9\textwidth]{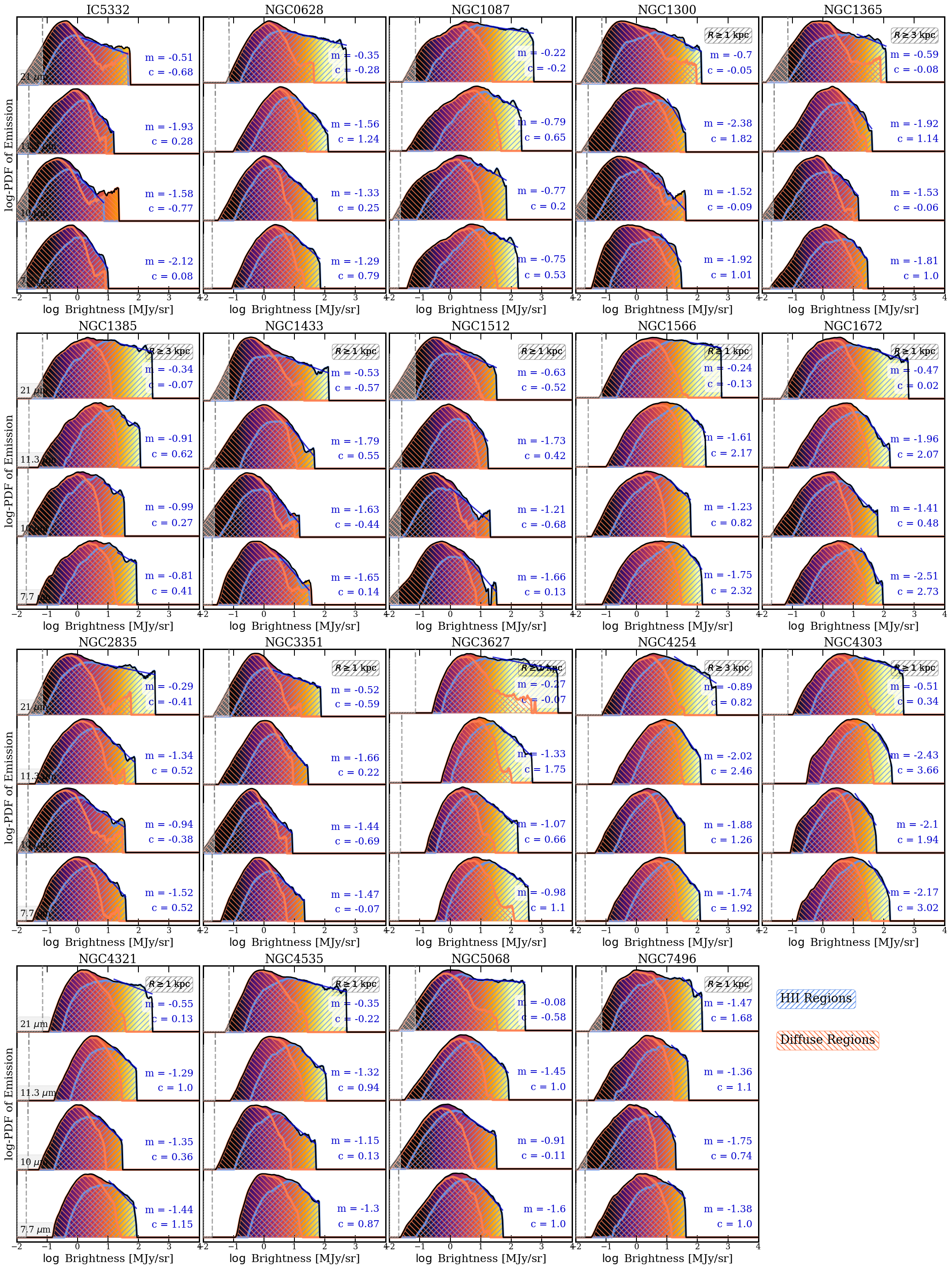}
    \caption{The disk PDFs (in black) of our sample of 19 galaxies at 7.7, 10, 11.3, and 21~$\mu$m decomposed by contributions from \ion{H}{2} regions (blue) and outside (red). These are the same PDFs as Figure \ref{fig:PDF_HII} but in log-stretch to highlight the power-law component. The best-fit power-law is shown in dark blue. The corresponding best-fit \ion{H}{2} region power-law index $m$ and log-log intercept $c$ for each PDF are included for each panel.}
    \label{fig:logPDF_HII}
\end{figure*}
% Saved Figure comments 
% jessica.sutter93: What would these look like if you didn't have the color-coded PDF behind the hatched HII/Diffuse region pdfs? Could this best fit line also be made more visible?  Maybe a dashed line in a bright color, over-plotted on the black so you can see both?

Figures \ref{fig:PDF_HII} and \ref{fig:logPDF_HII} expand this analysis to the full sample.
%, showing the PDFs and log-scaled PDFs of the disk at each mid-IR wavelength. 
The PDF contributions from \ion{H}{2} regions (in blue) and the ISM outside of \ion{H}{2} regions (in orange) are overlaid on each PDF and log-PDF for the disk of each target. 
%The figures show a clear correspondence between the \ion{H}{2} region masks and the two components of the PDF.
% Comparing the emission from within and outside of \ion{H}{2} region masks, 
\edit1{It is evident from Figure \ref{fig:PDF_HII} and \ref{fig:logPDF_HII} that this clear correspondence between the \ion{H}{2} region masks and the distinct log-normal and power-law components of each PDF (as shown in Figure \ref{fig:HIIdisk_ngc4303} for NGC4303) extends across our whole sample.} Decomposing the disks into \ion{H}{2} and non \ion{H}{2} regions effectively decomposes the overall PDFs into their respective power-law and log normal components for each target. The bright regions in the mid-IR images, the high-intensity power-law tails in the PDFs, and the \ion{H}{2} regions identified from VLT/MUSE H$\alpha$ emission are all the same to first order, as discussed further in Section \ref{sec:HII}. %HII regions consistently map out most of the bright sources in the mid-IR images, neatly accounting for the high-intensity power-law component. 
At shorter wavelengths, while most galaxies in Figure \ref{fig:logPDF_HII} show weaker but still distinguishable power-law components, the second high-\edit1{density} component in some galaxies could arguably also be fit by a second log normal. However, as we discuss in Section \ref{sec:globalprops} and \ref{sec:HII}, since the power-law index encodes information about the  luminosity function of \ion{H}{2} regions and \textit{is} pronounced at shorter wavelengths for a majority of our galaxies, we provide power-law fits to all 19 galaxies.

The emission from outside \ion{H}{2} regions corresponds to the log normal component of the PDF. Based on the similarity to the PDF shape expected for gas densities in a turbulent medium and the separation of these regions from strong heating sources, we suggest that this log normal component can be identified with the ``diffuse'' ISM. In these regions, we expect that variations in gas and dust column density drive much of the observed shape of the PDF. The observed correlation between mm-wave tracers of gas column density and mid-IR emission seen in the PHANGS-JWST first results support this interpretation \citep[e.g.,][]{2023LEROY,2023SANDSTROM}. As discussed further in Section \ref{sec:non-HII}, this log normal component of the PDF may constrain the column density distribution in the diffuse ISM, and thus perhaps the behavior of the turbulent ISM. Under our current definition, this includes some contribution from the cold dense part of the ISM that is highly turbulent with supersonic Mack numbers.
%thus broadly correspond to the log normal component of the intensity PDFs. %The next two sections include a more in-depth discussion of the power-law (HII) and log normal (non-HII) components of the disk PDF, followed by the PDFs of the centers of each target.

In summary, while details of the PDF, such as the exact mean and width of the log normal component, and the slope of the power-law distribution change somewhat across targets and show some systemic variation with wavelength, the overall shape of the PDFs appears similar across all 19 galaxies. This two-component behavior for galactic disks emerges across a wide range of galaxy morphologies.

% \textcolor{red}{Talk about how fitting separately is done here! Then move band-wise param comparison text up here instead of contrasts.}

% \textcolor{red}{Merge text from F1000W Filter Below.}

\begin{figure*}[htbp]
    \centering
    \includegraphics[width=1\textwidth]{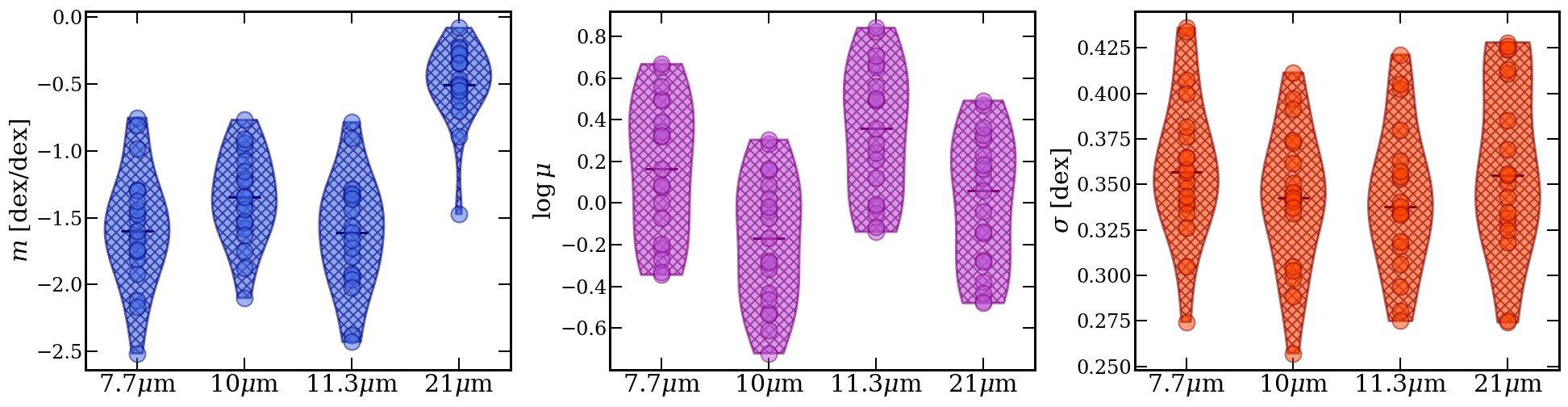}
    \caption{Summary of the galactic disk PDF parameters at each wavelength, including the \ion{H}{2} region power-law index ($m$) on the left-hand panel, and the diffuse component log normal mean ($\mu$), and variance ($\sigma$) on the middle and right hand panels respectively, as included in Table \ref{tab:decomposed}. The PDFs at 21~$\mu$m show significantly more pronounced power-law components, while the mean intensity of the diffuse log normal component is systematically higher in PAH-tracing 7.7 and 11.3~$\mu$m. The log normal PDF widths remain generally consistent across the four filters.}
    \label{fig:fit_summary}
\end{figure*}

\subsection{Characterizing the HII Region and Diffuse Component PDFs} \label{sec:characterizing_HII_diffuse}

\begin{deluxetable}{lcrcl}[th!]
\tabletypesize{\small}
\tablecaption{Summary of PDF Parameters \label{tab:summary}}
\tablewidth{0.5\textwidth}

\tablehead{
\colhead{Best-Fit Parameter} & \colhead{$\lambda$[$\mu$m]} &  \colhead{Median} &   \colhead{Q1/Q3} & \colhead{$\pm 1\sigma$}
}

\startdata
% the data here
\multirow[c]{4}{*}{Power-Law Index $m$}             & 7.7  & -1.60 & -1.78/-1.34 &  0.43  \\
                                                    & 10   & -1.35 & -1.56/-1.12 &  0.33  \\
                                                    & 11.3 & -1.61 & -1.93/-1.33 &  0.42  \\
                                                    & 21   & -0.51 & -0.57/-0.31 &  0.29  \\ \hline
\multirow[c]{4}{*}{Log Normal Mean $\mu$}           & 7.7  &  1.46 & 0.74/2.78   &  1.31  \\
                                                    & 10   &  0.68 & 0.36/1.14   &  0.54  \\
                                                    & 11.3 &  2.29 & 1.15/4.06   &  1.91  \\
                                                    & 21   &  1.14 & 0.53/1.87   &  0.86  \\ \hline
\multirow[c]{4}{*}{Log Normal Dispersion $\sigma$}  & 7.7  &  0.36 & 0.34/0.38   &  0.039  \\
                                                    & 10   &  0.34 & 0.32/0.37   &  0.038  \\
                                                    & 11.3 &  0.34 & 0.32/0.36   &  0.039  \\
                                                    & 21   &  0.36 & 0.33/0.41   &  0.048  \\ \hline
\enddata
\tablecomments{
Median, first/third quartiles, and standard deviation of the best-fit power-law index $m$ of \ion{H}{2} region PDFs, log normal mean $\mu$ in MJy/sr, and dispersion $\sigma$ in dex of diffuse PDFs at each filter.}
\end{deluxetable}

For the rest of this paper, we treat the PDFs of mid-IR intensity from disks as separable into diffuse and \ion{H}{2} regions. For each galaxy, we parameterize these two distinct components by using non-linear regression to fit a power-law to the \ion{H}{2} region and a log normal model to the diffuse component. The power-law index $m$, logarithmic intercept $c$, and corresponding standard errors for the \ion{H}{2} region component, and the log normal mean $\mu$, dispersion $\sigma$, and corresponding uncertainties for the diffuse ISM component are presented in Table \ref{tab:decomposed} and summarized in Figure \ref{fig:fit_summary} and Table \ref{tab:summary}.
%, which compares the fit parameters from the PDFs of \ion{H}{2} and and ``diffuse" regions across our full sample. 

As illustrated in Figure \ref{fig:fit_summary}, the power-law slope $m$ of the PDFs of \ion{H}{2} regions show a range of values between $-$2.5 and \edit1{$-$0.08}. At 21~$\mu$m, where the power-law component is most prominent, $m$ shows a relatively narrow range of values, with a median index of $-0.51 \pm 0.29 (\pm 1\sigma)$ , corresponding to tightly clustered indices around $\sim -0.5$ (see Table \ref{tab:summary} for all wavelengths). So the \ion{H}{2} region components consistently show shallow and extended power-law PDFs of intensities. %This reflects the distinct nature of the 21~$\mu$m band which largely traces thermal emission from larger grains, while the shorter wavelengths better trace smaller grains.

The widths of the log normal PDFs of the diffuse regions appear largely consistent across the four filters, with a median width (log normal dispersion) of $\sim 0.34-0.36$ dex in all four filters. The width also shows relatively little scatter from galaxy to galaxy, with a standard deviation of $\sim \pm0.039$ dex in log normal dispersion across 7.7--11.3$~\mu$m. The log normal widths at 21~$\mu$m show a somewhat wider spread compared to the other three filters, with a slightly larger standard deviation of $\pm 0.048$ dex.

In contrast to the width, the log normal mean $\mu$ shows some systematic variation across the four wavelengths. The diffuse component PDFs show higher mean intensities at 7.7 and 11.3~$\mu$m, where strong PAH emission complexes peak, and lower $\mu$ at 10 and 21~$\mu$m which are expected to better trace the dust continuum. The mean also shows a larger spread in the two PAH-tracing bands ($\pm 1.3-1.9$ MJy/sr, $\pm1\sigma$), compared to the two continuum-tracing bands ($\pm 05-0.8$ MJy/sr, $\pm1\sigma$).   %The 10~$\mu$m band thus shows some similarities with the PAH-tracing 7.7 and 11.3~$\mu$m, e.g., it lacks the same pronounced power-law tail seen at 21~$\mu$m. %However in the diffuse part of the ISM, the 10~$\mu$m does show an overall depression in mean intensity similar to the 21~$\mu$m filter, which primarily traces continuum emission.

\edit1{The galactic disk PDFs at 7.7 and 11.3 $\mu$m show similar widths and power-law indices, with slightly higher $\mu$ at 11.3 $\mu$m. The ratio of emission at 7.7 and 11.3 microns can be used to estimate the level of PAH ionization, since 7.7 microns is expected to better trace ionized PAHs, while 11.3 microns is expected to trace more neutral PAHs \citep[see e.g.,][and references therein]{2008TIELENS, 2008GALLIANO, 2016BOERSMA, 2021RIGOPOULOU}. \citet{2023CHASTENETionPAH} find a fairly uniform population of PAHs in the disks of the first four PHANGS-JWST galaxies (NGC0628, NGC1365, NGC7496, and IC5332) based on the variations observed in the 3.3, 7.7, and 11.3 $\mu$m features. The lack of significant variation in these ratios on average across galactic disks is consistent with our PDFs of galactic disks, which do not show significant variation between 7.7 and 11.3 micron across all 19 targets.}

% This illustrates the similarities between 7.7 and 11.3 $\mu$m PAH-tracing bands, the distinct nature of the thermal 21 $\mu$m PDFs, and the intermediate behaviour of the 10 $\mu$m band between PAH-tracing bands and thermal emission band. 
% Correlations between these parameters and global properties of galaxies are discussed in Section \ref{sec:globalprops}.

These general trends are in good agreement with dust models and previous observational results. These models attribute most of the 7.7 and 11.3~$\mu$m emission to PAH (small-grain) bending modes. Most of the 21 $\mu$m emission is attributed to continuum emission from either stochastically heated small grains in regions of low radiation pressure, or grains in thermal equilibrium in regions close to intense heating sources \citep{2011DRAINE}.

Interpreting the 10$~\mu$m emission is more complicated.
%Since the mean intensities in the diffuse part of the galactic disks fall at both 10 and 21 $\mu$m, with 10 $\mu$m showing significantly lower intensities than all three other filters, it is unlikely that 10 $\mu$m captures much PAH emission. 
The 10 $\mu$m shows significantly lower intensities than 7.7 or 11.3~$\mu$m, reflecting the lack of a strong PAH emission band centered in the filter. 
%Only the central regions are expected to have high enough opacity to produce any significant 10$\mu$m silicate absorption, and the centers are masked for this part of the analysis. 
One might be tempted to attribute the 10 $\mu$m emission primarily to continuum emission from stochastically heated very small dust grains. However, we find very different power-law slopes between 10 and 21~$\mu$m, with the 10~$\mu$m showing a steeper slope, similar to the PAH bands. The PAHs are suppressed in \ion{H}{2} regions, likely due to the destruction of the small-grain carriers. There is also the possibility of 9.7$~\mu$m silicate absorption in the 10$~\mu$m band \citep{2007SMITH}. This can uniquely affect the power-law component from \ion{H}{2} region emission, especially since \ion{H}{2} regions are known to have about twice the extinction compared to the more diffuse regions in the disks of normal star-forming galaxies \citep[e.g.][]{2001CALZETTI}.
Taking this all together, the 10~$\mu$m emission can best be described in our analysis as behaving ``like a PAH band but fainter.'' Either the 10$~\mu$m filter is sensitive to silicate absorption, 
%the wings of nearby strong PAH features extend into the 10~$\mu$m filter and produce much of the emission, undiagnosed PAH features contribute to the emission, 
it captures continuum emission from PAH grains, or the small grains that produce the 10~$\mu$m continuum emission are similarly sensitive to destruction as the PAHs themselves. %Future work including spectroscopic follow-up in the mid-IR will be necessary to diagnose the nature of the emission at 10$~\mu$m.

%indicate that F1000W may not be a good tracer of the dust continuum from larger grains, unlike F2100W. More similar power-law slopes and log normal widths between 10 $\mu$m and PAH-tracing 7.7 and 11.3 $\mu$m bands hints that 10 $\mu$m is a good tracer of small grains. %F1000W may thus be a better tracer of continuum emission from small dust grains. 

%The similarity in power-law slopes between PAH-tracing bands and F1000W may hence be attributed to the 

All of these possibilities seem plausible. We might expect the destruction of a wide range of small grains in \ion{H}{2} regions, which include PAHs \citep[see for example][and references therein]{2023EGOROV}. Meanwhile the similarity in log normal widths among all bands may similarly be attributed to PAHs being well-mixed with and generally representative of small grains, which are stochastically heated in the diffuse parts of the disks. Follow-up work involving high-resolution mid-IR spectroscopy in diffuse dusty regions of nearby star-forming galaxies will be necessary to establish the nature of the emission at 10~$\mu$m and its reliability as a small grain continuum tracer.

\subsection{Contrasting the PDFs of Galactic Centers, Disk HII Regions, and the Diffuse ISM} \label{sec:3ComponentContrast}

\begin{figure*}[htbp]
    \centering
    \includegraphics[width=1\textwidth]{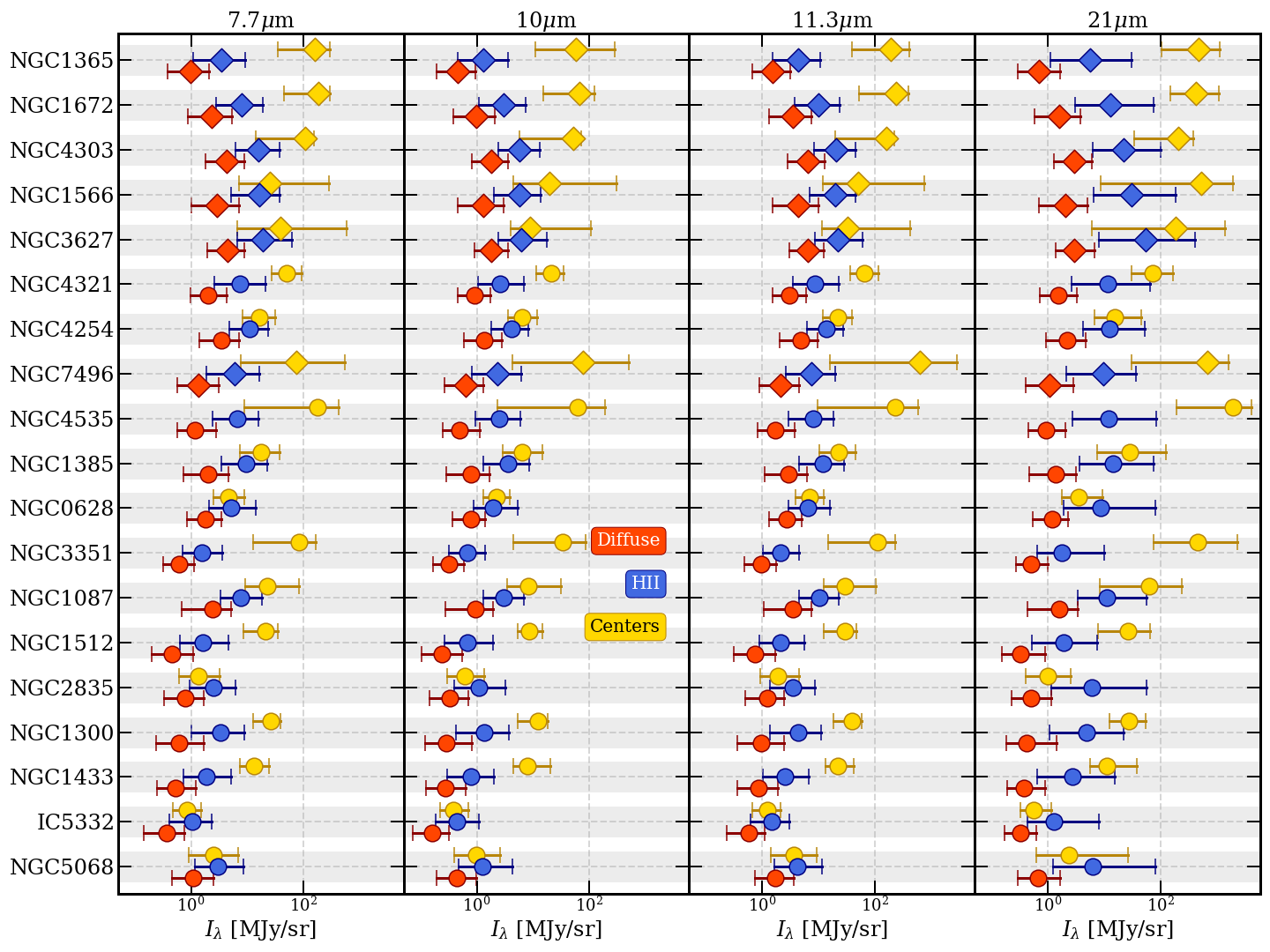}
    \caption{Median and 16-84$^{\rm{th}}$ range of mid-IR intensities in the diffuse ISM (red), \ion{H}{2} region (blue), and central components (yellow) at 7.7, 10, 11.3, and 21~$\mu$m. \edit1{AGN-classified galaxies (see Table \ref{tab:sample} for AGN classification) are represented with diamond markers.} Targets are arranged in decreasing order of SFR from top to bottom. The overall brightness of both the disk and centers generally increase with SFR. The \ion{H}{2} regions and centers (`bright' regions) show a significantly larger dynamical range of intensities at 21~$\mu$m.}
    \label{fig:intensity_contrasts}
\end{figure*}

\begin{figure*}[htbp]
    \centering
    \includegraphics[width=1\textwidth]{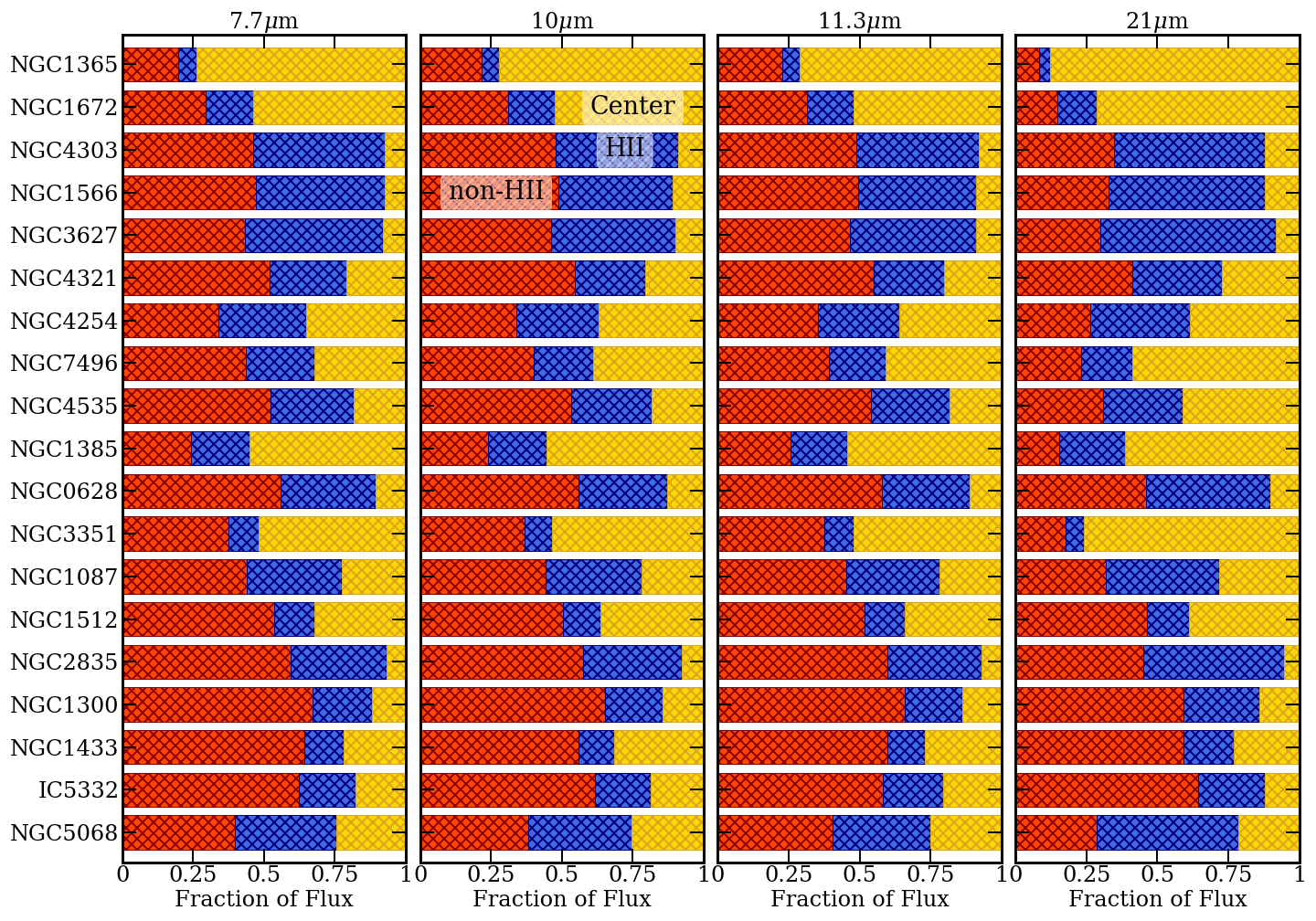}
    \caption{The fraction of total flux coming from the diffuse ISM (red), \ion{H}{2} regions (blue), and central regions (yellow) at 7.7, 10, 11.3, and 21~$\mu$m. Targets are arranged in decreasing order of SFR from top to bottom. While the contribution from the centers vary, a significant fraction of the flux in the disk comes from diffuse emission.}
    \label{fig:flux_fraction}
\end{figure*}

% - intensity contrasts
Above, we argue that the mid-IR PDF of each target naturally breaks into three relatively distinct regimes: galactic centers, star-forming regions in disks, and diffuse emission in disks. This section quantifies and compares the distribution of flux and intensities among these three regions for each target. We show the contrast between the variability of the galactic centers and the consistency of the emission from galactic disks.

Figure \ref{fig:intensity_contrasts} compares the 16--50--84$^{\rm{th}}$ percentile ranges of the PDFs of the galactic centers and the two disk components of each target. Galaxies are sorted from top to bottom in order of decreasing SFR. The overall mid-IR intensity of a galaxy generally increases with increasing SFR (bottom to top), and the centers (in yellow) contrast strongly with the rest of the disk (in blue and red).\edit1{\footnote{While the trend of increasing overall intensities with increasing SFR presented in Figure \ref{fig:intensity_contrasts} is not very strong, the trend is much weaker when arranging targets in order of stellar mass or sSFR. This is also reflected in Figure \ref{fig:spearman_21micron}, where overall intensity terms such as $\mu$ (log normal mean intensity), median intensity in centers, HII regions, and the diffuse ISM all correlate moderately with the SFR and $\Sigma_{\rm SFR}$, but show only weak correlations with $M_*$, and hence with sSFR.}} IC5332, NGC0628, NGC2835, and NGC5068 are exceptions, where the centers resemble the rest of the disks. \edit1{The centers of AGN-classified galaxies (diamond markers) generally show a larger separation in intensities between the disk and center components.} 

Within the disks, the \ion{H}{2} region PDFs show on average 0.5 dex higher median intensities compared to diffuse regions in the 7.7--11.3$~\mu$m range. At 21$~\mu$m, the median intensity of \ion{H}{2} regions \edit1{increases}, leading to 1 dex higher median intensities on average than diffuse regions. The widths of the diffuse PDFs of emission remain broadly consistent across the full sample at all four wavelengths, as previously noted in Section \ref{sec:characterizing_HII_diffuse}. %In the PAH-tracing 7.7, 10, and 11.3~$\mu$m bands, the \ion{H}{2} regions have stable, narrow PDF widths of order \textcolor{red}{X dex}. 
%Although the widths of the diffuse component PDFs remain stable across all four wavelengths, 
In the continuum-tracing 21 $\mu$m band,
%this no longer applies for the PDFs of 
\ion{H}{2} regions span a much broader range of intensities, translating to broader and more variable PDF widths, presumably because, unlike the PAH bands, the 21~$\mu$m captures emission from dust grains that survive in and near \ion{H}{2} regions and are therefore exposed to a wider range of radiation fields. Finally, across all bands, the centers show highly variable PDF widths, reflecting the wide variety of substructure and environments found in galaxy centers.
%where the emission covers a much larger dynamical range in intensities due to thermal heating from environments that span a range of radiation field intensities. 

%%%%%%%%%%%%%%%%%%%

% - flux fractions

The variation in both intensity and physical extent among the three region types
%between the centers and disks 
also leads to variation in the fraction of total flux coming from each region type. We plot these flux fractions for each region and each galaxy in Figure \ref{fig:flux_fraction}. Barred galaxies hosting massive bar-fed star-formation complexes in the central 1-3~kpc with a limited field of view, such as NGC1365, NGC1385, and NGC1672 show the highest central flux fractions.
%leading to a larger flux fraction from the center
% While the relative amount of flux between the center and disk depend on the orientation of our mosaics and FoV, drawing comparisons between the fraction of flux between \ion{H}{2} and ``diffuse" regions within the extended disks themselves remain generally agnostic to the FoV of our images. 
\edit1{NGC3351 is an exception that shows a comparably high intensity range for the central component despite a significantly lower SFR. This high intensity range may at least partially be attributed to NGC3351 being closer (10 Mpc) than NGC1365, NGC1385, and NGC1672 (17-20 Mpc away).}

Within the extended disks, the fraction of flux from within \ion{H}{2} regions for any given target remains roughly consistent across the four filters. On average 30\% of the flux in the disk at 7.7--11.3$~\mu$m, and 40\% of the disk flux at 21$~\mu$m can be attributed to \ion{H}{2} regions, while covering only 12\% of the total area in the imaged disk on average. This flux fraction is in reasonable agreement with previous determinations of similar quantities in these Cycle 1 data \citep{2023LEROY,2023BELFIORE}. This fraction has a weak anti-correlation with the star-formation rate and molecular gas surface density, as we discuss next in Section \ref{sec:globalprops}. 
Beyond these bright star-forming regions, the emission from diffuse regions also contributes a significant fraction of the total mid-IR flux in all four filters, on average 60-70\%. %On average, \textcolor{red}{X\%} of the total mid-IR flux in the disks of star-forming galaxies comes from the diffuse ISM.
%
%Hence, mid-IR emission at such high resolutions provide an alternative tracer for matter in extended low-density environments off the bright star-forming complexes in galaxy discs. This is a direct reflection of the abundance of turbulent substructure captured in the extended regions outside of \ion{H}{2} regions in our mid-IR images.

% \subsection{The F1000W Filter} \label{sec:F1000W}

% Furthermore, since our PDFs are ``weighed" by intensity, $\mu$ roughly marks the 50\textsuperscript{th} percentile in $I$ - half of the total diffuse emission from each target lies above $\mu$ and half of the total diffuse emission lies below $\mu$, which then translates directly to a 50\textsuperscript{th} percentile in luminosity from the volume-filling diffuse phase of the ISM. 
%%%%%%%%%%%% percentile I and flux fraction figures could be useful to quote some numbers here %%%%%%%%%%%%%

\subsection{Effect of Resolution} \label{sec:resolution}

\begin{figure*}[htbp]
    \centering
    \includegraphics[width=0.9\textwidth]{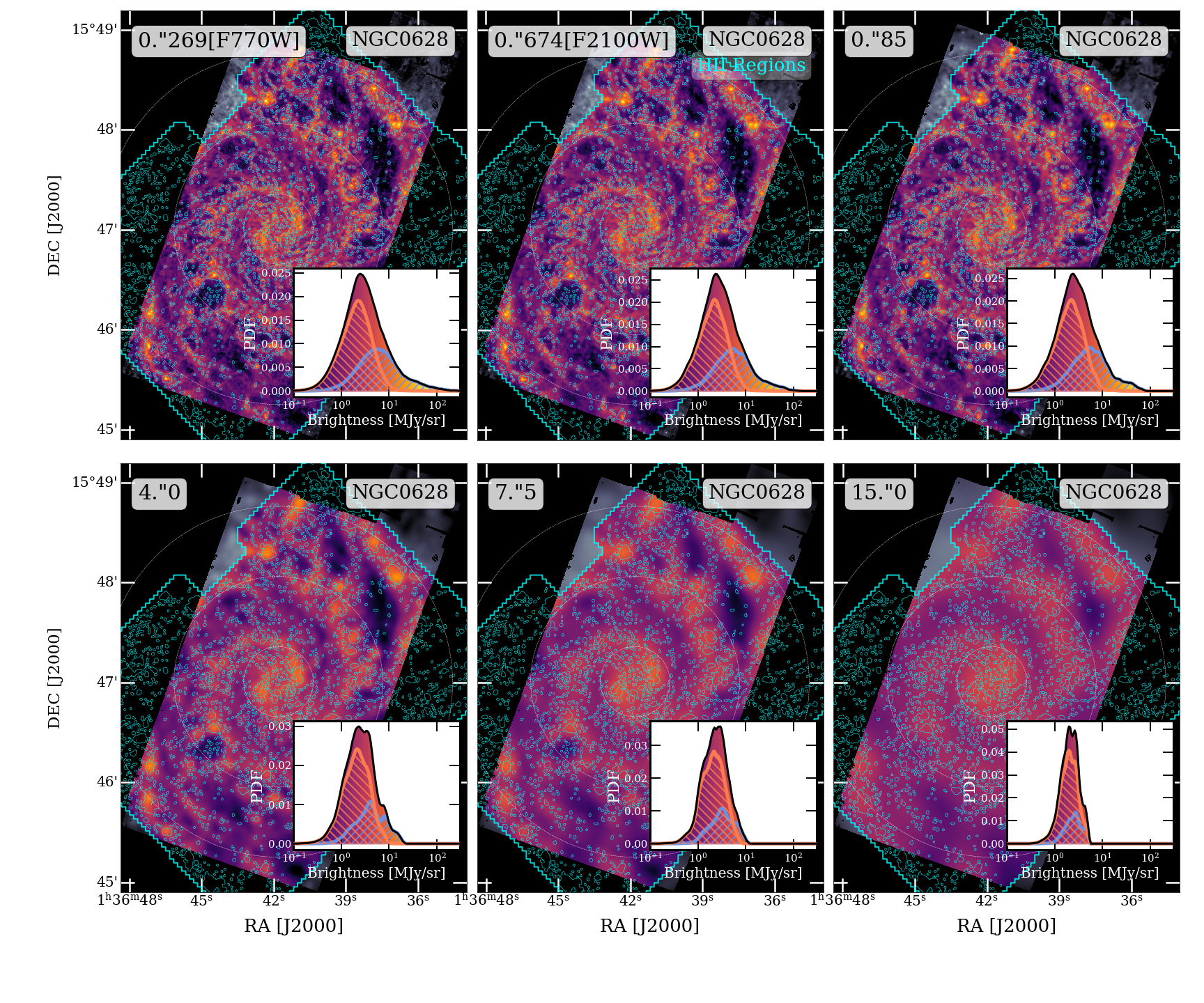}
    \caption{7.7~$\mu$m images and PDFs of the galactic disk of NGC0628 convolved to progressively lower resolutions --- $0\farcs269$ (the native resolution of the F770W filter), $0\farcs674$ (the native resolution of the F2100W filter), $0\farcs85$ (our working resolution), $4\farcs0$, $7\farcs5$, and $15\farcs0$. This corresponds to physical resolutions of 12, 32, 40, 190, 356, and 712 pc, respectively in NGC0628. The two-component PDF only emerges at high angular resolution when individual star-forming regions can be resolved from the surrounding diffuse emission.}
    \label{fig:resolution770}
\end{figure*}

\begin{figure*}[htbp]
    \centering
    \includegraphics[width=0.9\textwidth]{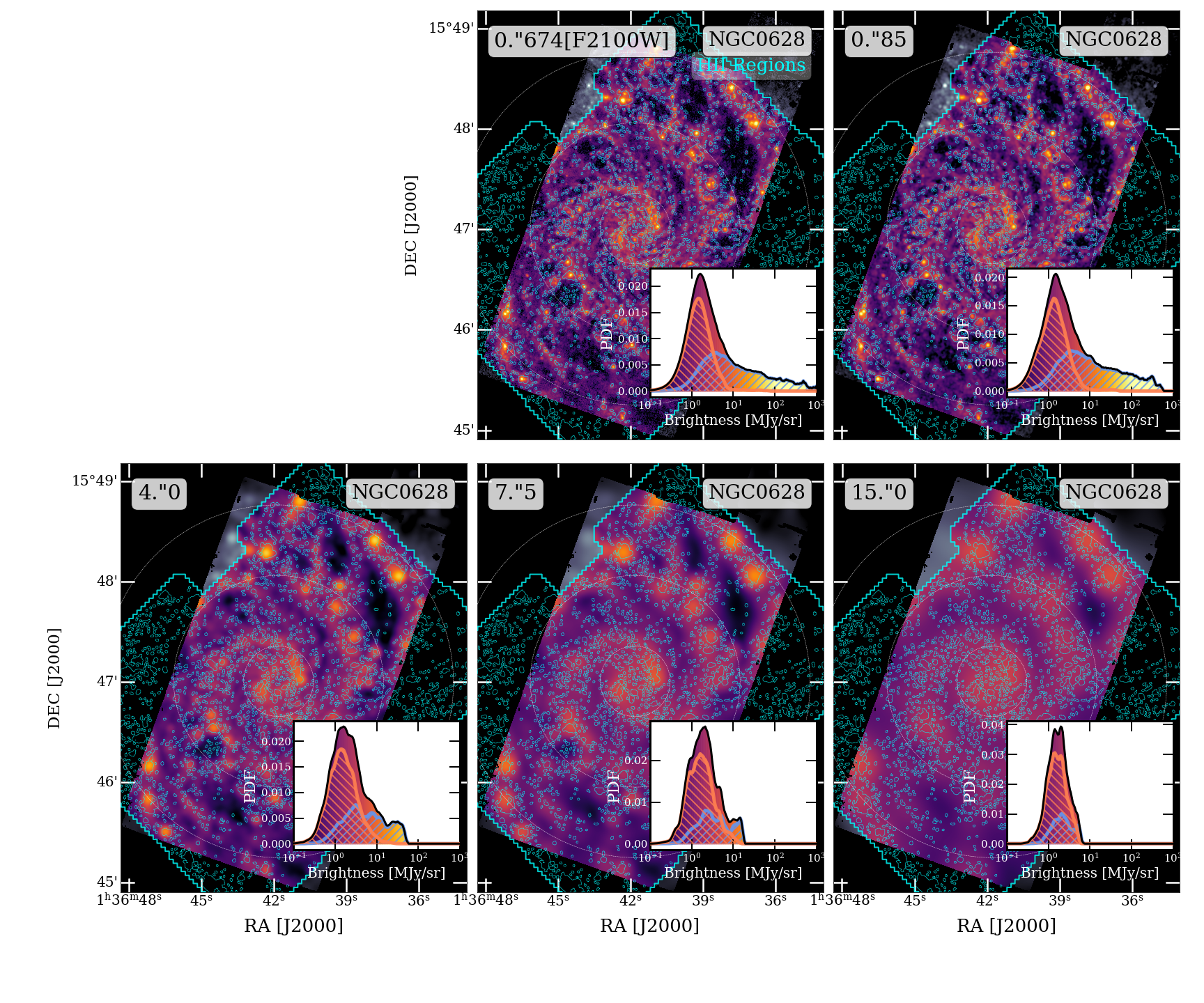}
    \caption{21~$\mu$m images and PDFs of the galactic disk of NGC0628 convolved to progressively lower resolutions, similar to Figure $\ref{fig:resolution770}$ --- $0\farcs674$ (the native resolution of the F2100W filter), $0\farcs85$ (our working resolution), $4\farcs0$, $7\farcs5$, and $15\farcs0$. This corresponds to physical resolutions of 32, 40, 190, 356, and 712 pc, respectively in NGC0628. While the power-law of very prominent at high resolution, the power-law component is rapidly diluted at lower resolutions that do not resolve individual star-forming regions in the disk.}
    \label{fig:resolution2100}
\end{figure*}

% Try: PDF params at native 7.7 micron vs 0.85"
Observing the visible power-law and log normal components in the PDFs requires separating compact mid-IR-bright star-forming regions from the surrounding diffuse ISM. It is therefore expected that the PDF, and in particular its clear separation into these distinct components, would have an appreciable resolution dependence. The extent to which we can distinguish the power-law and log normal components should depend on how well we can resolve individual \ion{H}{2} regions and star-forming complexes. We demonstrate this resolution dependence in Figure \ref{fig:resolution770}. Starting at the native F770W ($0\farcs269$) for a representative target, NGC0628, we convolve the native image with Gaussian kernels to progressively lower angular resolutions \citep[following][]{2011ANIANO}. We include convolutions to our lowest common native resolution at F2100W ($0\farcs674$), our slightly coarser current working resolution at $0\farcs85$, $4\farcs0$ comparable to the resolution of \textit{Spitzer} at $8$ and $24$ $\mu$m bands, $7\farcs5$ and $15\farcs0$, comparable to the resolution of the WISE bands 3 and 4. In NGC0628, these angular resolutions correspond to physical resolutions of 12, 32, 40, 190, 356, and 712 pc respectively.

With each successive smoothing, the separation between the log normal and power-law components of the PDF is blurred. As the resolution in the mid-IR drops below where we can physically resolve the mean separation between giant molecular clouds (GMCs) and \ion{H}{2} regions in star-forming disks, of order 100-300 pc \citep[see e.g.][]{2020CHEVANCE,2022KIM}, the power-law component is no longer visible. Below a physical resolution of a few hundred pc, the overall distribution of intensities collapses to a single narrow peak. This is especially evident in Figure \ref{fig:resolution2100} that shows convolved images of NGC0628 at F2100W, where the power-law component is most prominent at native F2100W (32 pc resolution). At lower resolution ($\gtrsim 200~$pc), the PDFs of \ion{H}{2} and diffuse regions overlap and occupy the same region in intensity space. This is analogous to the observed spatial decorrelation between molecular gas and high-mass star formation (CO and H$\alpha$) on the scale of giant molecular clouds (here: higher resolution, two-component PDF), and their tight correlation on galactic scales (here: lower resolution, single peaked PDF) \citep[e.g.][]{2010SCHRUBA,2010ONODERA,2019KRUIJSSEN,2020CHEVANCE,2022PAN,2022KIM}.
It is evident that the two-component PDF of galactic discs only emerges at high enough angular resolution needed to achieve very high physical resolution ($\lesssim200$~pc) necessary to spatially decorrelate star-formation from the surrounding gas in the star-forming disks of nearby galaxies. More sophisticated follow-up work using a constrained diffusion algorithm to separate the emission and compare the PDFs over different scales is currently underway. %Mederic Boquien's MS student project
%At the resolution of Spitzer or WISE, it would be difficult to distinguish the power-law tail.
% - We present the highest resolution gas column density maps at 0.3" using the native 7.7 micron filter data. The PDF parameters \textcolor{red}{change/do not change} significantly from the 0.85" data.

In addition, within the $0\farcs269$-$0\farcs85$ range (F770W, F2100W, and our common working resolution), PDF parameters such as the power-law slope, log normal mean and width remain stable within $<\pm 5\%$. Since the PDFs remain stable within the native F770W ($0\farcs269$) to $0\farcs85$ range, the fits we provide in Table \ref{tab:decomposed} can be reliably used to describe results at resolutions down to $0\farcs269$.

\subsection{Correlations with Global Properties} \label{sec:globalprops}

\begin{figure*}[htbp]
    \centering
    \includegraphics[width=1.0\textwidth]{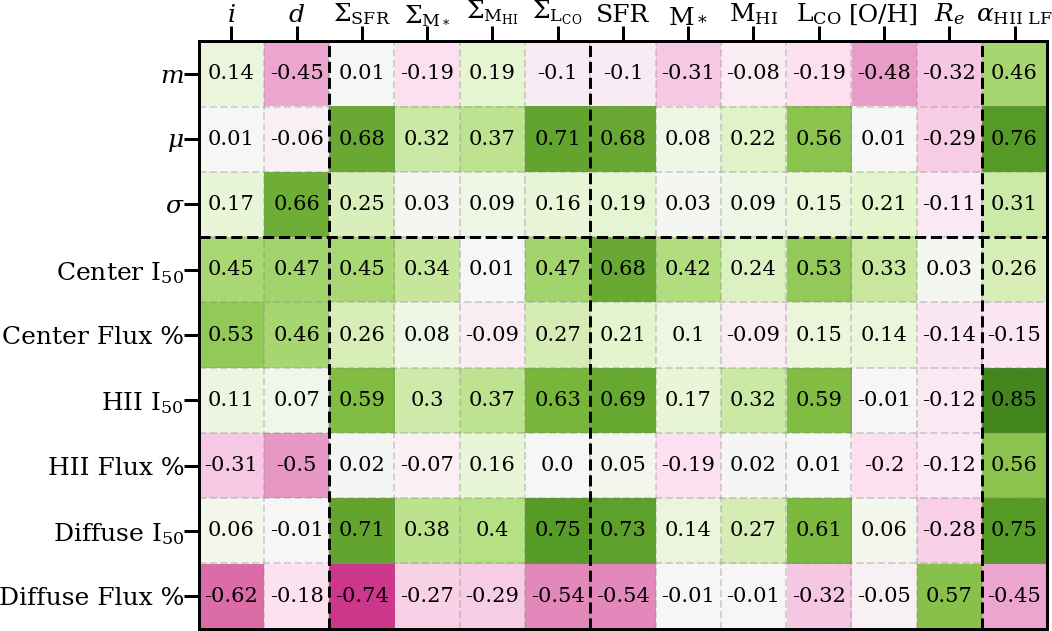}
    \caption{Spearman's rank correlation coefficients between the parameters of mid-IR PDFs of intensity at 21~$\mu$m and global galaxy properties. Stronger positive (negative) correlations are indicated in green (pink). We present rank correlations among the most important observation-specific parameters (galaxy inclination $i$, distance $d$), \edit1{intrinsic} properties (star-formation rate surface density $\Sigma_{\rm SFR}$, stellar mass surface density $\Sigma_{M_*}$, \ion{H}{1} mass surface density $\Sigma_{M_{HI}}$, and \textsc{CO} luminosity surface density $\Sigma_{L_{\rm CO}}$), and \edit1{extrinsic} properties (average SFR, total stellar mass $M_*$, \ion{H}{1} mass $M_{HI}$, \textsc{CO} luminosity $L_{\rm CO}$, average \textsc{[O/H]}, and effective radius $R_e$) as well as the \ion{H}{2} region Luminosity Function slopes $\alpha$ from \citet{2022SANTORO}, with PDF parameters (\ion{H}{2} region power-law index $m$, diffuse log normal mean $\mu$ and dispersion $\sigma$) and region-wise flux contributions (median intensity $I_{50}$ and \% of total flux in FoV for centers, \ion{H}{2} regions, and the diffuse ISM).}
    \label{fig:spearman_21micron}
\end{figure*}

Physically, the two components of the PDFs map to star-forming regions and the diffuse ISM, and the properties of these components might be expected to depend on conditions in the host galaxy (Section \ref{sec:centers} and \ref{sec:HII_nonHII}).
%, and the PDFs remain stable at our high angular (physical) resolutions (Section \ref{sec:resolution}), 
We therefore check for correlations between the parameters of the mid-IR PDFs and global properties of the galaxies.

To do this, we compute Spearman's rank correlation coefficients relating the PDF parameters for the 21~$\mu$m PDF (the \ion{H}{2} region power-law index $m$, the log normal mean $\mu$ intensity and dispersion $\sigma$), and also the median 21$~\mu$m brightness and flux fraction associated with the center, disk \ion{H}{2} regions, and diffuse ISM, to a range of different intensive and extensive global properties of the galaxies in our sample.
%$m$, $\mu$, $\sigma$ to both global %and local environmental properties of our sample. 
Figure \ref{fig:spearman_21micron} presents these correlation coefficients, $\rho$, relating each PDF parameter at 21~$\mu$m to each target property. Higher positive values of $\rho$ up to 1 suggest stronger correlations, and larger negative values down to $-1$ suggest stronger anti-correlations. \edit1{The correlation coefficients at 21~$\mu$m included in Figure \ref{fig:spearman_21micron} are representative of the correlations at 7.7-11.3 $\mu$m. Since the correlation coefficients are similar across the four wavelengths, and the power-law component is most prominent at 21~$\mu$m, we only include the correlations at 21~$\mu$m.}

Among intensive properties, the star-formation rate surface density $\Sigma_{\rm SFR}$ correlates most strongly with the diffuse log normal component - including the log normal mean and median pixel intensity, is strongly anti-correlated with the diffuse flux percentage, and shows a marginally weaker correlation with the median brightness of \ion{H}{2} regions. The CO luminosity surface density $\Sigma_{L_{\rm CO}}$ shows similar correlations with the diffuse and \ion{H}{2} region components \edit1{\citep{2021LEROY}}, while the stellar mass and \ion{H}{1} mass surface densities $\Sigma_{\star}$ and $\Sigma_{M_{\rm HI}}$ do not show particularly strong correlations \edit1{\citep{2022SUN}}.   

Among extensive properties, similar to $\Sigma_{\rm SFR}$ and $\Sigma_{L_{\rm CO}}$, the star-formation rate SFR and total CO luminosity $L_{\rm CO}$ show similar correlation patterns with the diffuse and \ion{H}{2} region components. We do not find strong correlations between mid-IR PDF components and stellar mass $M_*$, \ion{H}{1} gas mass $M_{\rm HI}$, mean metallicity [O/H], or effective size $R_e$.
% check this here: SUN2022
This is consistent with previously reported moderately strong correlations between averaged cloud-scale molecular gas properties and global galaxy properties \citep{2022SUN}. This indicates that the emission in the mid-IR is a good gas mass tracer, especially outside of \ion{H}{2} regions, as we discuss further in Section \ref{sec:non-HII}. %their averaging approach should resemble our PDF construction). 
 
Finally, the \ion{H}{2} region luminosity function slope $\alpha_{\rm HII~LF}$ \citep[from][]{2022SANTORO} shows some interesting correlations with the components of mid-IR PDFs. $\alpha_{\rm HII~LF}$ is the only property that the \ion{H}{2} region power-law index shows any significant correlation with. $\alpha_{\rm HII~LF}$ correlates moderately with the \ion{H}{2} region power-law index and flux contribution, and correlates strongly with the median \ion{H}{2} region brightness at 21$~\mu$m, diffuse log normal mean and median brightness in the diffuse ISM.

% We find a strong correlation between the mean intensity of the log normal component, the variance of the log normal component, and the flux fraction of the diffuse component with the average star-formation rate surface density, $\Sigma_{\rm SFR}$. We find a similar correlation with the CO(2-1) luminosity surface density, $\Sigma_{\rm LCO}$. We find little to no correlation with other global properties such as metallicity, stellar mass surface density, and atomic gas surface density. 

Note that the central flux contributions depend moderately on the size of the field of view for each target relative to the size of the galaxy, and hence the inclination $i$ and target distance $d$. %This may also contribute to the correlation with the inclination and distance to each galaxy. 
In addition, it becomes more difficult to separate the \ion{H}{2} region contribution and diffuse gas contribution for more distant galaxies, since the physical resolution is lower for our fixed $0\farcs85$ angular resolution. The moderate correlation seen between the diffuse log normal component dispersion $\sigma$ and target distance makes this evident, suggesting that galaxies at larger distances may show broadening in the log normal PDF component from some mixing of \ion{H}{2} region emission at lower physical resolution. This is consistent with the increased overlap in \ion{H}{2} region and diffuse component PDFs at lower resolutions (Section \ref{sec:resolution}). Similarly, in more highly inclined galaxies, a larger fraction of the diffuse gas may end up in front of or behind an \ion{H}{2} region, leading to emission that is more difficult to characterize as either star-forming or diffuse. However, this does not result in significant additional broadening (no correlation), since we use inclination-corrected intensities to calculate PDFs.

Building on these results, with the subsequent availability of data on 55 more galaxies from Cycle-2 JWST observations, a dedicated future paper will perform more robust comparison of multi-wavelength global properties with mir-IR emission PDFs across different scales. The paper will focus on interpreting PDF parameters in light of multi-wavelength information, and examining PDFs in different environments and on different spatial scales within each galaxy. The PDFs presented in this work are constructed across large (galactic) scales, which are considerably different than PDFs probing the turbulent driving scale near the disk scale height or within giant molecular clouds. Since the mid-IR is a good tracer of star-formation as well as turbulent gas (as discussed in Section 4.2, and 5.1, respectively), comparing mid-IR PDFs across different scales will also provide a key point of comparison between observations and sophisticated hydrodynamic simulations of the ISM. Finally, we devote the remainder of this paper to further analyze and discuss the \ion{H}{2} and diffuse components of the disk PDFs.
%these correlations will become more robust and may become useful as future predictors of global or local environmental environmental properties.  

\section{Mid-IR Emission from the \textsc{Hii} Region Component} \label{sec:HII}

%As the emission from \ion{H}{2} regions contributes the majority of the high-intensity power-law part of the PDF, the slope of the power-law component is a result of adding up individual bright sources and the few diffuse traced well by \ion{H}{2} region maps. 

As introduced in Section \ref{sec:PDFs}, the emission from mid-IR bright \ion{H}{2} regions combines to yield the high-intensity power-law component of the PDFs of galactic disks. Table \ref{tab:decomposed} and Figure \ref{fig:fit_summary} provide the parameters of the power-law component for our full sample. 
%The overall power-law slopes for each galaxy are consistent with the slopes for \ion{H}{2} regions only, 
Our results above show first that that the optically identified \ion{H}{2} region are able to trace areas of high mid-IR luminosity surprisingly well \citep[in excellent agreement with][]{2023HASSANI}. In detail, the power-law component has some key characteristics. First, it varies by band, appearing most prominent and extended at 21~$\mu$m. Second, the intensities observed provide a direct probe into the distribution of bolometric luminosity surface densities in \ion{H}{2} regions (indicated by the correlation of $m$ with $\alpha_{\rm{HII~LF}}$), both of which we discuss further.

% saved comment:
% rowan.smith: have you considered that in the HII regions that these bands might actually be picking up the CNM? In this paper https://ui.adsabs.harvard.edu/abs/2023MNRAS.524..873S I see the CNM has a lognormal pdf, but at higher column density that the full HI distribution.

% - There isn't 1 threshold above which power-law behaviour dominates for all galaxies, so the transition point is dependent on overall brightness and morphology

%Since the power-law component primarily arises from young star-forming regions, it is a tracer of dust that is heated by young stars. It is important to note here that at our resolution, it is possible to resolve individual star clusters and hence begin to resolve the behaviour of dust in these extreme environments. 

\subsection{The Power-law Component at 21 $\mu$m} \label{sec:F2100W_HII_slope}

The PAH-tracing 7.7 and 11.3~$\mu$m and the small-grain tracing 10~$\mu$m filter (see Section \ref{sec:characterizing_HII_diffuse} for a discussion on the complex origin of the 10$~\mu$m emission) show some key differences from the 21~$\mu$m band. PAH-tracing bands consistently show much steeper power-law slopes than the continuum-dominated 21~$\mu$m, as seen in Figure \ref{fig:PDF_HII} and \ref{fig:logPDF_HII}, and summarized in Figure \ref{fig:fit_summary}, while the PDFs at 21~$\mu$m show significantly more pronounced power-law tails that span a larger dynamic range in intensities. 

This difference in the 21~$\mu$m power-law may arise from two processes. First, a wide range of small dust grains, including PAHs, are destroyed in high radiation pressure dominated regions, which include \ion{H}{2} regions \citep[e.g.,][]{2023CHASTENET,2023EGOROV}. Second, this destruction of very small grains couples with the increased thermal emissivity of the surviving larger dust grains at longer wavelengths beyond 20~$\mu$m \citep{2011DRAINE}. The former suppresses the mid-IR emission from \ion{H}{2} regions in PAH-tracing bands leading to steeper power-law slopes across 7.7 to 11.3~$\mu$m, while the latter enhances the thermal emission from \ion{H}{2} regions at 21~$\mu$m. Together these lead to a more pronounced and extended power-law component with a shallower index at $21$~$\mu$m compared to the other bands.

%This is consistent with the destruction of PAHs near young star clusters, which is where the power-law component originates from, as well as the increased emissivity of dust grains at longer wavelengths.

% Note that the PDFs at 10 $\mu$m, which are expected to trace the dust continuum \citep{2007DRAINE, 2007DRAINELI} show steeper power-law slopes that more consistent with PAH-dominated filters instead of the other continuum-tracing F2100W filter. This intermediate behaviour of the 10 $\mu$m filter adds complexity to interpreting the 10 $\mu$ m PDFs, which are discussed further in Section \ref{sec:F1000W}. %may require further investigation to understand whether this hints at different PAH features or is due to the PSFs of the two adjacent strong PAH emission features bleeding into the 10 $\mu$m filter.

% \textcolor{red}{Merge text on Luminosity Functions from below.}

% \subsection{HII Region and Cluster Luminosity Functions} \label{sec:luminosityfunctions}

\begin{figure}[htbp]
    \centering
    \includegraphics[width=0.45\textwidth]{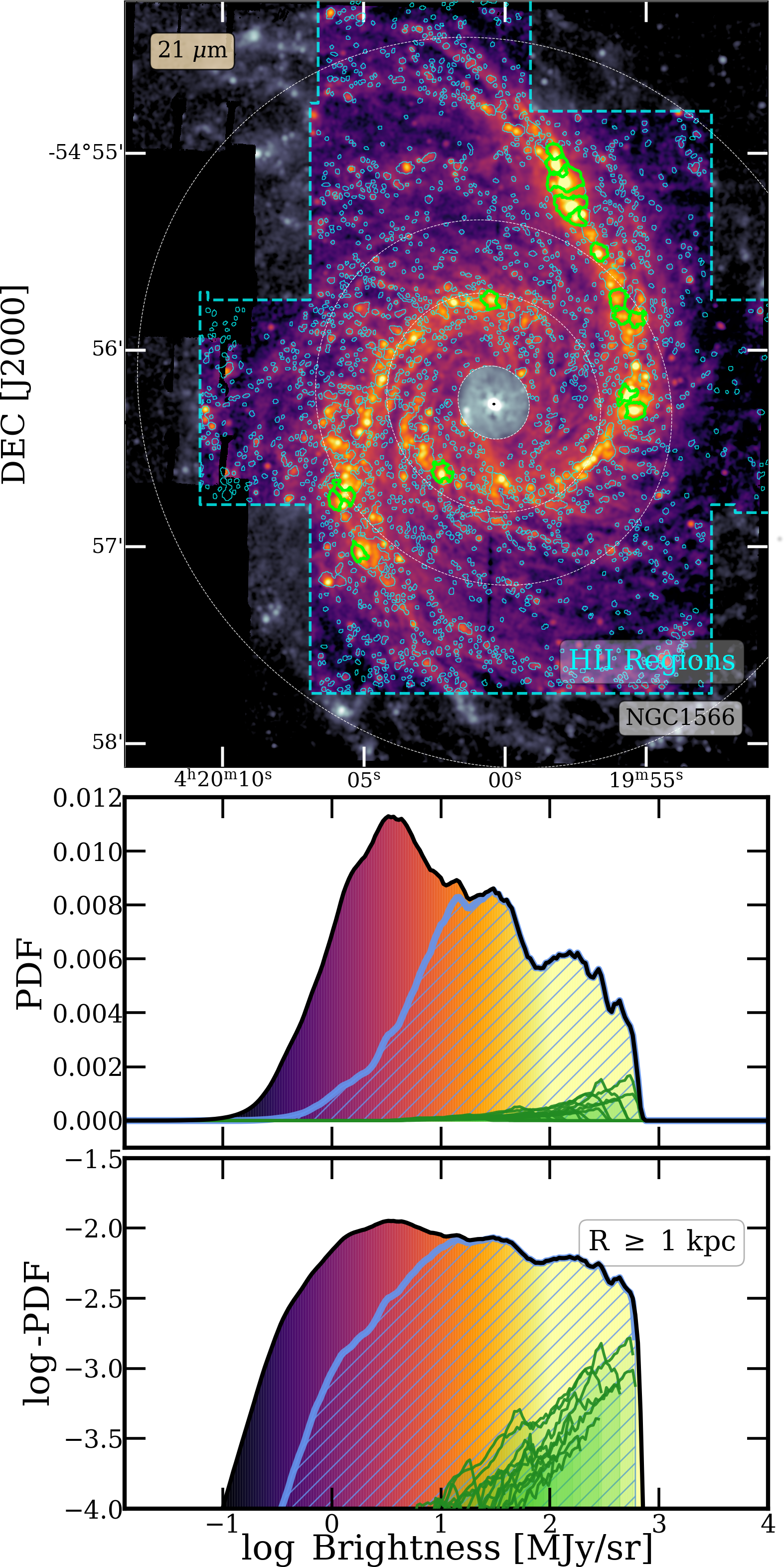}
    \caption{NGC1566 at 21~$\mu$m with all VLT/MUSE-identified \ion{H}{2} regions outlined in blue (top), with the corresponding PDF (middle) and log-PDF in intensity (bottom). A few of the largest \ion{H}{2} regions in NGC1566 are highlighted in green. The overall PDF and log-PDF of intensities at 21~$\mu$m (black), from all \ion{H}{2} regions (blue), and each of the highlighted individual \ion{H}{2} regions (green) are shown for comparison.}
    \label{fig:hii_regs_NGC1566}
\end{figure}
% figure comment: Show a cumulative version of all the hii regions as well.

% \begin{figure*}[htbp]
%     \centering
%     \includegraphics[width=0.9\textwidth]{Figures/Santoro22_slope_comparison.png}
%     \caption{No correlation between slope of \ion{H}{2} luminosity functions $\alpha$ and power-law slopes $m$ from the PDFs of intensity at 7.7, 10, 11.3, and 21 $\mu$m. Corresponding Spearman's correlation coefficients are included. (need to +2 v0p8 galaxies)}
%     \label{fig:hii_LF_slope}
% \end{figure*}

Young massive star clusters \citep[e.g,][]{2010PORTEGEISZWART,2019KRUMHOLZ}, \ion{H}{2} regions \citep{1993BANFI, 2021MASCOOP, 2022SANTORO}, and giant molecular clouds \citep{2001VASQUEZSEMEDENI, 2005ROSOLOWSKY, 2015BURKHART} all show power-law distributions in luminosity. 
The high physical resolution of our mid-IR images allows us to distinguish substructure at the 20--80 pc physical scale. This is roughly the physical resolution at which one or a few resolution elements make up individual \ion{H}{2} regions. Since our resolution allows us to resolve the vast majority of individual \ion{H}{2} regions and OB associations, the power-law component from the \ion{H}{2} region PDFs at the continuum-tracing 21~$\mu$m filter is closely related to -- but is not identical to -- the \ion{H}{2} region luminosity function.
%In particular, the power-law slopes from our PDFs of intensity are analogous to the power-law component in PDFs of column density due  to the competition between turbulence and self-gravity in the disks of spiral galaxies, which is explored further in Section \ref{sec:columndensity}.

Since the power-law component traces the emission from \ion{H}{2} regions, the overall power-law slope of the PDF results from summing up the emission from individual \ion{H}{2} regions. As an illustration, Figure \ref{fig:hii_regs_NGC1566} shows the PDF contributions at 21~$\mu$m from a few of the most massive \ion{H}{2} regions in NGC1566 (in green) overlaid on the overall power-law component (in blue). Each massive \ion{H}{2} region shows its own characteristic shape and spans a range of intensities in PDF-space. 
%The `sawtooth' shape of the PDF of individual mid-IR-bright \ion{H}{2} regions is characteristic of individual bright point-source objects. 
The pixel-wise nature of the PDFs constitutes a key point of distinction between the power-law in PDFs and the \ion{H}{2} region luminosity-function for each galaxy. %While numerous `sawtooth'-shaped PDFs stack up to create the overall power-law slope in PDF-space, each \ion{H}{2} region contributes only a single point in luminosity to the overall luminosity function slope. 
%There is an additional layer of complication 
Due to the size-luminosity relation for \ion{H}{2} regions, where brighter \ion{H}{2} regions tend to be larger in size (i.e., each region spans more pixels) and fewer in number, while fainter regions tend to be smaller (span fewer pixels) and more numerous, the PDF power-law is dominated by larger and fewer bright \ion{H}{2} regions. This is unlike the luminosity function, where each \ion{H}{2} region contributes a single luminosity value (number counts) irrespective of any difference in sizes. So the power-law components in the intensity PDFs capture to first order a convolution of the luminosity function and the size-luminosity relationship for \ion{H}{2} regions. This is reflected in Figure \ref{fig:spearman_21micron}, where the 21~$\mu$m PDF power-law indices correlate moderately (Spearman's rank correlation $\rho = 0.44$) with \ion{H}{2} region luminosity function slopes measured in the optical \citep{2022SANTORO}. We find little to no correlation with the PDF power-law indices at shorter wavelengths 7.7, 10, and 11.3~$\mu$m ($\rho=0.18, 0.14, 0.21$ respectively).

\subsection{Bolometric Luminosity Surface Density and Radiation Pressure} \label{sec:hii_luminosity_surface_dens}

% 21 $\mu$m as a Proxy for Bolometric Luminosity Surface Density
\begin{figure*}[htbp]
    \centering
    \includegraphics[width=1\textwidth]{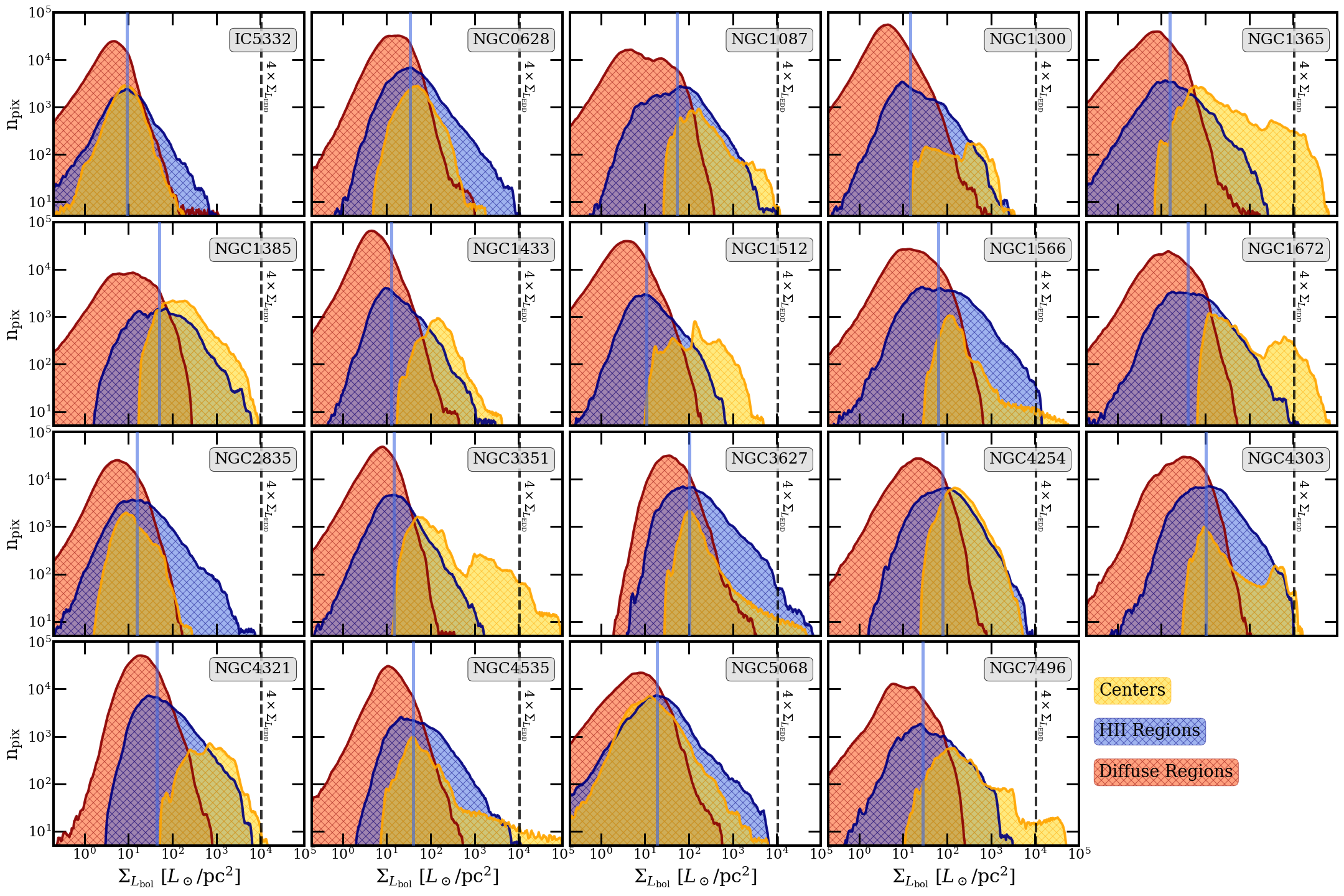}
    \caption{Histograms of pixel-wise bolometric luminosity surface density $\Sigma_{L_{\rm bol}}$ from the center (yellow), diffuse component (red), and \ion{H}{2} regions (blue) for each of our targets. The median $\Sigma_{L_{\rm bol}}$ for \ion{H}{2} regions (vertical blue line) remains roughly consistent between targets. The vertical dashed line shows the estimated Eddington luminosity surface density in star-forming regions, scaled from $\mathcal{F}_{L_{\rm bol}}$ to $\Sigma_{L_{\rm bol}}$ units for direct comparison.} 
    \label{fig:sigma_L_bol}
\end{figure*}

% - Equations on how to go from 21 micron intensity to $\Sigma_{L_{\rm bol}}$ + what assumptions go in

The intensity of emission in the mid-IR directly traces a fraction of the UV and optical starlight that is reprocessed by dust grains. This is true especially at longer wavelengths like 21~$\mu$m, which captures primarily thermal emission from larger dust grains. The bolometric luminosity surface density $\Sigma_{L_{\rm bol}}$ at each pixel can thus be estimated from the observed inclination-corrected mid-IR specific intensity $I_\nu$ values as, 
\begin{equation} \label{eq:Sigma_L_bol}
    \Sigma_{L_{\rm bol}} = f_{\nu_{bol}} 4 \pi \nu I_\nu \cos{i},
\end{equation}
where $f_{\nu_{bol}}$ is the bolometric intensity correction factor such that $I_{bol} = f_{\nu_{bol}} \nu I_\nu$. We assume $f_{{21 \mu\text{m}}_{bol}} \approx 5$ for \ion{H}{2} regions\footnote{While $f_{{24 \mu\text{m}}_{bol}} \approx 10-20$ on galactic scales at 24$~\mu$m \citep[\textit{Spitzer,}][]{2013GALAMETZ}, this factor is expected to be lower in \ion{H}{2} regions because local heating by young massive (ZAMS) stars leads to higher dust temperatures with more of the SED emerging in the mid-IR. On the scale of \ion{H}{2} regions, if one extrapolates the mid-IR SFR estimators by \citet{2007CALZETTI} or \citet{2023BELFIORE} to the fully IR case and adopts a fiducial Starburst99 ZAMS, then$f_{{21 \mu\text{m}}_{bol}} \approx 3-5$. Among our four filters, this estimate is most accurate at F2100W since both 21~$\mu$m (JWST) and 24~$\mu$m (\textit{Spitzer}) are similar tracers of thermal emission from larger dust grains.}.
While the metallicity, dust composition, overall interstellar radiation field (ISRF) strength, and exact geometry of the region can lead to variations in $\Sigma_{L_{\rm bol}}$, Equation \ref{eq:Sigma_L_bol} is a good first order approximation. 
% The inclination corrected luminosity surface densities from the disk of the galaxy would thus be
% \begin{equation}
%     \Sigma_{L_{\rm bol}} = f_{\nu_{bol}} 4 \pi \nu I_\nu \cos{i},
% \end{equation}
% where $i$ is the inclination of the galaxy. 

Scaled to typical 21~$\mu$m intensities in \ion{H}{2} regions for our galaxies, 
\begin{equation}
    \dfrac{\Sigma_{L_{\rm bol}}}{L_\odot/\text{pc}^2} = 223.11 \Bigl (  \dfrac{I_{21 \mu \text{m}}}{10~\text{MJy/sr}} \Bigr ) \Bigl (  \dfrac{\lambda}{21~\mu\mathrm{m}} \Bigr )^{-1} \Bigl (  \dfrac{f_{\nu_{bol}}}{5} \Bigr ) .
\end{equation}

% - Histograms of $\Sigma_{L_{\rm bol}}$ for each target: Figure \ref{fig:sigma_L_bol}
% - Quote typical $\Sigma_{L_{\rm bol}}$ values in \ion{H}{2} regions

Figure \ref{fig:sigma_L_bol} shows the distribution of $\Sigma_{L_{\rm bol}}$ for each target in the centers, disk \ion{H}{2} regions, and diffuse ISM assuming $f_{{21 \mu\text{m}}_{bol}} = 5$. Modulo large differences in SFR (such as IC5332 and NGC5068 which are dwarf galaxies), the median luminosity of \ion{H}{2} regions in the disks of star-forming galaxies \edit1{remains} generally consistent across our sample. \ion{H}{2} regions show typical 21~$\mu$m intensities of about $10 $~MJy sr$^{-1}$, and typical bolometric luminosity surface densities of order $225~L_\odot \text{pc}^{-2}$. 
%This implies that a significant fraction of the total luminosity from \ion{H}{2} regions is reprocessed into the mid-IR. % Most \ion{H}{2} regions are bright in the mid-IR. What about the ones that are mid-IR-dark? 

% \textcolor{red}{- How similar/different are these values to Milky Way values for SF regions?} -> Matt Povich etc.

% \subsection{Radiation Pressure in \ion{H}{2} Regions} \label{sec:radiationpressure}

% \textcolor{red}{Merge these two sections to share a common figure. Merge the text as well.}

% \begin{figure*}[htbp]
%     \centering
%     \includegraphics[width=1\textwidth]{Figures/flux_eddington_luminosity_hists.png}
%     \caption{Histograms of bolometric flux through the plane of the disk $\mathcal{F}_{L_{\rm bol}}$ colored by emission from the center (orange), diffuse component (red), and \ion{H}{2} regions (blue) for each of our targets. An estimate for the Eddington luminosity in a typical star-forming region (dashed black line) is included for comparison. Some regions in the centers and some of the most massive \ion{H}{2} regions come close to or even exceed Eddington luminosity.}
%     \label{fig:Eddington_L_bol}
% \end{figure*}

As a tracer of thermal dust emission, the intensity of emission at~21 $\mu$m likely varies at least approximately linearly with the full bolometric intensity emerging from the young stellar population
%expected to depend linearly on the ambient radiation field strength 
\citep{2007DRAINE}. This allows us to extend the analysis of $\Sigma_{L_{\rm bol}}$ to conduct preliminary force-balance estimates. We can estimate the pixel-wise bolometric flux $\mathcal{F}_{L_{\rm bol}}$ through the plane of the disk as, 
\begin{equation} \label{eq:F_L_bol}
    \mathcal{F}_{L_{\rm bol}} = f_{\nu_{bol}} \pi \nu I_\nu \cos{i}.
\end{equation}
%This quantity is similar to the luminosity surface density $\Sigma_{L_{\rm bol}}$ calculated in Section \ref{sec:hii_luminosity_surface_dens}, with some key differences. Instead of the average luminosity seen through a spherical surface, $\mathcal{F}_{L_{\rm bol}}$ 
This captures the luminosity from a uniformly bright patch of sky (here, each pixel) that is incident on a plane parallel to the disk, measuring the total force or radiation pressure $F_{\rm rad}$ directed out of the plane of the disk of the galaxy. For a plane disk in a galaxy, this reduces to $\Sigma_{L_{\rm bol}}=4 \mathcal{F}_{L_{\rm bol}}$. Scaled to typical values at 21 $\mu$m for a face-on galaxy, 
\begin{equation}
    \dfrac{\mathcal{F}_{L_{\rm bol}}}{L_\odot/\text{pc}^2} = 55.76 \Bigl (  \dfrac{I_\nu}{10 \text{MJy/sr}} \Bigr ) \Bigl (  \dfrac{\lambda}{21 \mu m} \Bigr )^{-1} \Bigl (  \dfrac{f_{\nu_{\rm bol}}}{5} \Bigr ) \Bigl (  \dfrac{\cos{i}}{1} \Bigr ) .
\end{equation}
% Since the conversion factor from mid-IR to TIR intensities is better constrained and modeled, we can use the ratio
% \begin{equation}
%     \dfrac{\nu I_\nu}{\Sigma_{TIR}} = \dfrac{1}{f_{\nu_{TIR}}4\pi}
% \end{equation}
% to estimate the total IR intensity from each pixel, assuming the conversion factor $f_{\nu_{TIR}}$ is stable within the range of environments we probe. (\textcolor{red}{cite}) 

% Figure \ref{fig:Eddington_L_bol} summarizes the distribution of $\mathcal{F}_{L_{\rm bol}}$ in the centers, within \ion{H}{2} regions, and ``diffuse" regions in the disk for each of our targets. 

The force due to radiation pressure can be expressed as, 
\begin{equation} \label{eq:F_rad}
    F_{\rm rad} = \dfrac{\kappa_{\rm gas}}{c}\dfrac{L_{\rm bol}}{4\pi R^2} = \dfrac{\kappa_{\rm gas}}{c}\mathcal{F}_{L_{\rm bol}},
\end{equation} 
\edit1{where $\kappa_{gas}$ is the opacity of gas along the line of sight.}
Since accounting for whether the photons are absorbed or scattered will give a correction factor $<2$, using Equation \ref{eq:F_L_bol}, we estimate the radiation pressure in an arbitrary patch of the disk (here, each pixel) as follows, 
\begin{equation}
    P_{\rm rad} = \dfrac{F_{\rm rad}}{\kappa_{\rm gas}} = \dfrac{\mathcal{F}_{L_{\rm bol}}}{c}.
\end{equation}
$\mathcal{F}_{L_{\rm bol}}$ thus traces the radiation pressure in different environments. Finally, comparing $\mathcal{F}_{L_{\rm bol}}$ with the Eddington luminosity of star-forming regions can indicate whether radiation pressure is the dominant mechanism of feedback. This is expected to be especially important in the central molecular zones and some of the most massive star-forming complexes along the bars of star-forming galaxies, as hosted by many galaxies in our sample.

The Eddington luminosity for a dusty region is given by $L_{\rm EDD} = \dfrac{4\pi G M_{\rm tot} c}{\kappa_{\rm gas}}$. The Eddington luminosity surface density can thus be expressed as follows,
\begin{equation}
    \Sigma_{L_{\rm EDD}} = \dfrac{4\pi G c}{\kappa_{\rm gas}} \Sigma_{M_{\rm tot}}.
\end{equation}
Scaled to a typical value in star-forming regions \citep{2022SUN,2023BLACKSTONE},
\begin{equation} \label{eq:F_Eddington_lumin}
    \dfrac{\Sigma_{L_{\rm EDD}}}{L_\odot/\text{pc}^2} = 2612.163 \Bigl (  \dfrac{\Sigma_{M_{\rm tot}}}{100 M_\odot/\text{pc}^2} \Bigr ) \Bigl (  \dfrac{\kappa_{\rm gas}}{500 \text{cm}^2 / g} \Bigr )^{-1}  .
\end{equation}

Comparing $\Sigma_{L_{\rm bol}}$, and hence $\mathcal{F}_{L_{\rm bol}}$, in different environments with this typical estimate of $\Sigma_{L_{EDD}}$ in star-forming regions, Figure \ref{fig:sigma_L_bol} shows that regions of diffuse emission lie comfortably below the Eddington limit, while only some of the most massive \ion{H}{2} regions and centers come close to the typical Eddington luminosity. All six galaxies in our sample that host central AGN (NGC1365, NGC1566, NGC1672, NGC3627, NGC4303, and NGC7496) have centers that exceed Eddington luminosity. The only outliers are NGC3351 and NGC4535, that do not host AGNs (Table \ref{tab:sample}) but show a significant amount of central emission above typical Eddington luminosities. While the actual values of Eddington luminosity surface densities in the centers may be significantly higher than the estimate in Equation \ref{eq:F_Eddington_lumin} due to higher $\Sigma_{M_{tot}}$ in the galactic centers, it is still evident that there is significant contribution from additional feedback mechanisms beyond radiation pressure in the centers and some of the most massive star-forming regions in our sample. Future work will evaluate the correlations of force balance in star-forming regions as well as quiescent ``diffuse'' parts of the galactic disk more closely by leveraging multi-wavelength information, stellar catalogs, and incorporating results on $q_{\rm PAH}$, $D/G$, and local ionized and molecular gas fraction.

\section{Mid-IR Emission from the Diffuse ISM} \label{sec:non-HII}

\begin{figure}[tbp]
    \centering
    \includegraphics[width=0.5\textwidth]{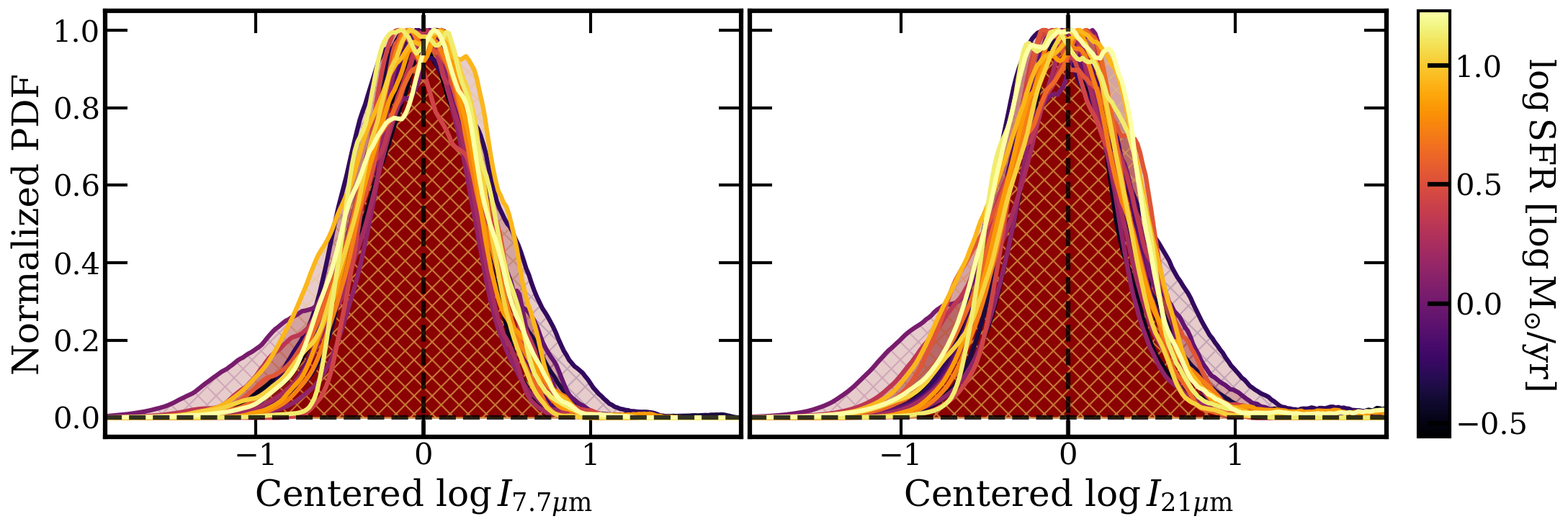}
    \caption{Stacked PDFs of emission from ``diffuse'' regions normalized by mean intensity and amplitude at 7.7 and 21~$\mu$m. The widths of the PDFs remain consistent across targets, with the exception of NGC1300 with a broad PDF.}
    \label{fig:all_lognorms}
\end{figure}

Simply looking at the structure of the lower-density material in mid-IR images, i.e., away from the bright \ion{H}{2} regions and galaxy centers, reveals an inhomogeneous and apparently highly turbulent medium, which is detected almost everywhere in our images (see Appendix \ref{sec:all_images} for 4-band images of the full sample).
%As the power-law component falls off at lower intensities, 
In Section \ref{sec:HII_nonHII}, we found that the PDFs of mid-IR emission from diffuse regions appear log normal. 
Table \ref{tab:decomposed} and Figure \ref{fig:fit_summary} present the best-fit log normal mean and variance fit to the diffuse emission for each band and galaxy in our sample. 
%Figures \ref{fig:PDF_HII} and \ref{fig:logPDF_HII} show the PDFs of emission from outside the VLT/MUSE \ion{H}{2} regions in red for each of our targets. The fit parameters for the log normal PDFs from the ``diffuse" regions, including the mean $\mu$ and standard deviation $\sigma$ are included in Table \ref{tab:decomposed}, as well as the associated uncertainties for our fits. 
%For emission from diffuse regions, the mean brightness $\mu$ varies systemically across the four wavelengths. 

The best-fit peak intensity, $\mu$, varies from band-to-band in a consistent way, with higher values in the PAH-dominated 7.7 and 11.3~$\mu$m bands compared to the continuum dominated 10 and 21~$\mu$m bands. On average, the median ratio of $\mu$ at $7.7$:$10$:$11.3$:$21$ $\mu$m is 1.0:0.46:1.57:0.78 (Table \ref{tab:summary}). These ratios appear at least roughly consistent with the expected SED for emission from dust illuminated by a moderate interstellar radiation field (ISRF) \citep[e.g.,][]{2007DRAINE,2011DRAINE,2018GALLIANO}, and they agree well with band ratio measurements in papers targeting the first four PHANGS-JWST galaxies \citep[][]{2023CHASTENET,2023LEROY} and with band ratio analysis of the full sample currently underway (J. Sutter et al. in preparation).

%While the mean $\mu$ varies from band to band and galaxy to galaxy, 
Meanwhile, the widths of the diffuse component PDFs, captured by the standard deviation $\sigma$, remain impressively consistent across different filters and targets, as shown in Figure \ref{fig:all_lognorms}. The common log normal shape across bands is consistent with the idea that the different mid-IR bands mostly vary together in response to changing dust columns and radiation fields. The uniformity across targets may point to something deeper, a common distribution of ISM column density in the diffuse medium across a diverse set of galaxies. 

As we discuss in the following subsections, with the assumption that mid-IR emission traces gas column density, this is in line with the gravo-turbulent theory of star formation (Section \ref{sec:convirtonh}), which predicts log-normal gas (column) density PDFs in the non-self-gravitating regions of the ISM (Section \ref{sec:columndensity}), with a width that is related to the RMS turbulent Mach number in simple isothermal models (Section \ref{sec:machnumbers}). 
We focus on this interpretation that the shape of the mid-IR PDF from the diffuse disk region may reflect the underlying shape of the gas column density PDF. We briefly review the motivation and mechanics for translating mid-IR intensity to gas column, then we discuss the implications of our measurements for the gas column distribution and the Mach number.  %Our observations reach physical resolutions of 20--80~pc, 
%our images resolve detailed substructure both in star-forming regions as well as in diffuse parts of galaxies 
\edit1{The dust likely traces both the atomic and the molecular gas with sensitivity to fairly low column densities given the high resolution \citep[see][]{2023SANDSTROM}.}
%beyond the Local Group for the first time \citep[e.g., ][]{2023SANDSTROM}. 
We emphasize the novelty of this measurement: these are some of the first measurements to sample the shape of the PDFs of gas column density at such high resolution for galaxies outside the Local Group.

%However, despite the large diversity in substructure and individual morphologies of galaxies, diffuse regions consistently add up to produce a log normal PDF.

\begin{figure*}[htbp]
    \centering
    \includegraphics[width=1\textwidth]{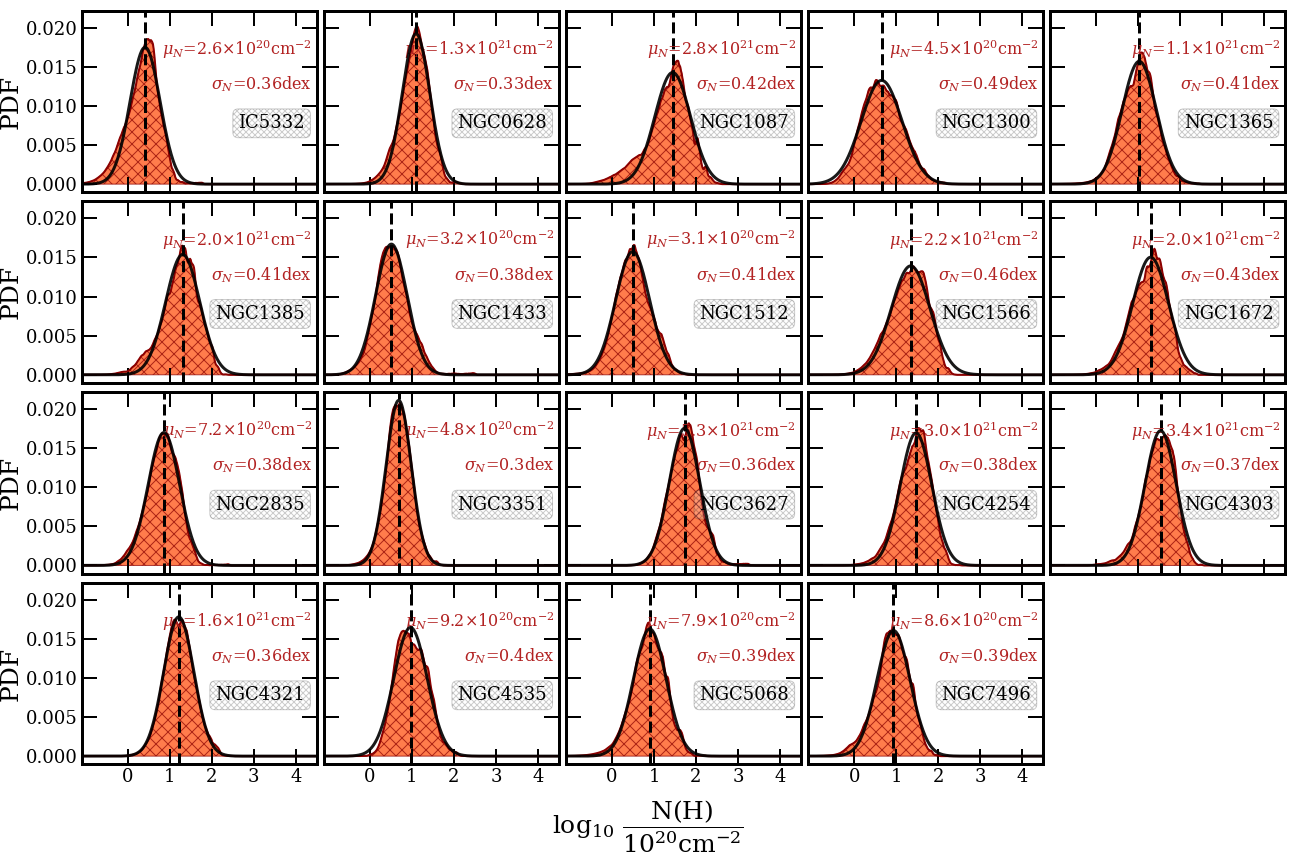}
    \caption{PDFs of gas column density N(H) in diffuse regions (hatched red) derived from 7.7$~\mu$m intensities (Equation \ref{eq:IR_CO} and \ref{eq:CO_NH}) for each galaxy. Log normal fits to each of the PDFs with the fit means are overlaid in black for comparison. The best-fit log normal mean N(H) ($\mu_N$) and dispersion ($\sigma_N$) are included, as in Table \ref{tab:decomposed}.}
    \label{fig:NH}
\end{figure*}

% - How the column density relates to Intensity and our log normal PDF values
%In the mid-IR, also more broadly, the column density $N$ is proportional to the intensity $I$ directly, given a distance. So, a log normal PDF of intensities at such high angular resolution that captures the signatures of turbulent flow, directly translates to a log normal PDF of column densities in the diffuse ISM of these star-forming galaxies. 

\subsection{Conversion from mid-IR Intensity to Gas Column Density}
\label{sec:convirtonh}

Mid-IR intensities reflect both the radiation field heating the dust and the amount of dust present, which in turn depends on the amount of gas present \citep[e.g.,][]{1999DALE}. In regions of very intense heating, the observed mid-IR intensity mostly reflects the heating term and depends on the amount of radiation absorbed by the dust. However, this situation mostly holds within \ion{H}{2} regions and the centers of galaxies. In the more diffuse regions away from \ion{H}{2} regions, the diffuse mid-IR emission should be a good tracer of column density, as the radiation field varies more weakly than the gas column density. This appears \edit1{to be borne out} by a very good local correlation between CO emission, tracing molecular gas column density, and mid-IR emission on $\sim 100{-}200$~pc scales \citep[see][]{2023LEROY,2023SANDSTROM,2023WHITCOMB}. In this case, we might expect the log normal PDFs that we observe to trace the underlying gas column density distribution and to reflect the convolution of turbulence and self-gravity in the diffuse ISM \citep{2017BURKHART}.

In regions of relatively weak radiation fields (away from \ion{H}{2} regions), 
%the intensity of mid-IR emission from dust depends on the projected column density of dust, modulo significant variations in the dust-to-gas ratio, ISRF, and PAH-fraction. So 
the emission in the mid-IR translates to projected gas column densities when the strength of the local IRSF, dust-to-gas ratio ($D/G$) and PAH-fraction ($q_{\rm{PAH}}$) are known. 
Based on the \citet{2007DRAINELI, 2007DRAINE, 2010COMPIEGNE} dust models, as summarized in \citet{2023LEROY},

% \begin{align} 
%     \dfrac{I_\nu^X}{\text{MJy~sr}^{-1}} &\approx 0.022 R_{24}^X \Bigl( \dfrac{N(\rm{H})}{10^{20} \text{cm}^{-2}} \Bigr) \Bigl( \dfrac{D/G}{0.01} \Bigr) \Bigl( \dfrac{U}{U_0} \Bigr) \Bigl( \dfrac{q_{\rm PAH}}{0.046} \Bigr),
% \end{align}

\begin{align} \label{eq:DraineDust_IR_NH}
    \dfrac{I_\nu^X}{\text{MJy~sr}^{-1}} &\propto \Bigl( \dfrac{N(\rm{H})}{10^{20} \text{cm}^{-2}} \Bigr) \Bigl( \dfrac{D/G}{0.01} \Bigr) \Bigl( \dfrac{U}{U_0} \Bigr) \Bigl( \dfrac{q_{\rm PAH}}{0.046} \Bigr),
\end{align}
where $I_\nu^X$ is the mid-IR intensity for filter $X$ and $N(\rm{H})$ is the projected column of density of hydrogen along the line of sight. The additional parameters are $D/G$, the dust-to-gas mass ratio with a typical value of 0.01 from dust models of star-forming galaxies \citep{2007DRAINE, 2013SANDSTROM, 2020ANIANO} and $q_{\rm PAH}$, the fraction of total dust mass in PAHs, which is approximately 0.046 in Milky Way-like galaxies \citep{2007DRAINE, 2007DRAINELI}. The factor of $q_{\rm PAH}$ becomes relevant for PAH-tracing bands such as F770W and F1130W. 
% $R_{24}^X \equiv \Bigl( \dfrac{I_\nu^X}{I_\nu^\text{MIPS24}} \Bigr)$ converts from the MIPS 24 $\mu$m to respective JWST-MIRI filters, which is the typical band ratio as seen in dust models \citep{2011DRAINE}, as presented in \cite{2023LEROY}. 
$U/U_0$ expresses the mean local ISRF in solar neighborhood units from \citet{1983MATHIS}. Directly estimating $N(\rm{H})$ from dust modeling thus involves several moving parts, which are currently being investigated for the full sample of 19 galaxies (see e.g., J. Sutter et al. in prep). 
Future work will use these results to investigate $N(\rm{H})$ PDFs at smaller scales across a diversity of physical environments. 
%While ongoing efforts to better constrain some of these variables in local galaxies are underway, 
In the current paper, we present the PDFs of average column density in the diffuse regions of star-forming disks. Since our PDFs are constructed on large (galactic) scales, we take an alternative and more empirical approach that leverages the strong correlation between mid-IR intensities and CO(2-1) intensities.   

% text below will ger updated depending on whether we use Mid-IR vs CO or CO+HI correlations.
%
\citet{2023LEROY} find that mid-IR emission at F770W, F1000W, F1130W, and F2100W \edit1{correlates} strongly with $I_{\rm CO(2-1)}$ at the 20--80~pc resolution for the first four of our targets observed (NGC0628, NGC1365, NGC7496, and IC5332) and these correlations extend to our full sample of targets (Leroy et al. in prep.). %, we use these intensity corrleations to calculate $N(H)$. 
We use the fiducial empirical correlations between mid-IR intensities and $I_{\rm CO(2-1)}$ that include all 19 galaxies in our sample.
The intensities at 7.7~$\mu$m show the best correlation with $I_{\rm CO(2-1)}$ where the correlation is almost linear, followed by 11.3, 10, and 21~$\mu$m respectively (\citet{2023LEROY}, Leroy et al. in prep), 
\begin{equation} \label{eq:IR_CO}
    I_{\mathrm{CO}(2-1)} = 0.912 I_{7.7\mu m}^{1.06},
\end{equation}
\noindent with the fit conducted primarily at column densities where molecular gas is expected to dominate the total gas mass. Since CO(2-1) is a widely used and robust tracer of molecular gas column density in galaxy disks (modulo subtleties with conversion factors), we subsequently estimate the projected gas column density of H using CO(2-1) intensities in units of K km s$^{-1}$ as
\begin{equation} \label{eq:CO_NH}
    \dfrac{N(\rm H)}{10^{20} \text{cm}^{-2}} \approx 6.2 \left( \dfrac{\alpha_{\rm CO(1-0)}}{\alpha_{\rm CO(1-0)}^{MW}} \right) \left( \dfrac{0.65}{R_{21}} \right) \left( \dfrac{I_{\rm CO(2-1)}}{\text{K km s}^{-1}} \right),
\end{equation}
which translates to
\begin{equation} \label{eq:NH_SigmaMgas}
    \dfrac{\Sigma_{M_{\rm gas}}}{M_\odot \text{pc}^{-2}} \approx  1.089 \dfrac{N(\rm H)}{10^{20} \text{cm}^{-2}},
\end{equation}
assuming $N(\text{H}_2)\approx 0.5 N(\text{H})$ for a typical $R_{21} \equiv I_{\rm CO(2-1)}/I_{\rm CO(1-0)} = 0.65$ \citep{2021DENBROK, 2021YAJIMA, 2022LEROY} and a typical Galactic CO(1-0)-to-H$_2$ conversion factor of $\alpha_{\rm CO(0-1)}^{MW} = 4.35~M_\odot \text{pc}^{-2} ~ (\text{K km s}^{-1})^{-1}$ \citep[e.g,][]{2013BOLATTO}.

% and then the text after should be fine.

\subsection{Column Density PDF from mid-IR emission}
\label{sec:columndensity}

\begin{deluxetable}{lccc}[th!]
\tabletypesize{\small}
\tablecaption{Summary of $N(\rm H)$-PDF Parameters \label{tab:NH_summary}}
\tablewidth{0.5\textwidth}

\tablehead{
\colhead{Best-Fit Parameter} &  \colhead{Median} &   \colhead{Q1/Q3} & \colhead{$\pm 1\sigma$}
}

\startdata
% the data here
Log Normal Mean $\mu_N$             &  $11$ & $6/21$   &  $12$  \\ 
Log Normal Dispersion $\sigma_N$    &  0.39 & 0.36/0.41   &  0.04  \\ 
Mach Number $\mathcal{M}$           &  10.21 & 8.44/11.86 &  3.63  \\ 
\enddata
\tablecomments{
Median, first/third quartiles, and standard deviation of the best-fit log normal mean $\mu_N$ in units of $10^{20}$ cm$^{-2}$ and dispersion $\sigma_N$ in dex of column density PDFs in the diffuse ISM, and derived isothermal turbulent Mach number $\mathcal{M}$.}
\end{deluxetable}

The PDFs of projected gas column densities $N(\text{H})$ for each of our targets are presented in Figure \ref{fig:NH}. The PDFs estimate typical projected gas column densities of order $\sim 10^{21} \text{cm}^{-2}$ in the diffuse ISM of nearby galactic disks, which is consistent with current estimates for the Milky Way \citep[see e.g.][]{2023SMITH}. Table \ref{tab:decomposed} includes the best-fit log normal mean $\mu_N$ and dispersion $\sigma_N$ for the $N(\text{H})$ PDFs. \edit1{These parameters are summarized in Table \ref{tab:NH_summary}.} \edit1{These estimates use empirical correlations between $I_{\text{CO(2-1)}}$ and F770W over a range of column densities where we expect $N(\text{H}) \sim 2 N(\text{H}_2)$. If we use similar empirical correlations from Leroy et al. (in prep.) that also include low-resolution 21~cm HI imaging to estimate $N(\mathrm H) = N($\ion{H}{1}$) + 2 N(\text{H}_2)$, we find $\sim0.05$ dex lower $\sigma_N$ and $\sim0.5\times10^{20}~\text{cm}^{-2}$ higher $\mu_N$ on average, compared to the fiducial correlations using only $I_{\text{CO(2-1)}}$ to estimate the gas, which yield the results reported in Table \ref{tab:decomposed}. The differences between the two cases can be treated as realistic uncertainties on our estimates of gas column density.}
%The PDFs of gas column in diffuse regions for our galaxies estimated empirically are in agreement with previous observations of log normal gas column density PDFs in the diffuse turbulent-driven ISM in the Milky Way. 
The log normal mean $\mu_N$ and dispersion $\sigma_N$ of the $N(\text{H})$ PDFs remain consistent to within $5\%$ and $7\%$ respectively within the $0\farcs269-1\farcs0$ range for all our galaxies (Appendix \ref{sec:resolution_stability}). 

The overall log normal shape of the PDFs of gas column from mid-IR intensities are also in agreement with the well-studied log normal shape of the PDF of column densities in the turbulent gas phase of the local ISM \citep{2001VASQUEZSEMEDENI, 2010BRUNT, 2010FEDERRATH, 2015BURKHART, 2016THOMPSON}. In addition to observations in the Local Group and nearby galaxies \citep[e.g.,][]{2018CORBELLI}, high-resolution simulations of turbulence and feedback-driven outflows predict a log normal PDF of $N({\rm{H}})$ in turbulent diffuse gas. The log normal shape of the PDFs emerges in simulations of both isothermal turbulence, as well as simulations that account for realistic heating and cooling in the neutral (CNM) and molecular phases \citep[see e.g.,][]{2010GLOVER}.
So the key to a log normal gas (column) density PDF is the absence (or non-dominance) of gravity, which allows supersonic turbulence to dominate the dynamics (see, e.g., \citet{2008FEDERRATH}, or \citet{2014GIRICHIDIS} for analytic models). The physical origin of the log normal shape at low densities is still debated \citep[see, e.g.,][]{2015LOMBARDI}, due to the difficulty in mapping out nearby molecular clouds to low extinction ($A_V$).
%It is important to note that the PDFs of local molecular clouds are not well-described by a log normal, but are instead power-laws \citep[e.g.,][]{2015LOMBARDI}. 

Finally, while the PDFs of molecular clouds show similar log normal and power-law components as the overall mid-IR intensity PDFs of disks, the power-law in molecular clouds emerges due to the dominance of self-gravity in dense cores, not due to heating from bright young stars, which is a key distinction. 
%The log normal shape of the PDFs of diffuse regions (outside of \ion{H}{2} regions) are in agreement with previous results from high-resolution observations of turbulent gas within the Milky Way \textcolor{red}{(look up citations)} as well as hydrodynamical simulations of turbulent ISM \textcolor{red}{(look up citations)}. 

% - Estimate luminosity surface brightness: 21 $\mu$m to bolometric intensity correction
% - $\nu I_\nu$ (at 21 micron) to $L_\odot (bol) /pc^2$  to $\Sigma_{tot}$ conversion
% - Histogram of implied luminosity surface densities
% - Typical $\nu I_\nu / \Sigma_{TIR}$ ratio values
% - Estimate Eddington Flux and from luminosity surface density: $F_{EDD} = \dfrac{4\pi Gc}{\kappa_{gas}} \Sigma_{TOT}$

\subsection{Mach Numbers of Isothermal Turbulence} \label{sec:machnumbers}

In the diffuse parts of the galactic disks, we can estimate mean turbulent Mach numbers $\mathcal{M}$ using the dispersion $\sigma_N$ of the PDFs of $N(\rm{H})$ \citep[see][and references therein]{2014KRUMHOLZ}. While this involves assumptions of how the temperatures in the diffuse ISM vary spatially, our $N(\rm{H})$ PDFs capture gas far from strong heating sources (away from centers and \ion{H}{2} regions). These parts of the galaxies are primarily cold neutral or molecular gas dominated, which show a similar dependence of $\mathcal{M}$ on $\sigma_N$ to the isothermal case. We provide upper bounds of isothermal turbulent Mach numbers, where we attribute the entire width of the $N(\rm{H})$ PDF to turbulent broadening. 
Following \citet{2010BRUNT} and \citet{2016THOMPSON}, 
%assuming that the full width of the gas column density PDFs from Section \ref{sec:columndensity} can be attributed to isothermal turbulent motion, 
we can estimate these upper bounds for isothermal ($\gamma = 1$ in $P \propto \rho^\gamma$) turbulent Mach numbers, where the dispersion $\sigma_\rho$ of 3D (volumetric) gas density PDFs \edit1{translates} to Mach numbers of isothermal turbulence $\mathcal{M}$ as \citep{2001OSTRIKER,2002PADOAN,2003LI,2016THOMPSON}, % and also Nordlund99 and Ostriker99?

\begin{align}
    \label{eq:volmach}
    \sigma^2_{\ln \rho} \approx \ln (1+  \mathcal{M}^2/4).
    % \sigma_{\log N} \approx 0.43 \sqrt{\ln(1 + \mathcal{M}^2/4)}
\end{align}

\noindent which neglects any contribution from the magnetic field and assumes 
%mixed turbulence -- 
a natural mix of compressive and solenoidal driving modes \citep[see][]{2012MOLINA}. The spatial scale associated with the Mach numbers we derive from the PDF widths correspond to the galactic scales included in Table \ref{tab:sample}, which allow us to be agnostic about the turbulent driving scale.

Equation \ref{eq:volmach} describes the volume density distribution. Following \citet{2016THOMPSON}, this relationship transforms to an analogous expression relating the dispersion of the projected two dimensional gas column density PDFs, $\sigma_N$, to $\mathcal{M}$ given by
\begin{align}
    \sigma^2_{\ln N} \approx \ln (1+ R \mathcal{M}^2/4), \text{ or, }\\
    \sigma_{\log N} \approx 0.43 \sqrt{\ln(1 + R \mathcal{M}^2/4)}.
\end{align}
Here, $R$ is essentially a conversion factor between 3D and 2D column densities of the form \citep{2016THOMPSON}
\begin{align} \label{eq:R_machnum_alpha}
    R = \dfrac{1}{2} \Biggl ( \dfrac{3-\alpha}{2-\alpha} \Biggr ) \Biggl [ \dfrac{1- \mathcal{M}^{2(2-\alpha)}}{1- \mathcal{M}^{2(3-\alpha)}} \Biggr ],
\end{align}
where $\alpha$ is the index of the underlying 3D density power-spectrum. For $\mathcal{M}\gg1$ (supersonic turbulence), $\alpha \approx 2.5$ \citep[for a summary, see][and references therein]{2014KRUMHOLZ}. Since most astrophysical systems, including turbulent gas in the ISM of galaxies, have $M \gg 1$, we take $\alpha=2.5$ as our standard value. % or: (Kim & Ryu 2005; Beresnyak et al. 2005)
Using $\alpha=2.5$ simplifies Equation \ref{eq:R_machnum_alpha} to an analytic form, using which $\mathcal{M}$ for isothermal turbulence can be estimated as 
\begin{align} \label{eq:sigma_to_mach_final}
    \mathcal{M} \approx 8 [\exp{(\sigma^2_{\log_{10}{N}}/0.18)}-1].
\end{align}

Note that although the location of the peak in the column density PDF depends on whether one is weighting by area or by mass, the dispersion of the PDF is insensitive to this choice \citep[see e.g.][]{2016THOMPSON}. Thus Equation \ref{eq:R_machnum_alpha} should apply to our mass-weighted column density PDFs.

We consider that these inferred Mach numbers are most likely to be upper limits. The widths of the measured PDFs may be due to a variety of factors other than turbulence. For example, the gas column density may be shaped by bulk flows and galaxy dynamics or large scale regional variations in the gas surface density (e.g. due to the exponential disk). 
Any non-circular flow or motion in the disk environment can contribute to observed velocity dispersions, which is present even where the rotation curve shear is zero \citep[see e.g.,][]{2018MEIDT}.

Alternatively our inference of gas column density may be too simplistic, so that we conflate, e.g., variations in the radiation field with column density variations. 
Formally, our estimate does not \textit{have} to be a lower limit; for example, if variations in the gas column density and radiation field anti-correlate, then the width of the \edit1{observed} PDF might be suppressed, and as mentioned above we have neglected any impact of magnetic fields. 

% In addition, while locally in dense molecular shocked regions in supersonic isothermal magneto-turbulence the gas can exhibit high Mach numbers ($\sim10-50$) \citep{2018MOCZ,2019BURKHART}, and our diffuse regions may include some molecular gas contribution, our diffuse regions in galactic disks are primarily neutral gas (\ion{H}{1}) dominated.

We also note that Equations \ref{eq:volmach} through \ref{eq:R_machnum_alpha} formally apply to isothermal turbulence but that the dust is likely mixed with a wide range of ISM phases spanning some temperature range (e.g., the cold molecular phase has $T\sim 20$~K while the warm neutral medium has $T\sim 5,000$~K). If most of the dust in these regions is with associated cold neutral atomic gas \citep[$T \lesssim 250$~K][]{2018MURRAY} or molecular gas ($T \lesssim 50$~K), we expect these relations to at least approximately hold.

With these caveats in mind, applying Equation \ref{eq:sigma_to_mach_final} to the widths in Table \ref{tab:decomposed} suggests likely upper bounds to the mean Mach numbers 
%of isothermal turbulence in the diffuse ISM estimated using our best-fit values of log normal $\sigma$ from column density PDFs. $\mathcal{M}$ shows values 
of order $\mathcal{M} = 5{-}15$. \edit1{The estimated mean Mach numbers are summarized in Table \ref{tab:NH_summary}.}
These are in good agreement with what one would estimate contrasting the sound speeds implied by $T \sim 20$~K molecular gas or $T \sim 200$~K cold atomic gas with the observed velocity dispersion of $\sim 3{-}10$~km~s$^{-1}$ for molecular gas \citep[e.g.,][]{2020SUN,2021ROSOLOWSKY} or $\sim 10$~km~s$^{-1}$ for atomic gas \citep[e.g.,][]{2009TAMBURRO}. Thus despite a fair number of caveats, it seems reasonable that the width of the approximately log normal column density distribution that we measure does appear roughly consistent with expectations from isothermal turbulence.

\section{Summary} \label{sec:summary}

In this paper, we present for the first time the PDFs of mid-IR intensity at an angular resolution of $0\farcs3-0\farcs85$ (corresponding to a physical resolution of 20--80~pc) for a representative sample of nearby low-inclination star-forming galaxies with multi-wavelength coverage (Section \ref{sec:data}). We compare the PDFs across the first 19 PHANGS galaxies observed with JWST-MIRI at 7.7~$\mu$m, 10~$\mu$m, 11.3~$\mu$m, and 21~$\mu$m (Section \ref{sec:PDFs}). Together this wavelength range captures stochastic emission from small dust grains including PAHs, as well as thermal emission from the dust continuum. We decompose these PDFs into emission from the galactic centers and galactic disks (Sections \ref{sec:centers}, \ref{sec:extended_disks}). We then parameterize and find the components of the PDFs of the galactic disks using MUSE H$\alpha$-identified \ion{H}{2} regions (Section \ref{sec:HII_nonHII}) as follows: %Finally, decoupling the emission from extreme central enviraonments and the rest of the extended disks for galaxies with central AGNs, intense nuclear star clusters, etc. reveals consistent PDFs for all 16 of our star-forming galaxies, summarized as follows:

\begin{enumerate}
    
    \item The disks of nearby star-forming galaxies show consistent PDF shapes in the mid-IR with an approximately log normal component that peaks at lower intensities, and a power-law tail that emerges at higher intensities (Section \ref{sec:HII_nonHII}). 
    
    \item Based on optically identified \ion{H}{2} regions, the power-law component traces emission from \ion{H}{2} regions, while the log normal component results from the emission from the diffuse ISM (Figures \ref{fig:PDF_HII} and \ref{fig:logPDF_HII}).

    \item We provide the relative flux fractions and typical range of mid-IR intensities for the galactic centers, disk \ion{H}{2} regions, and diffuse emission for all 19 galaxies (Section \ref{sec:3ComponentContrast}). 

    \item We compute correlations between the intensity PDF components and global properties of our targets (Section \ref{sec:globalprops}). The diffuse log normal component correlates strongly with the star-formation rate and gas surface density, while the \ion{H}{2} region power-law index shows a moderate correlation with \ion{H}{2} region luminosity function slopes. 
    
    \item The \ion{H}{2} region component (power-law) is most prominent (shallowest) and extended at 21~$\mu$m, which traces thermal emission from dust grains that survive in \ion{H}{2} regions (Section \ref{sec:F2100W_HII_slope}). The power-law slope captures a convolution of the \ion{H}{2} region luminosity function and size-luminosity relation. 
    
    \item We use the emission at 21~$\mu$m to estimate the distribution of luminosity surface densities, and show that most \ion{H}{2} regions remain well below the typical Eddington luminosity surface density in star-formation regions (Section \ref{sec:hii_luminosity_surface_dens}). We find typical 21$\mu$m intensities in \ion{H}{2} regions of about 10~MJy/sr, and bolometric luminosity surface densities of about 225~$L_\odot/$pc$^2$. 

    \item Diffuse region PDFs (log normal) show higher mean intensities at 7.7~$\mu$m and 11.3~$\mu$m due to PAH emission, and an overall depression at the continuum-tracing 10~$\mu$m and 21~$\mu$m filters (Section \ref{sec:HII_nonHII}). 
    
    \item The shape of the diffuse PDF remains strikingly log normal across our full sample of galaxies that span the star-forming main sequence (Figure \ref{fig:all_lognorms} and \ref{fig:fit_summary}; Table \ref{tab:summary}). While the mean varies as a function of environment and band, the log normal dispersion (width) remains generally consistent in the mid-IR across our full sample ($0.3 \lesssim \sigma \lesssim 0.4$ dex).     
    
    \item We leverage the tight correlation between 7.7$~\mu$m emission and tracers of gas column density to construct unprecedented high resolution ($0\farcs269$) PDFs of projected gas column density in the diffuse ISM of nearby galaxies (Section \ref{sec:convirtonh}). We find median column densities of order $10^{21}\text{cm}^{-2}$ (Table \ref{tab:decomposed}, Figure \ref{fig:NH}). 
    
    \item Assuming that the full width of the gas column density PDFs can be attributed to isothermal turbulence, we measure isothermal turbulent Mach numbers (upper limits) of order 8--15 in diffuse regions (Section \ref{sec:machnumbers}).  
    
\end{enumerate}

\section{Acknowledgements} \label{sec:acknowledgements}

\edit1{We thank the anonymous referee for their thoughtful and constructive feedback which greatly improved the paper.}

This work was carried out as part of the PHANGS collaboration.

DP gratefully acknowledges support from the NSF GRFP.

AKL gratefully acknowledges support by grants 1653300 and 2205628 from the National Science Foundation, by award JWST-GO-02107.009-A, and by a Humboldt Research Award from the Alexander von Humboldt Foundation.

LAL acknowledges support by the the Heising-Simons Foundation and the Simons Foundation. This work was performed in part at the Simons Foundation Flatiron Institute's Center for Computational Astrophysics during LAL's time as an IDEA Scholar.

JS acknowledges support by the Natural Sciences and Engineering Research Council of Canada (NSERC) through a Canadian Institute for Theoretical Astrophysics (CITA) National Fellowship.

MB gratefully acknowledges support from the ANID BASAL project FB210003 and from the FONDECYT regular grant 1211000.

JC acknowledges funding from the Belgian Science Policy Office (BELSPO) through the PRODEX project “JWST/MIRI Science exploitation” (C4000142239).

MC and LR gratefully acknowledge funding from the DFG through an Emmy Noether Research Group (grant number CH2137/1-1). COOL Research DAO is a Decentralized Autonomous Organization supporting research in astrophysics aimed at uncovering our cosmic origins.

EWK acknowledges support from the Smithsonian Institution as a Submillimeter Array (SMA) Fellow and the Natural Sciences and Engineering Research Council of Canada.

TGW acknowledges funding from the European Research Council (ERC) under the European Union’s Horizon 2020 research and innovation programme (grant agreement No. 694343).

FB would like to acknowledge funding from the European Research Council (ERC) under the European Union’s Horizon 2020 research and innovation programme (grant agreement No.726384/Empire).

ER acknowledges the support of the Natural Sciences and Engineering Research Council of Canada (NSERC), funding reference number RGPIN-2022-03499.

KMS and JS acknowledge the support of JWST-GO-02107.006-A.

This work uses observations made with the NASA/ESA/CSA James Webb Space Telescope. The data were obtained from the Mikulski Archive for Space Telescopes at the Space Telescope Science Institute, which is operated by the Association of Universities for Research in Astronomy, Inc., under NASA contract NAS 5-03127 for JWST. These observations are associated with program 2017. The specific observations analyzed can be accessed via \dataset[ 10.17909/9bdf-jn24]{http://dx.doi.org/10.17909/9bdf-jn24}.

Also based on observations collected at the European Southern Observatory under ESO programmes 094.C-0623 (PI: Kreckel), 095.C-0473,  098.C-0484 (PI: Blanc), 1100.B-0651 (PHANGS-MUSE; PI: Schinnerer), as well as 094.B-0321 (MAGNUM; PI: Marconi), 099.B-0242, 0100.B-0116, 098.B-0551 (MAD; PI: Carollo) and 097.B-0640 (TIMER; PI: Gadotti).

\vspace{5mm}
\facilities{JWST,VLT/MUSE}

\software{astropy \citep{ASTROPY13,ASTROPY18}}

\bibliography{sample631}{}

\appendix
\section{Full Sample MIRI Images and PDFs} \label{sec:all_images}

\begin{figure*}[htbp] \label{fig:ic5332}
    \centering
    \includegraphics[width=1\textwidth]{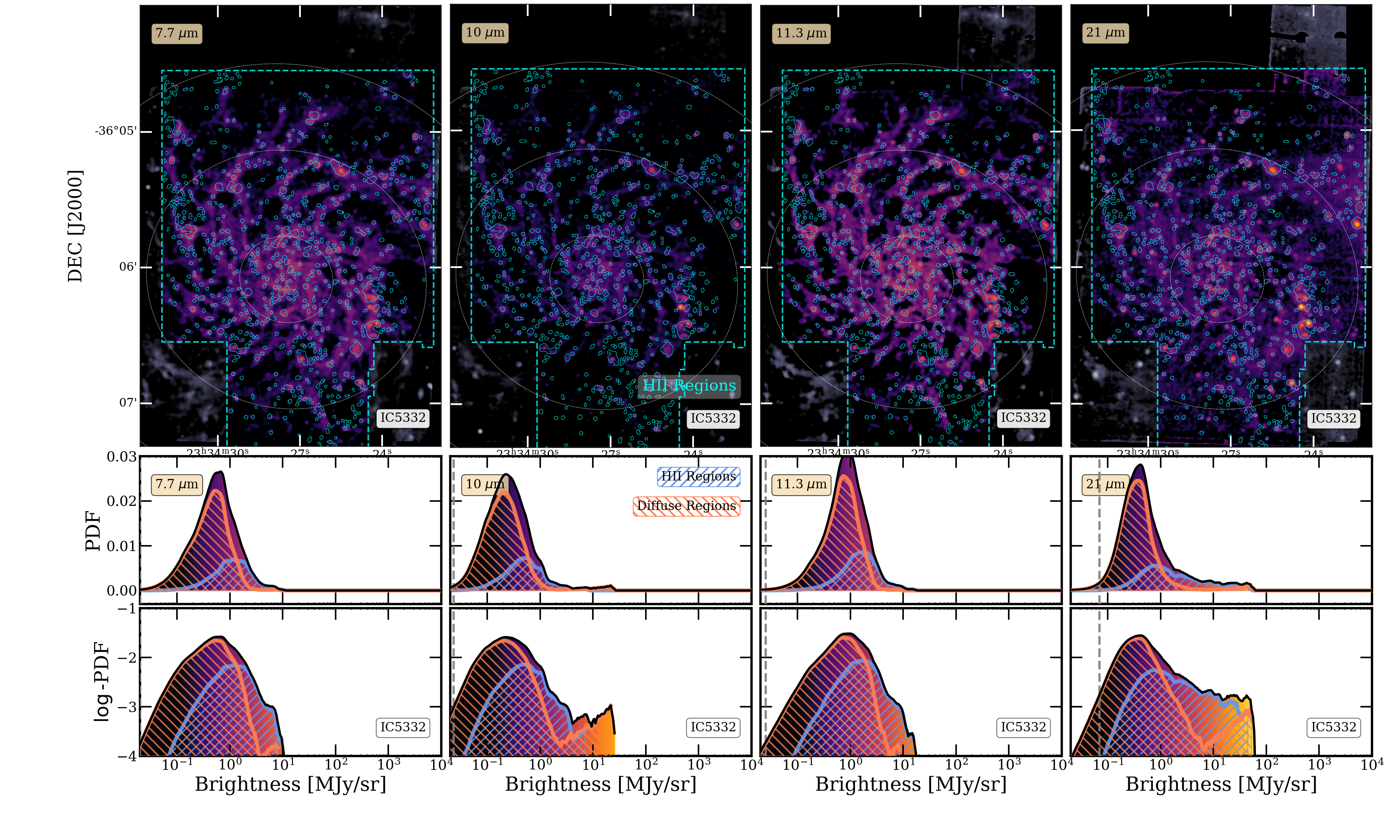}
    \caption{Mid-IR images and PDFs of emission of the galactic disk at 7.7~$\mu m$, 10~$\mu m$, 11.3~$\mu m$, and 21~$\mu m$ respectively for IC5332.}
\end{figure*}

This Appendix presents the mid-IR images and corresponding PDFs of the full PHANGS-JWST Cycle-1 sample of 19 galaxies \citep{2023LEE}. This includes Figures \ref{fig:ic5332} (18) to \ref{fig:ngc7496} (36), where the top panels show JWST-MIRI images of the galactic disks at 7.7, 10, 11.3, and 21$~\mu$m, convolved to $0\farcs85$, with VLT-MUSE-identified \ion{H}{2} regions overlaid in cyan. Our analysis is restricted to the footprint of VLT-MUSE H$\alpha$ mapping for each target \citep{2022EMSELLEM}. The middle panels show the PDFs of the entire footprint (centers and disks) in fainter dashed lines, and the PDFs of the galactic disks overlaid in bold. The PDFs of emission from \ion{H}{2} regions and the diffuse ISM are shown as blue and orange hatched regions, respectively. Finally, the bottom panels show the same PDFs in log-stretch. As argued in Section \ref{sec:HII_nonHII}, the PDFs of galactic disks are separable into a power-law component from \ion{H}{2} regions, and a log-normal component from diffuse regions. 

\begin{figure*}[htbp]
    \centering
    \includegraphics[width=1\textwidth]{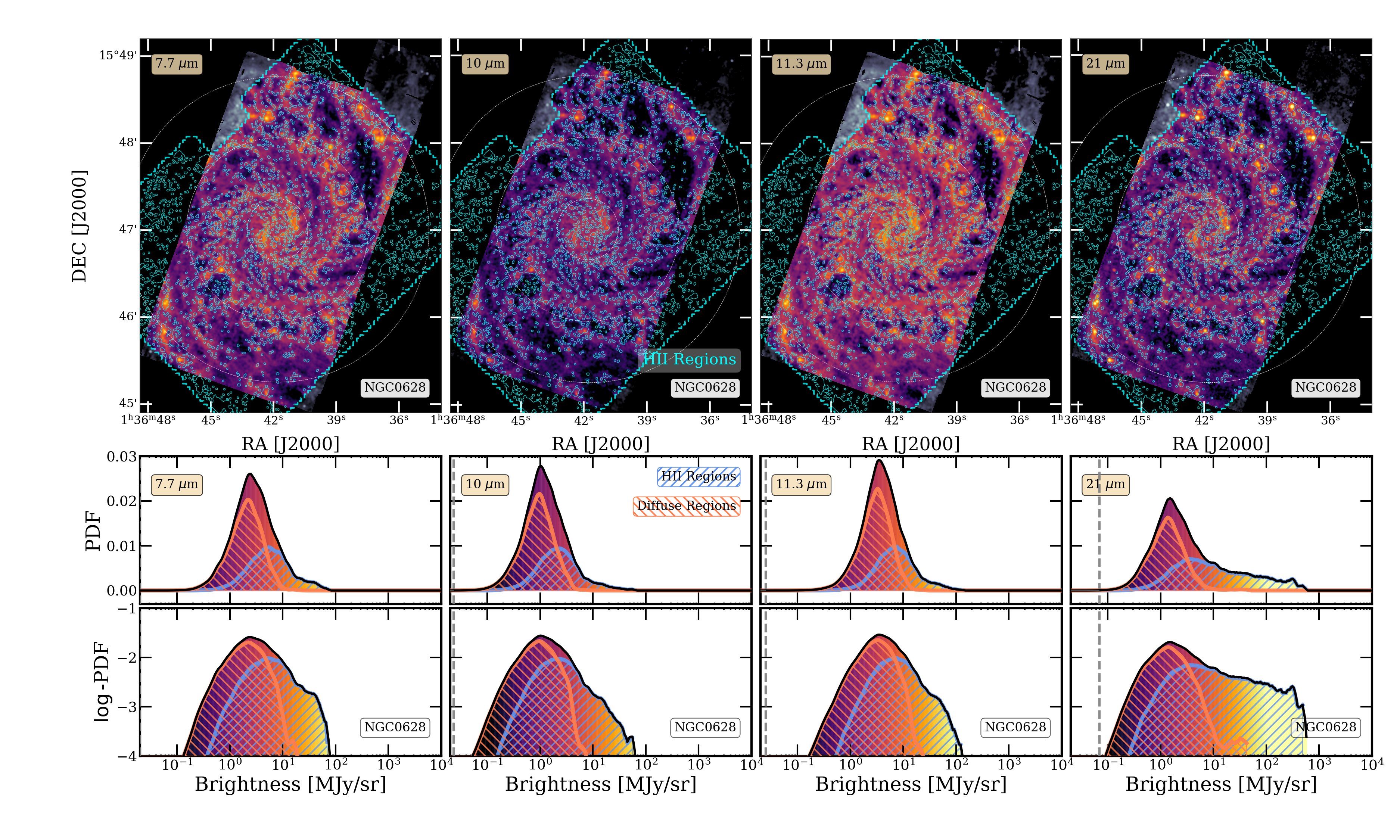}
    \caption{Mid-IR images and PDFs of emission of the galactic disk at 7.7~$\mu m$, 10~$\mu m$, 11.3~$\mu m$, and 21~$\mu m$ respectively for NGC0628.}
\end{figure*}

\begin{figure*}[htbp]
    \centering
    \includegraphics[width=1\textwidth]{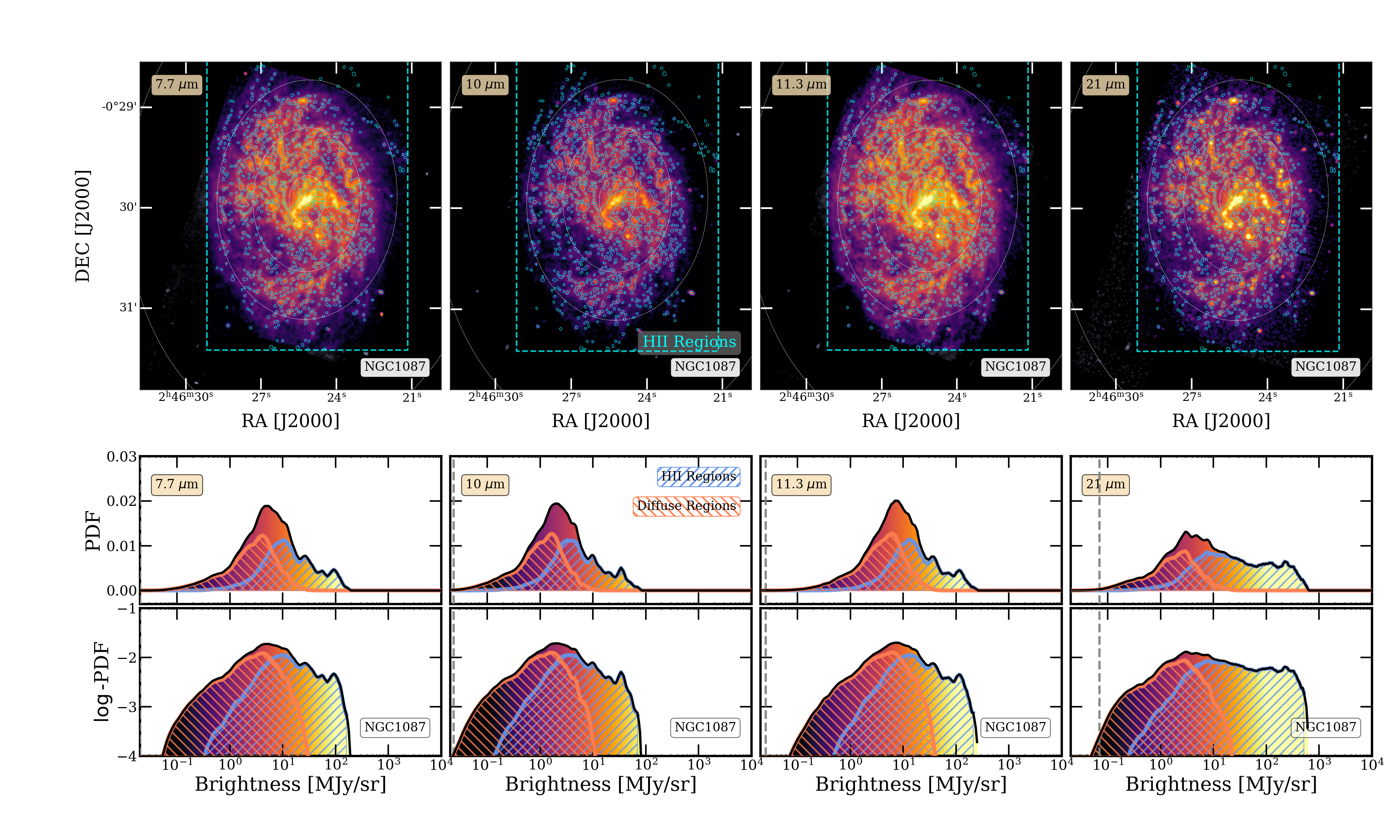}
    \caption{Mid-IR images and PDFs of emission of the galactic disk at 7.7~$\mu m$, 10~$\mu m$, 11.3~$\mu m$, and 21~$\mu m$ respectively for NGC1087.}
\end{figure*}

\begin{figure*}[htbp]
    \centering
    \includegraphics[width=1\textwidth]{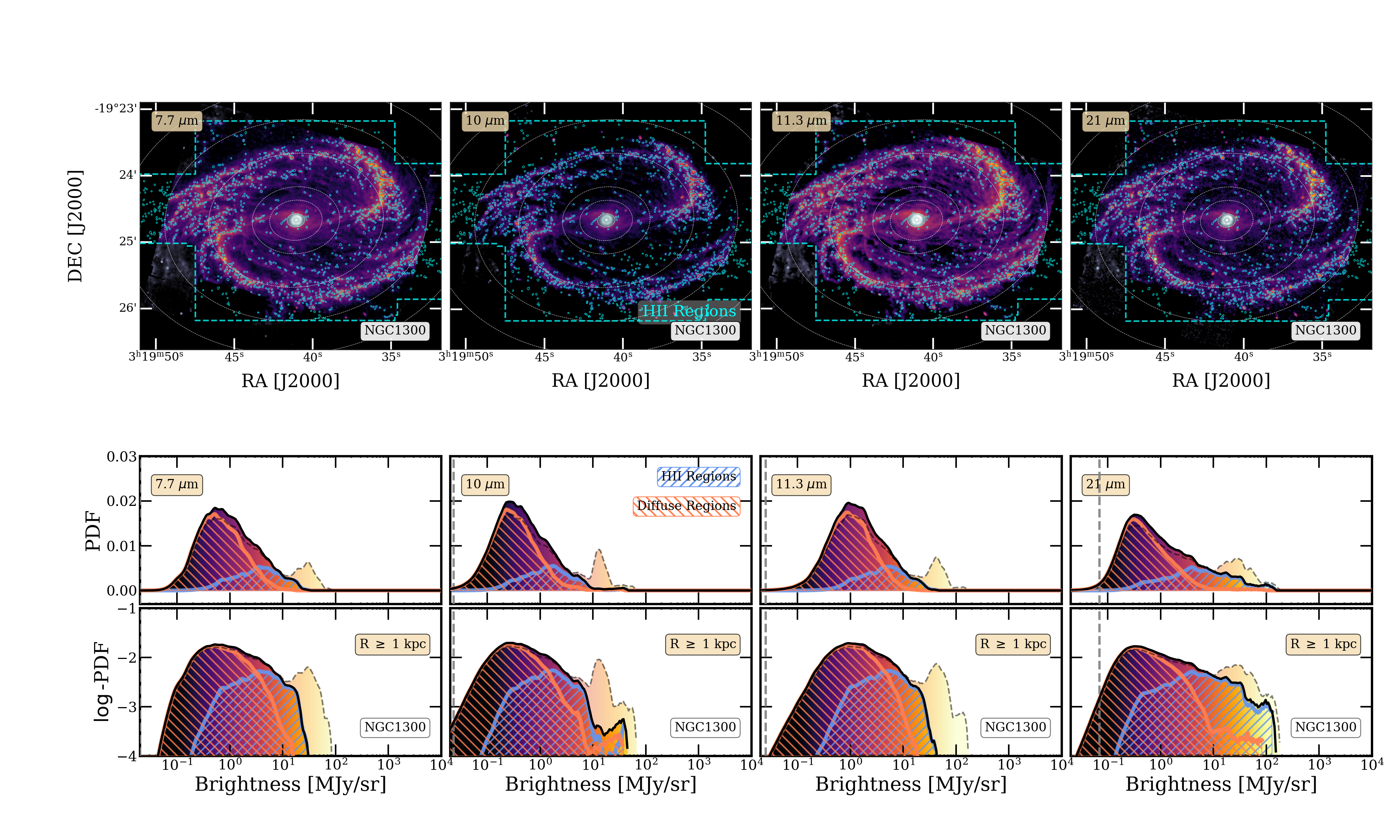}
    \caption{Mid-IR images and PDFs of emission of the galactic disk at 7.7~$\mu m$, 10~$\mu m$, 11.3~$\mu m$, and 21~$\mu m$ respectively for NGC1300.}
\end{figure*}

\begin{figure*}[htbp]
    \centering
    \includegraphics[width=1\textwidth]{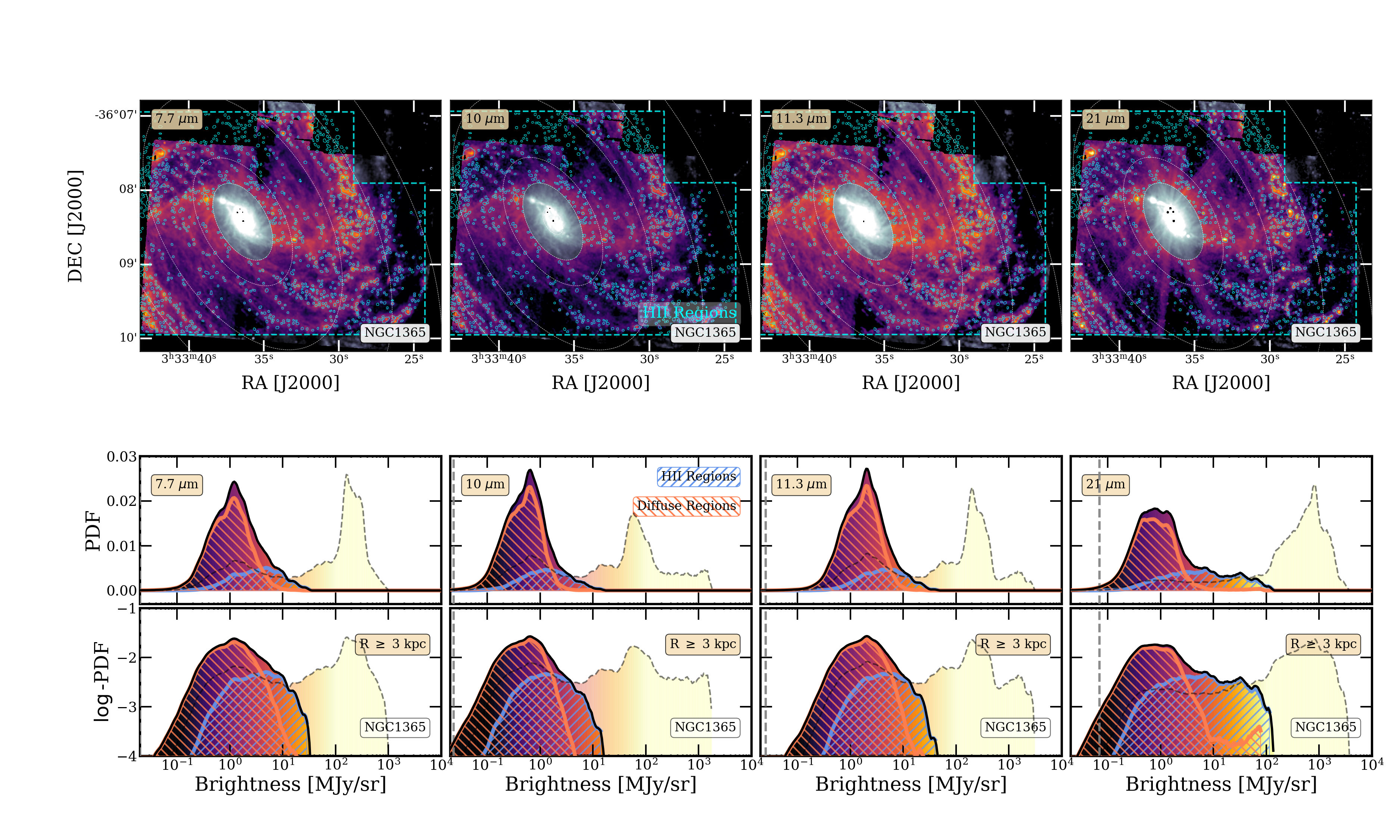} \label{fig:ngc1365}
    \caption{Mid-IR images and PDFs of emission of the galactic disk at 7.7~$\mu m$, 10~$\mu m$, 11.3~$\mu m$, and 21~$\mu m$ respectively for NGC1365.}
\end{figure*}

\begin{figure*}[htbp]
    \centering
    \includegraphics[width=1\textwidth]{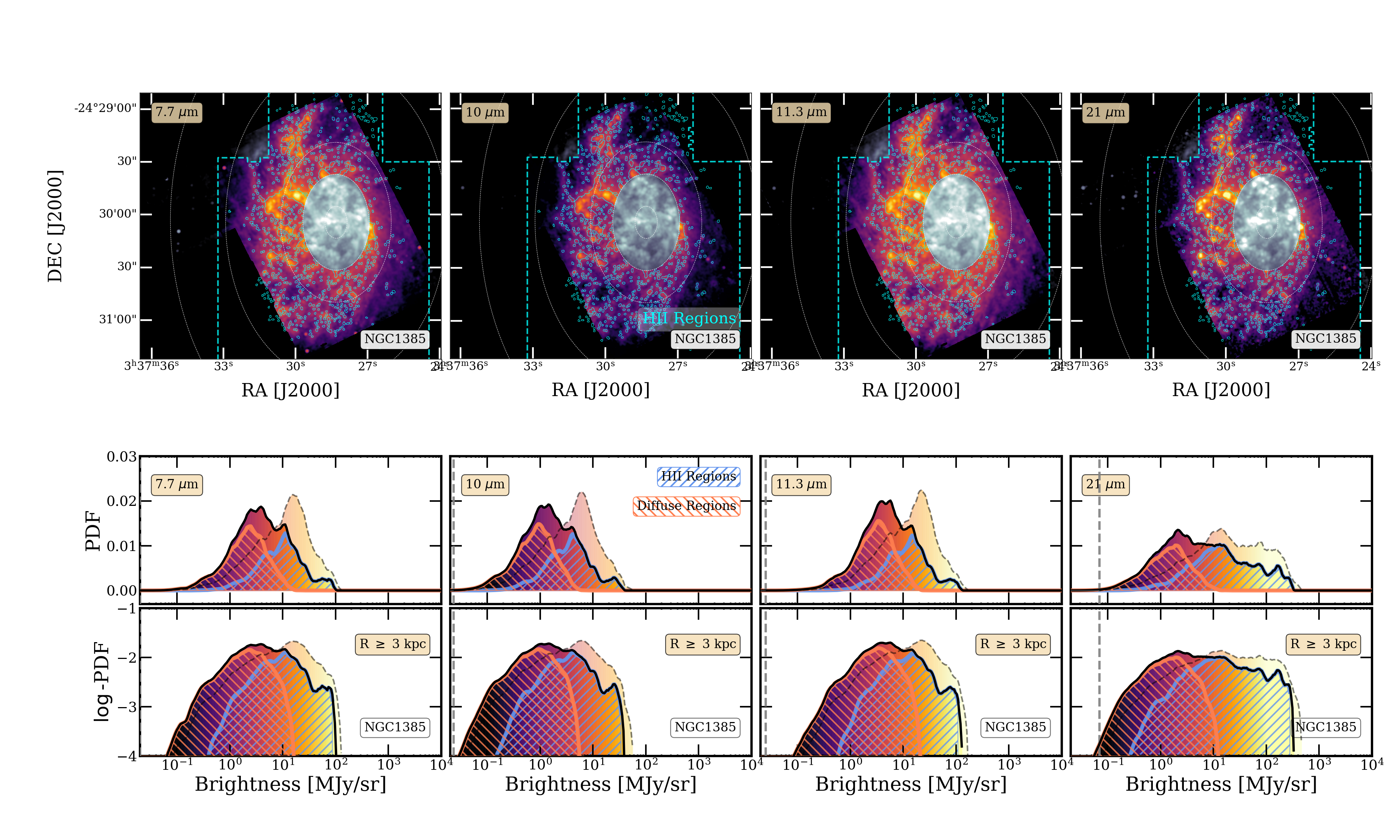} \label{fig:ngc1385}
    \caption{Mid-IR images and PDFs of emission of the galactic disk at 7.7~$\mu m$, 10~$\mu m$, 11.3~$\mu m$, and 21~$\mu m$ respectively for NGC1385.}
\end{figure*}

\begin{figure*}[htbp]
    \centering
    \includegraphics[width=1\textwidth]{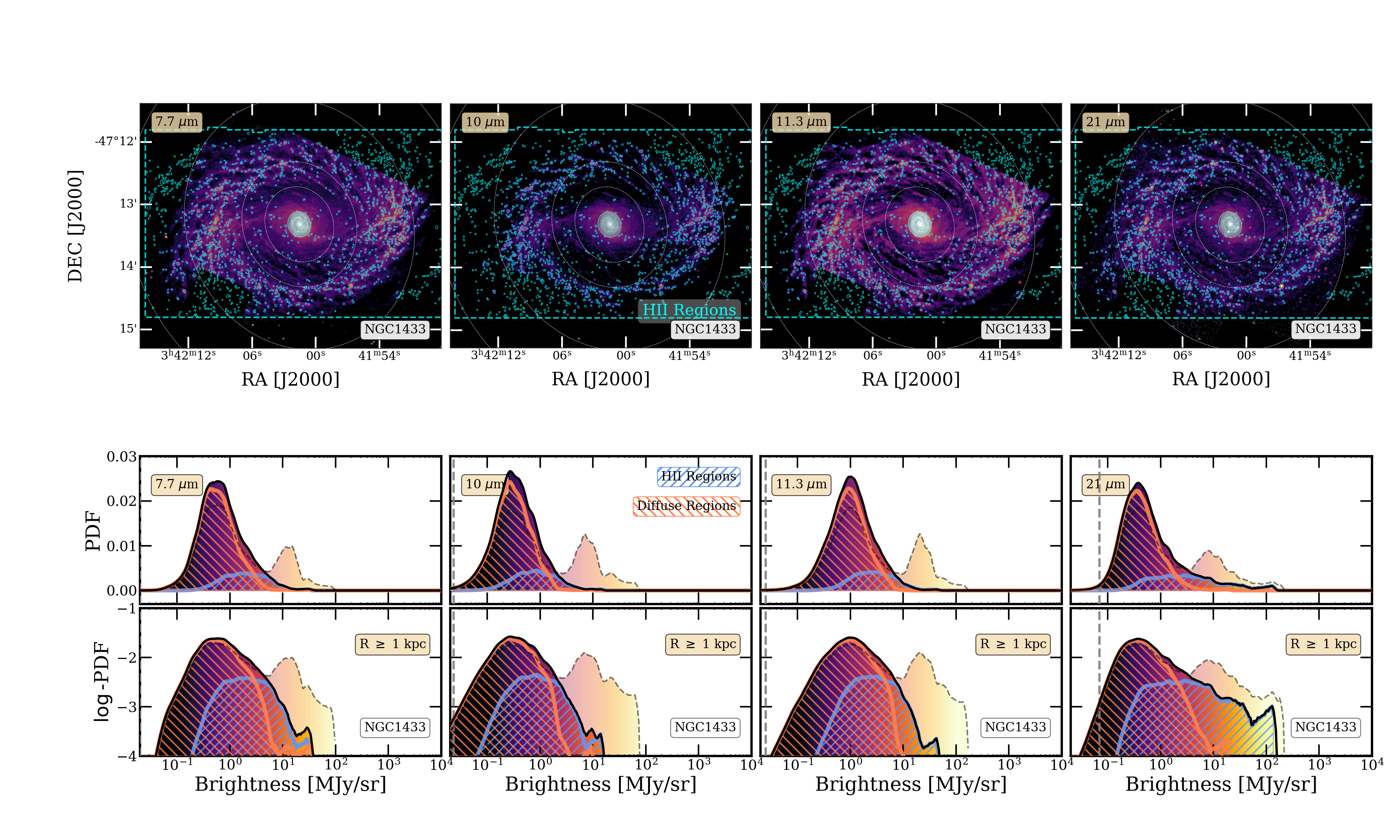}
    \caption{Mid-IR images and PDFs of emission of the galactic disk at 7.7~$\mu m$, 10~$\mu m$, 11.3~$\mu m$, and 21~$\mu m$ respectively for NGC1433.}
\end{figure*}

\begin{figure*}[htbp]
    \centering
    \includegraphics[width=1\textwidth]{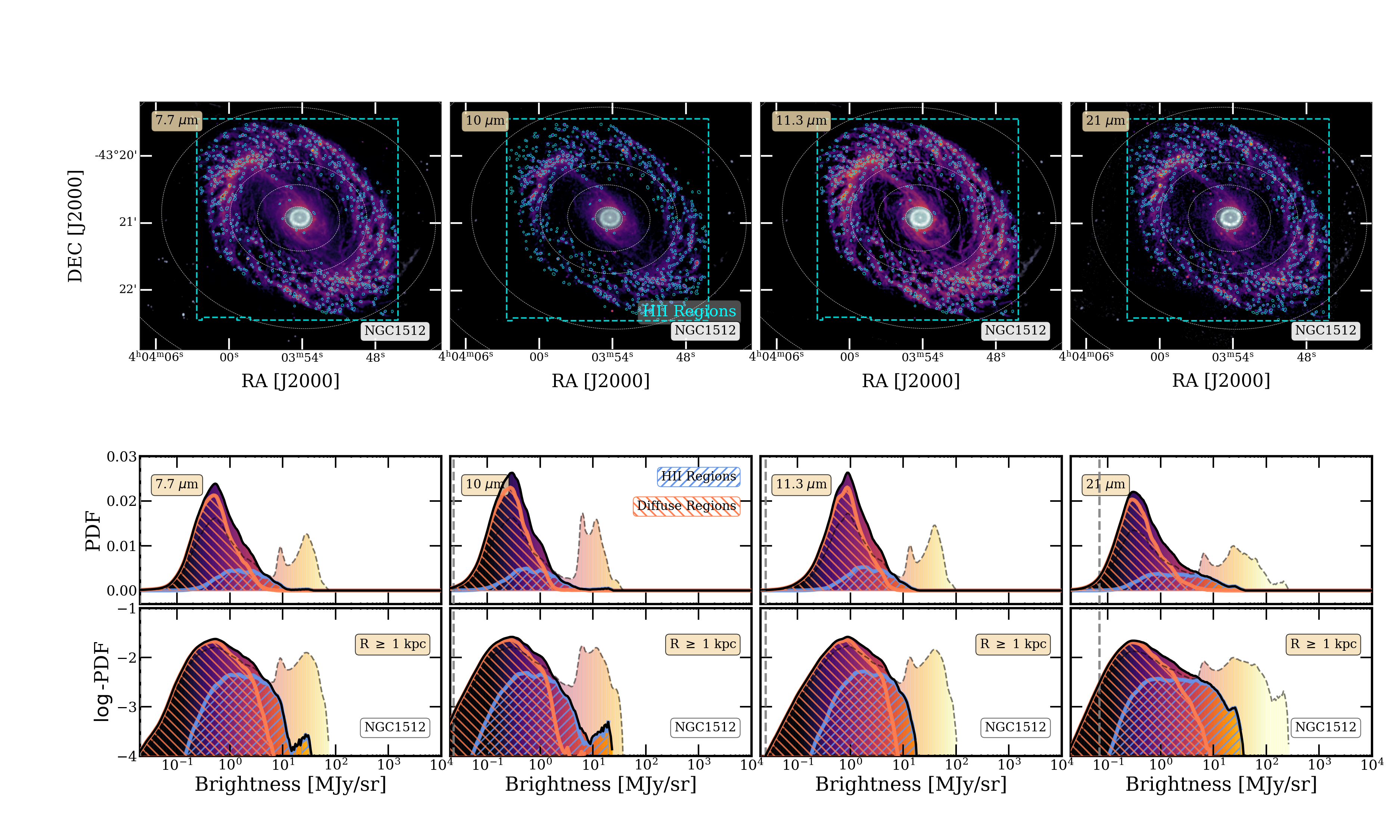}
    \caption{Mid-IR images and PDFs of emission of the galactic disk at 7.7~$\mu m$, 10~$\mu m$, 11.3~$\mu m$, and 21~$\mu m$ respectively for NGC1512.}
\end{figure*}

\begin{figure*}[htbp]
    \centering
    \includegraphics[width=1\textwidth]{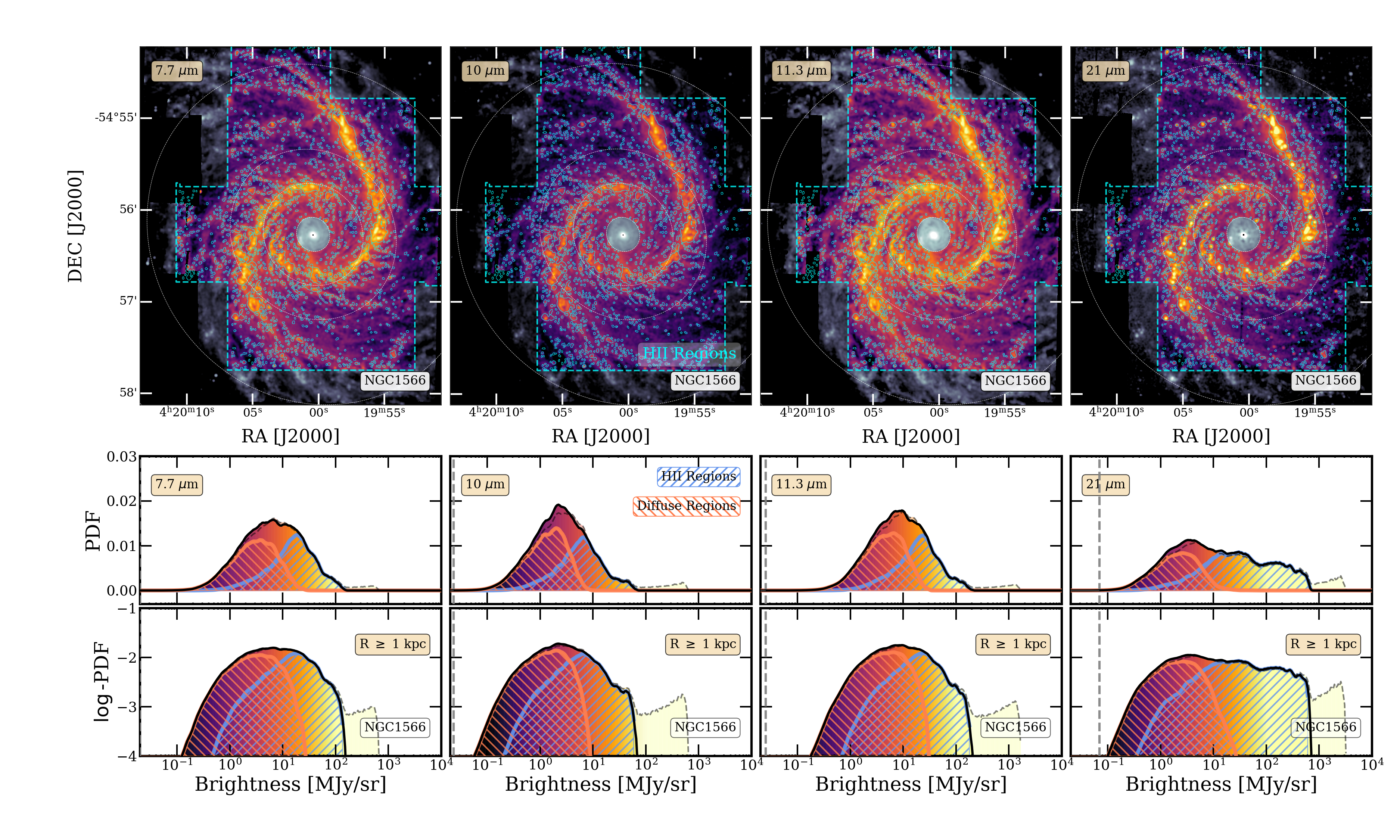}
    \caption{Mid-IR images and PDFs of emission of the galactic disk at 7.7~$\mu m$, 10~$\mu m$, 11.3~$\mu m$, and 21~$\mu m$ respectively for NGC1566.}
\end{figure*}

\begin{figure*}[htbp]
    \centering
    \includegraphics[width=1\textwidth]{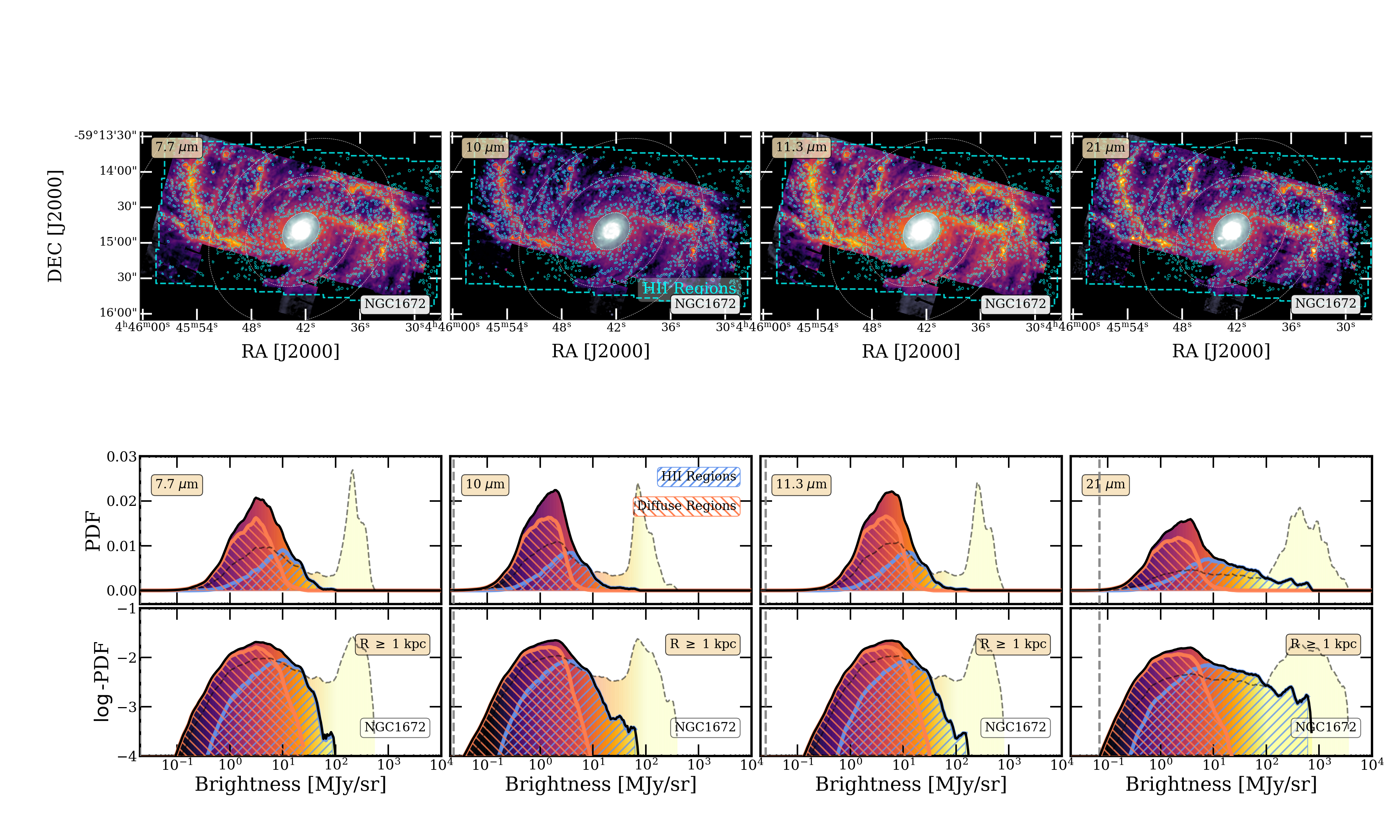}
    \caption{Mid-IR images and PDFs of emission of the galactic disk at 7.7~$\mu m$, 10~$\mu m$, 11.3~$\mu m$, and 21~$\mu m$ respectively for NGC1672.}
\end{figure*}

\begin{figure*}[htbp]
    \centering
    \includegraphics[width=1\textwidth]{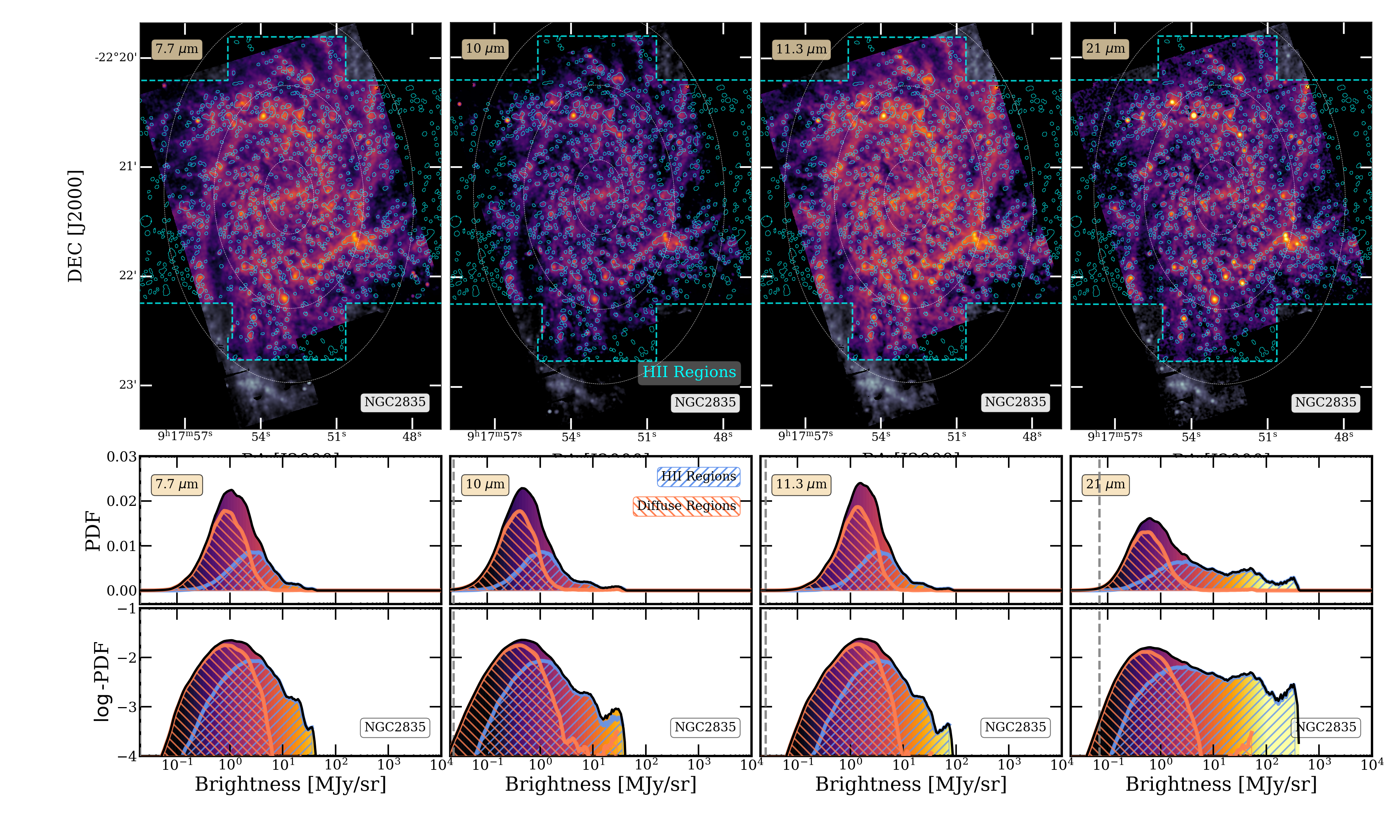}
    \caption{Mid-IR images and PDFs of emission of the galactic disk at 7.7~$\mu m$, 10~$\mu m$, 11.3~$\mu m$, and 21~$\mu m$ respectively for NGC2835.}
\end{figure*}

\begin{figure*}[htbp]
    \centering
    \includegraphics[width=1\textwidth]{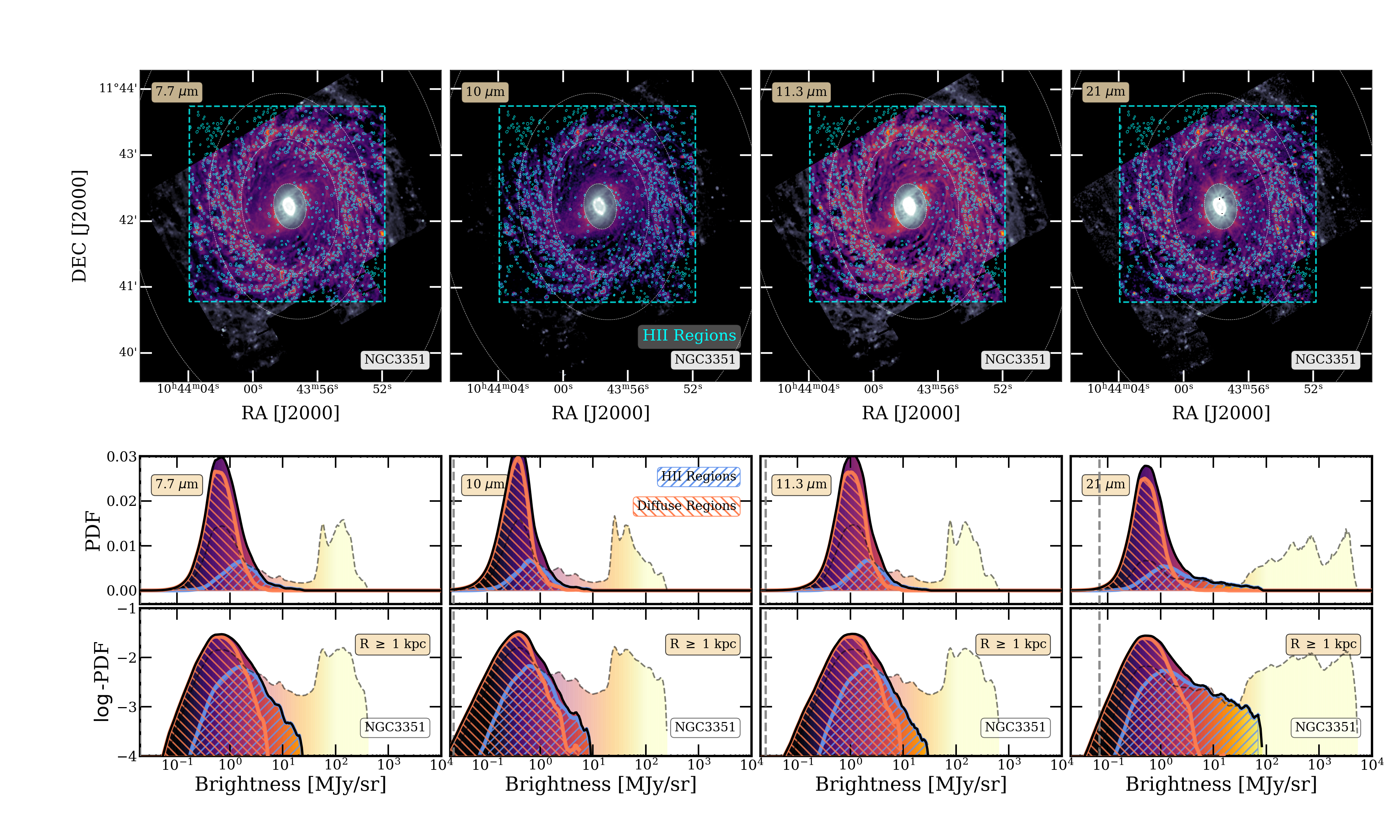}
    \caption{Mid-IR images and PDFs of emission of the galactic disk at 7.7~$\mu m$, 10~$\mu m$, 11.3~$\mu m$, and 21~$\mu m$ respectively for NGC3351.}
\end{figure*}

\begin{figure*}[htbp]
    \centering
    \includegraphics[width=1\textwidth]{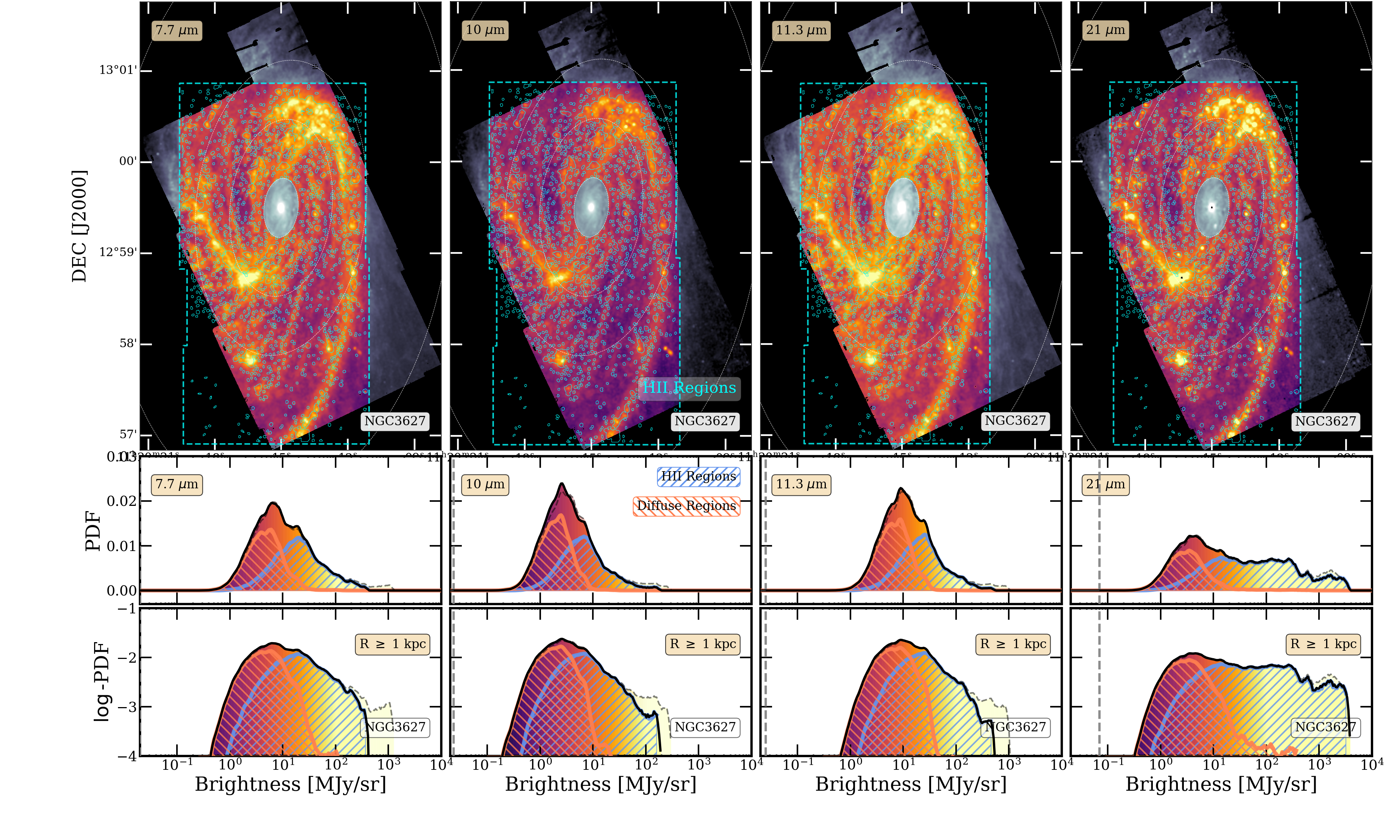}
    \caption{Mid-IR images and PDFs of emission of the galactic disk at 7.7~$\mu m$, 10~$\mu m$, 11.3~$\mu m$, and 21~$\mu m$ respectively for NGC3627.}
\end{figure*}

\begin{figure*}[htbp]
    \centering
    \includegraphics[width=1\textwidth]{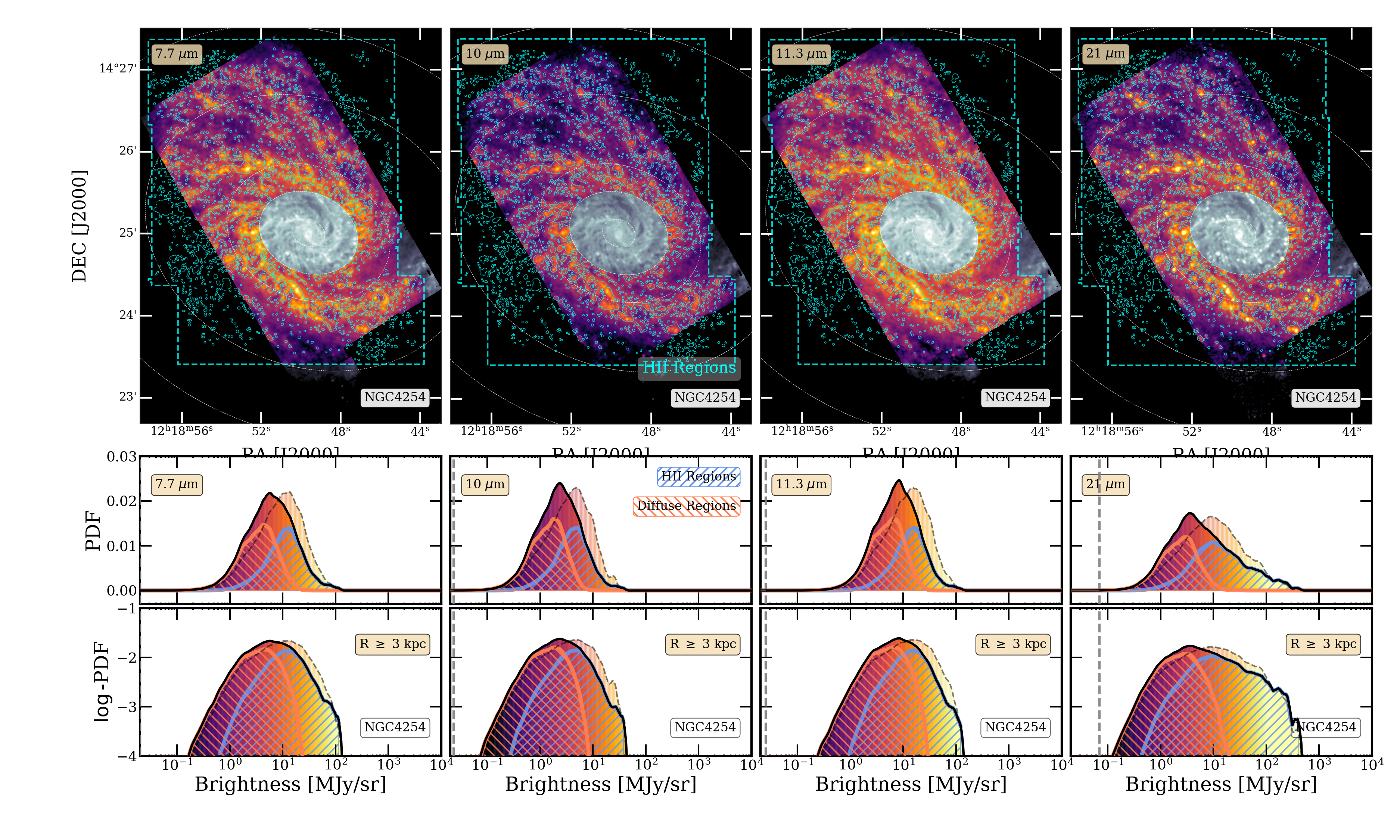}
    \caption{Mid-IR images and PDFs of emission of the galactic disk at 7.7~$\mu m$, 10~$\mu m$, 11.3~$\mu m$, and 21~$\mu m$ respectively for NGC4254.}
\end{figure*}

\begin{figure*}[htbp]
    \centering
    \includegraphics[width=1\textwidth]{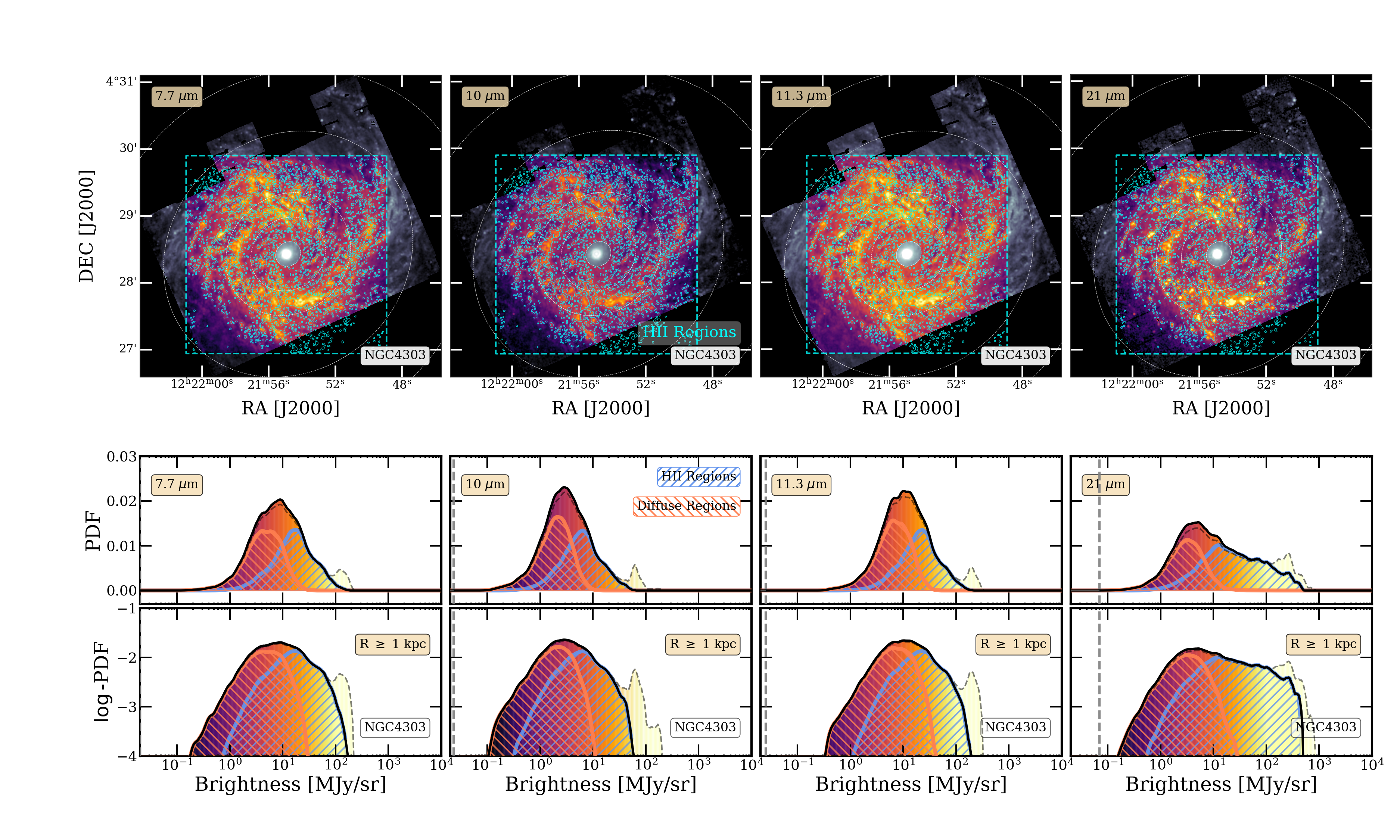}
    \caption{Mid-IR images and PDFs of emission of the galactic disk at 7.7~$\mu m$, 10~$\mu m$, 11.3~$\mu m$, and 21~$\mu m$ respectively for NGC4303.}
\end{figure*}

\begin{figure*}[htbp]
    \centering
    \includegraphics[width=1\textwidth]{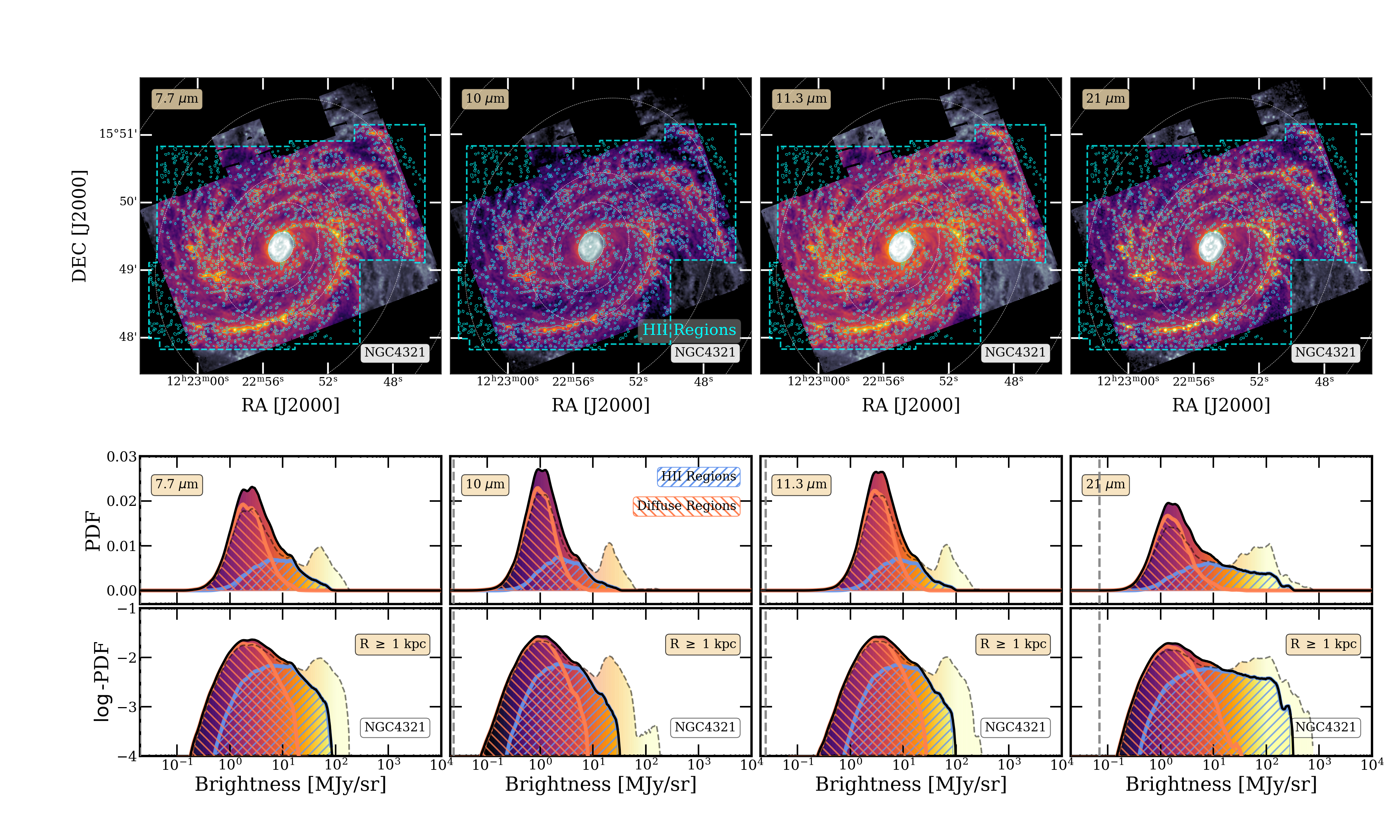}
    \caption{Mid-IR images and PDFs of emission of the galactic disk at 7.7~$\mu m$, 10~$\mu m$, 11.3~$\mu m$, and 21~$\mu m$ respectively for NGC4321.}
\end{figure*}

\begin{figure*}[htbp]
    \centering
    \includegraphics[width=1\textwidth]{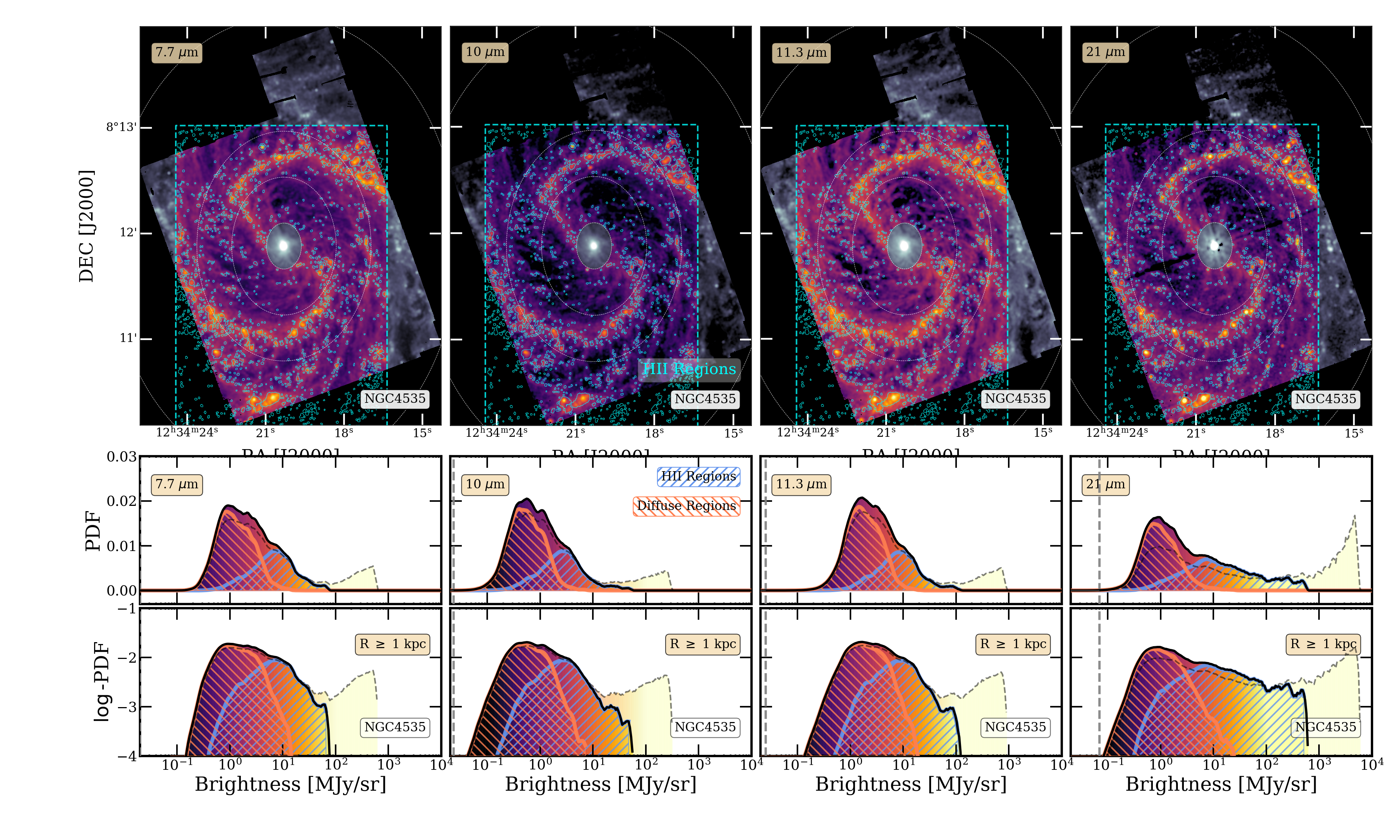}
    \caption{Mid-IR images and PDFs of emission of the galactic disk at 7.7~$\mu m$, 10~$\mu m$, 11.3~$\mu m$, and 21~$\mu m$ respectively for NGC4535.}
\end{figure*}

\begin{figure*}[htbp]
    \centering
    \includegraphics[width=1\textwidth]{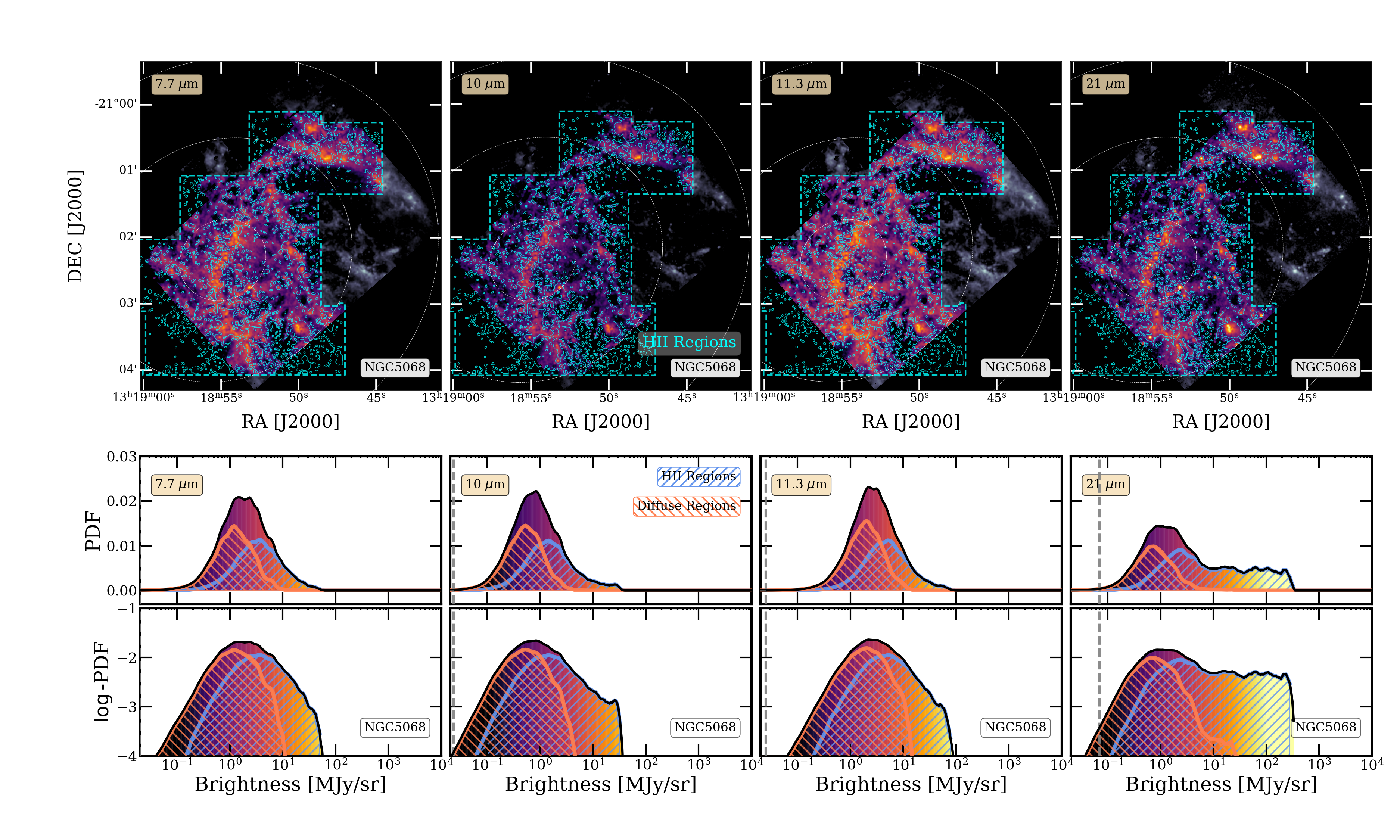}
    \caption{Mid-IR images and PDFs of emission of the galactic disk at 7.7~$\mu m$, 10~$\mu m$, 11.3~$\mu m$, and 21~$\mu m$ respectively for NGC5068.}
\end{figure*}

\begin{figure*}[htbp] \label{fig:ngc7496}
    \centering
    \includegraphics[width=1\textwidth]{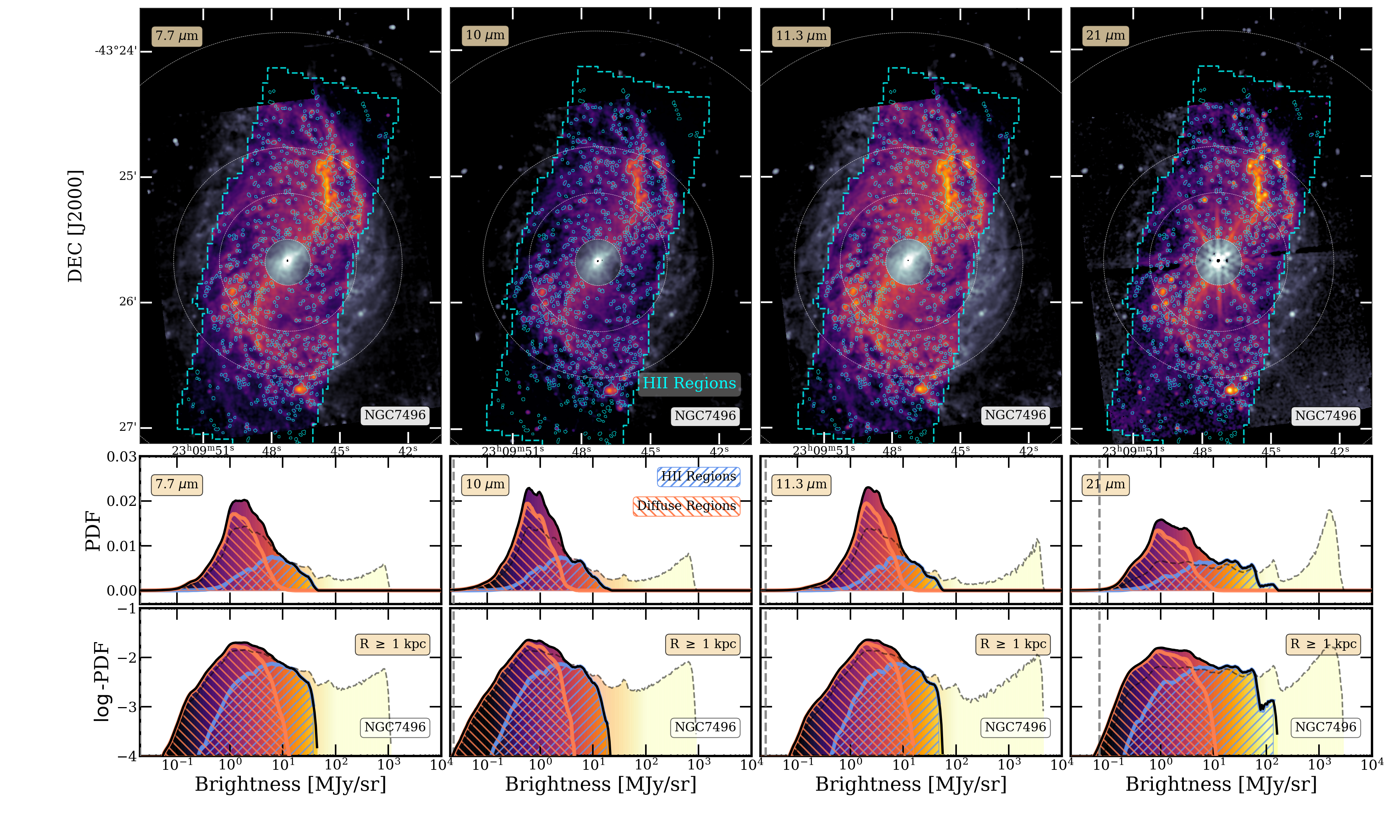} \label{fig:ngc7496}
    \caption{Mid-IR images and PDFs of emission of the galactic disk at 7.7~$\mu m$, 10~$\mu m$, 11.3~$\mu m$, and 21~$\mu m$ respectively for NGC7496.}
\end{figure*}

% \newpage

\section{Resolution and PDF Stability} \label{sec:resolution_stability}

This Appendix checks the stability of the diffuse component PDFs with change in resolution. We mask out the contributions from \ion{H}{2} regions at native F770W resolution ($0\farcs269$) and check how stable the log normal component is with change in resolution.
We compare the log normal mean $\mu$ and dispersion $\sigma$ across the images of diffuse emission convolved from the native F770W resolution to successively blurred Gaussian PSFs down to $8\farcs0$ with uniform $0\farcs02$ steps.

\begin{figure*}[htbp]
    \centering
    \includegraphics[width=0.85\textwidth]{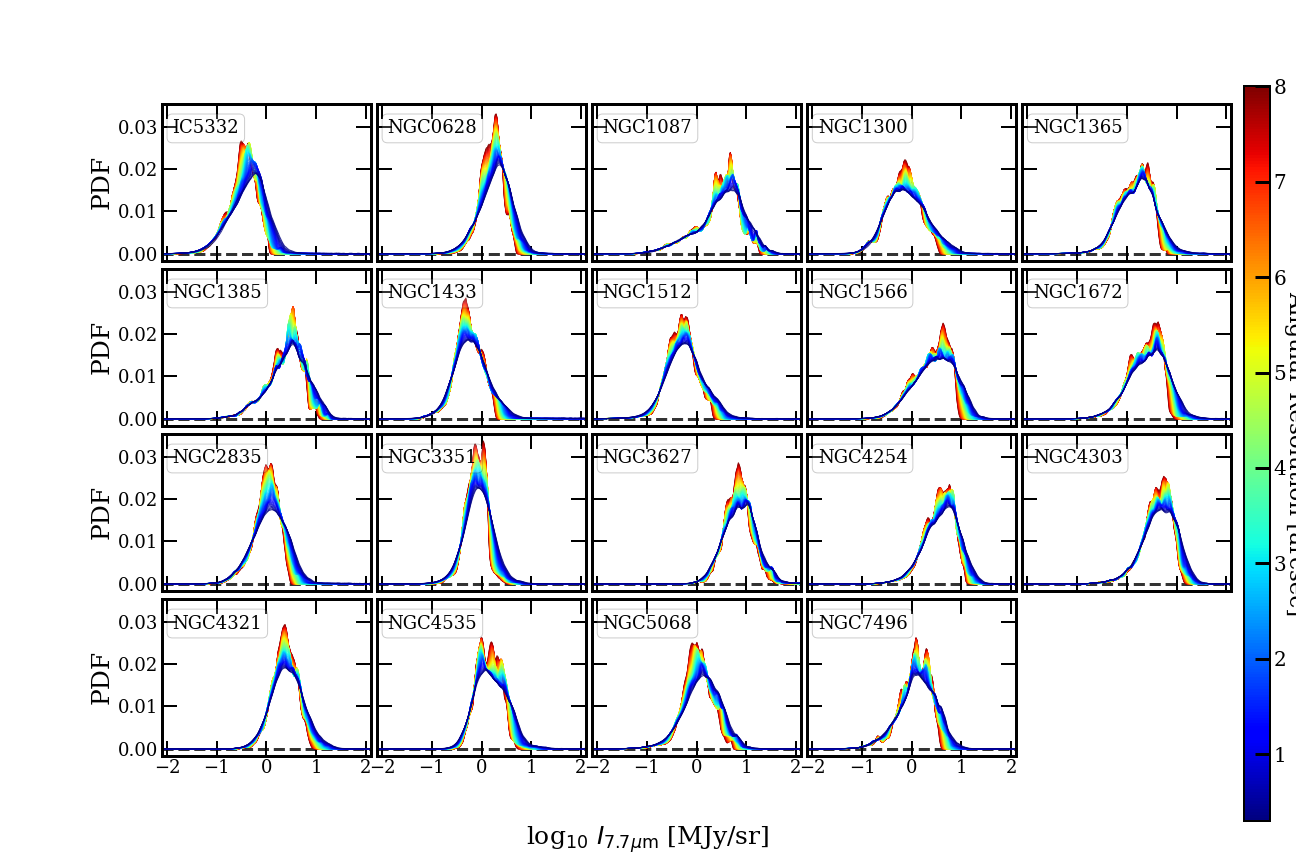}
    \caption{PDFs of $I_{\rm F770W}$ across different resolutions from native F770W ($0\farcs269$) down to $8\farcs0$ for each target. PDFs are colored by the FWHM of the Gaussian PSF at each resolution. With lower resolution, the overall PDFs become narrower and shift to lower intensities.}
    \label{fig:resolution_test_I_F770W}
\end{figure*}

\begin{figure*}[htbp]
    \centering
    \includegraphics[width=0.85\textwidth]{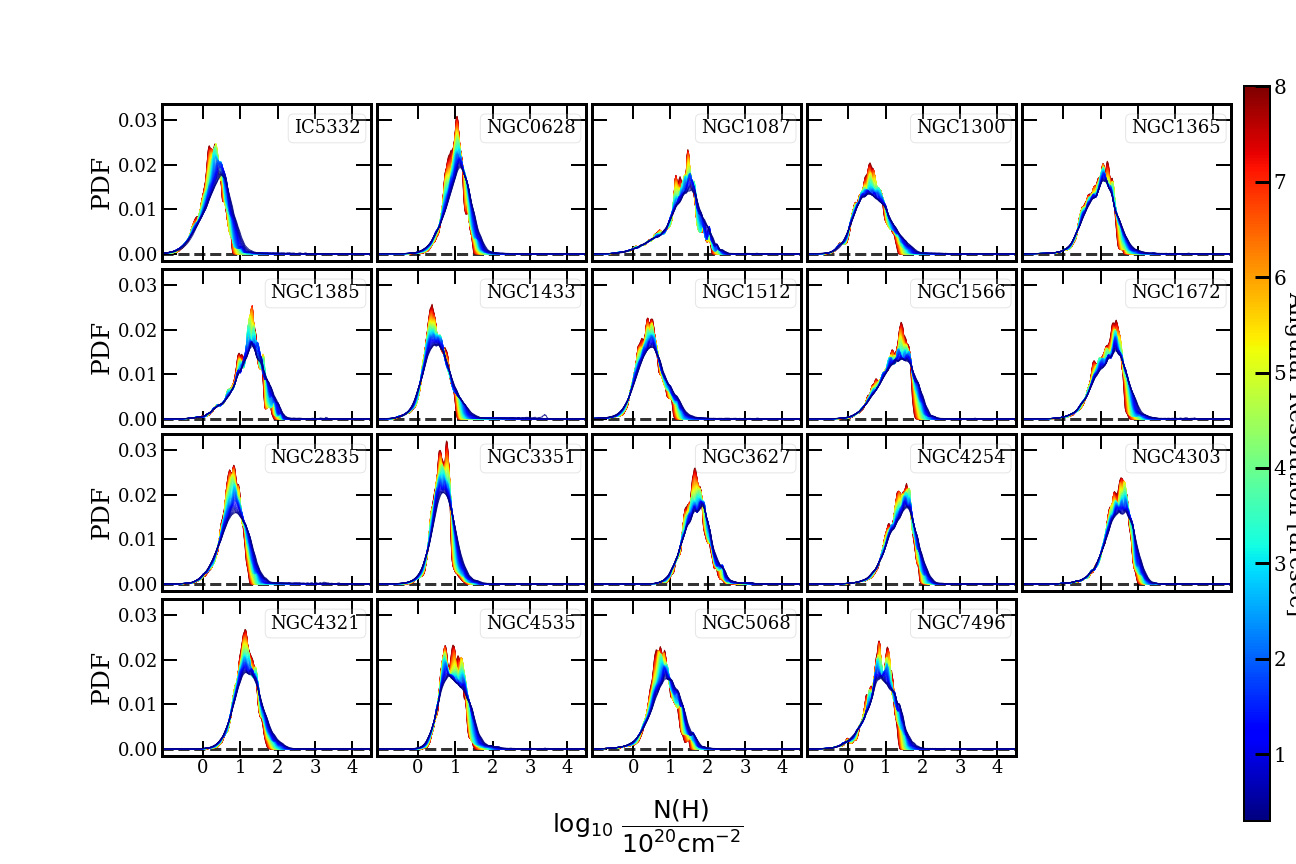}
    \caption{PDFs of $N(\rm H)$ across different resolutions from native F770W ($0\farcs269$) down to $8\farcs0$ for each target. PDFs are colored by the FWHM of the Gaussian PSF at each resolution. With lower resolution, the overall PDFs become narrower and shift to lower column densities.}
    \label{fig:resolution_test_NH}
\end{figure*}

\begin{figure*}[htbp]
    \centering
    \includegraphics[width=0.8\textwidth]{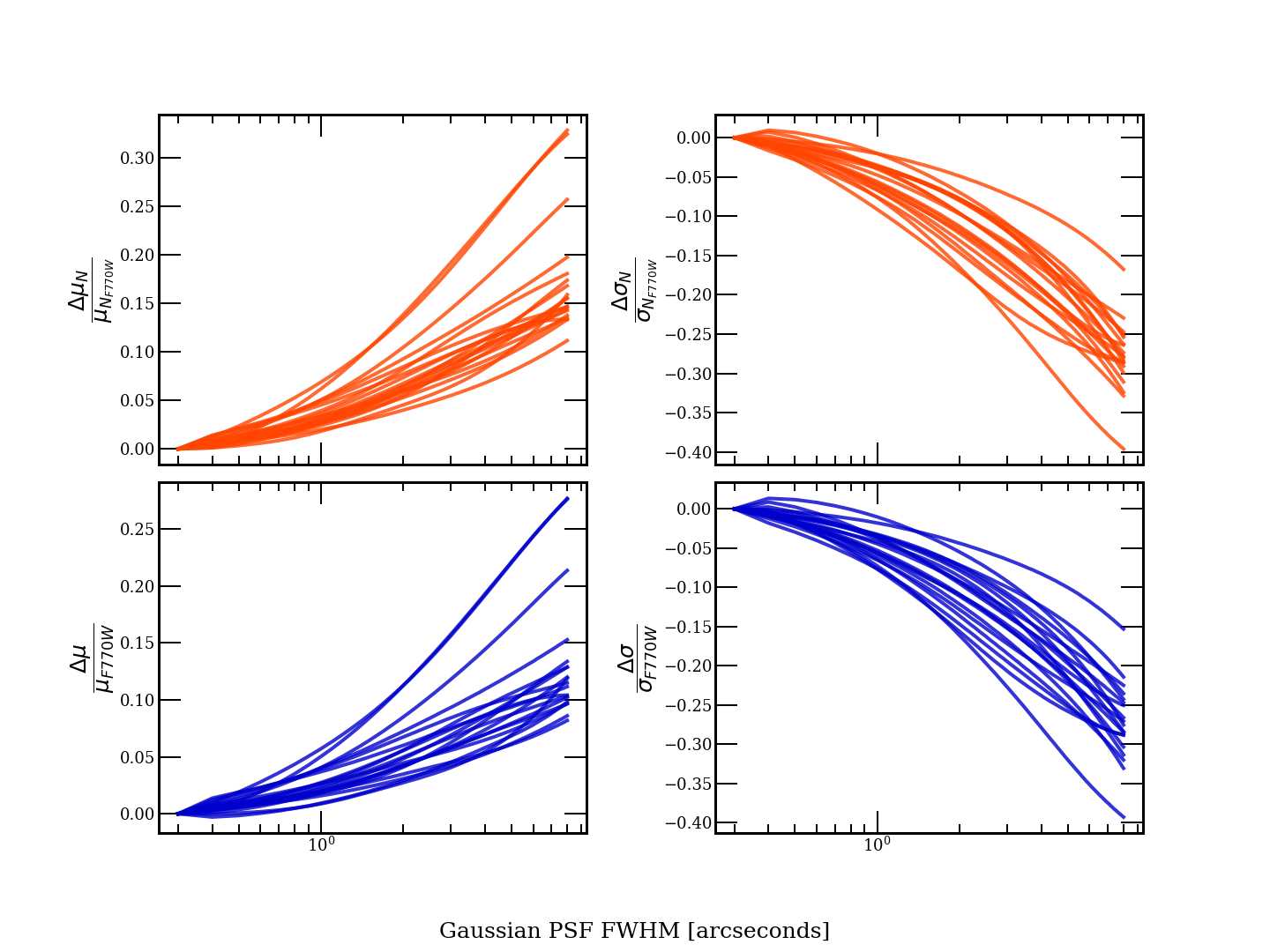}
    \caption{Summarizing the fractional variation in diffuse column density PDF mean ($\mu_N$) and dispersion ($\sigma_N$), and diffuse F770W emission PDF mean ($\mu$) and dispersion ($\sigma$).}
    \label{fig:resolution_change_params}
\end{figure*}

As seen in Figure \ref{fig:resolution_test_I_F770W}, while the diffuse intensity PDFs remain generally stable with changes in resolution down to $4\farcs0$, with large changes in resolution, the log normal component gradually becomes less smooth, the mean shifts to lower intensities, and the PDFs become narrower. Figure \ref{fig:resolution_test_NH} shows the same trend for the PDFs of $N(\rm H)$. 

Both PDFs only show $< 4\%$ change in log normal mean ($\mu,~\mu_N$) and $< 7\%$ change in log normal dispersion ($\sigma,~\sigma_N$) in the $\sim 0\farcs3-1\farcs0$ range, as summarized in Figure \ref{fig:resolution_change_params}. So our diffuse log normal PDFs of intensity and gas column density remain reliably stable over the $0\farcs269-0\farcs85$ resolution range where we present results. 

% \bibliographystyle{aasjournal}

\suppressAffiliationsfalse
\allauthors


\end{document}